\documentclass[12pt]{article}
\pdfoutput=1

\usepackage{siunitx}
\usepackage{booktabs}
\usepackage{putex}
\usepackage{graphicx}
\usepackage{caption}
\usepackage{amsmath}
\usepackage{amssymb}
\usepackage{array}
\usepackage{bm}
\usepackage{multirow}
\usepackage{mathtools}
\usepackage{comment}
\usepackage{subcaption}
\usepackage{epstopdf}
\usepackage{enumerate}
\usepackage{cite}
\usepackage{youngtab}
\usepackage{tensor}
\usepackage{slashed}
\usepackage[aligntableaux=center]{ytableau}
\usepackage[utf8]{inputenc}
\usepackage{rotating}
\usepackage{bigfoot}
\usepackage[
      colorlinks=true,
      linkcolor=blue,
      urlcolor=blue,
      filecolor=black,
      citecolor=red,
      linktocpage=true
      ]{hyperref}
\usepackage{dsfont}

\newcommand{\abs}[1]{\left\lvert #1 \right\rvert}
\def\Tr{\mop{Tr}}
\newcommand {\be} {\begin {equation}}
\newcommand {\ee} {\end {equation}}

\newcommand {\bes} {\begin {equation*}}
\newcommand {\ees} {\end {equation*}}

\newcommand{\es}[2]{%
  \begin{equation}
    \begin{aligned}
      #2
    \end{aligned}
    \phantomsection\label{#1}%
  \end{equation}%
}

\newcommand{\CP}{\mathbb{CP}}
\newcommand{\Z}{\mathbb{Z}}

\newcommand{\R}{\mathbb{R}}

\newcommand{\cA}{{\mathcal A}}
\newcommand{\cB}{{\mathcal B}}

\newcommand{\bea}{\begin{equation}\begin{aligned}}
\newcommand{\eea}[1]{\label{#1}\end{aligned}\end{equation}}

\newcommand{\beq}{\begin{equation}}
\newcommand{\eeq}{\end{equation}}

\newcommand{\ov}{\over}
\newcommand{\Det}{{\rm Det}}
\def\le{\left}
\def\ri{\right}

\def\ie{\begin{equation}\begin{aligned}}
\def\fe{\end{aligned}\end{equation}}

\numberwithin{equation}{section}

\def\<{\langle}
\def\>{\rangle}

\usepackage{tikz}
\usepackage[compat=1.1.0]{tikz-feynman}
\usetikzlibrary{decorations.pathmorphing, decorations.markings}
\usepackage{subcaption}

\begin{document}

\preprint{}

\institution{imperial}{Abdus Salam Centre for Theoretical Physics, Imperial College London, London SW7 2AZ, UK}
\institution{oxford}{Departement de physique, Universite de Montreal, Montreal (Quebec), H3C 3J7, Canada }

\title{ Monopoles, duality, and energy splitting in Chern-Simons QED$_3$}

\authors{Shai M.~Chester\worksat{\imperial} and \'Eric Dupuis\worksat{\oxford}}

\abstract{
We compute the scaling dimensions of monopole operators in QED$_3$ with $N$ 2-component complex fermions and Chern-Simons level $k$ to subleading order in the limit where $k,N$ are large and $k/N$ is fixed, which generalizes previous results for scalar QED$_3$ (sQED$_3$) coupled to $N$ complex scalars. In both cases, at leading large $N$ many monopole operators are degenerate, and we compute their energy splitting. We use these results to check the duality between QED$_3$ with $N=1,k=-3/2$ and sQED$_3$ with $N=1,k=2$, whose $U(1)$ monopole symmetry is enhanced to $SO(3)$. We find the scaling dimensions in each dual theory match to high precision for all charges, the lowest charge spin 1 monopoles are close to 2 as expected for the emergent $SO(3)$ current, and the degenerate second lowest charge monopoles match a prediction from a fuzzy sphere calculation, including the energy splitting. We also find similar evidence for dualities between QED$_3$ with $N=1,k=-(2m+1)/2$ and one scalar coupled to two $U(1)$ gauge fields, which has an effective description as sQED$_3$ with $N=1,k=\frac{m+1}{m}$ and $q_s=mq_f$ for $m>1$. Finally, we also find similar evidence for Son's duality between QED$_3$ with $N=1,k=0$ and a free fermion, and clarify a previously discussed match between sQED$_3$ with $N=k=1$ dual to a free fermion.
}
\date{}

\maketitle

\tableofcontents
\newpage

\section{Introduction}\label{sec:intro}

When two quantum field theories flow to the same conformal field theory (CFT) in the IR, we call them IR dual. In 2d such dualities are common, but for $d>2$ this phenomenon was first proposed in the high energy community back in the nineties for 4d $\mathcal{N}=1$ gauge theories \cite{Seiberg:1994pq}, and then 3d $\mathcal{N}=2$ gauge theories \cite{Aharony:1997gp}. These dualities could be checked using observables like the sphere partition functions that can be computed exactly using supersymmetric localization in 4d \cite{Pestun:2007rz} and 3d \cite{Kapustin:2010mh}. However, the first $d>2$ duality actually predates these dualities by two decades, and arose from the condensed matter community. The so-called particle-vortex duality relates scalar quantum electrodynamics in 3d (sQED$_3$) with one complex scalar, to the critical $O(2)$ model \cite{Peskin:1977kp,PhysRevLett.47.1556}. 

In particle-vortex duality, charge $q$ operators in the critical $O(2)$ model, which can be formed from fields in the Lagrangian in the standard way, are mapped to charge $q$ monopole operators in sQED$_3$. These monopole operators are not formed from fields in the gauge theory Lagrangian, but instead are charged under a topological global $U(1)$ symmetry whose conserved current and charges are 
\es{topDef}{
j^\mu=\frac{1}{8\pi}\epsilon^{\mu\nu\rho}F_{\nu\rho}\,, \qquad q=\frac{1}{4\pi}\int_{\Sigma}F \,,
}
where $F_{\nu\rho}$ is the $U(1)$ gauge field strength\footnote{Not to be confused with the topological $U(1)$ global symmetry.}, Greek indices are spacetime indices, $q$ is restricted by Dirac quantization to values $q\in\mathbb{Z}/2$, and $\Sigma$ is a closed two-dimensional surface. The current $j^\mu$ is conserved due to the Bianchi identity, which makes local monopole operators special to 3d. The particle vortex duality was verified by simulating each theory on the lattice, and finding similar scaling dimensions for operators with $q=0,1,2,3$ \cite{Kajantie:2004vy,Karthik:2018rcg,Hasenbusch:2019jkj}.\footnote{In fact, \cite{Kajantie:2004vy,Karthik:2018rcg} only studied the $\mathbb{R}$ gauge theory, instead of $U(1)$ gauge theory, where monopole operators are not local operators. The earlier lattice study \cite{Kajantie:2004vy} thus only computed the $q=0$ non-monopole operator, while \cite{Karthik:2018rcg} studied non-local operators in $\mathbb{R}$ gauge theory that can be used to estimate the scaling dimension of local monopole operators in the $U(1)$ gauge theory.}

More recently, other non-supersymmetric dualities in 3d were proposed where the matter fields are fermions on one side, and scalars on the other side \cite{PhysRevB.89.235116,PhysRevB.48.13749}, and are thus called 3d bosonization. For instance, consider the duality \cite{Seiberg:2016gmd,Karch:2016sxi}\footnote{In the introduction, we work in Lorentzian signature for simplicity, while in the main text we will use Euclidean signature.}
\es{seed}{
\mathcal{L}_\text{QED$_3$}[\psi,B]-\frac{BdB}{8\pi} \qquad\Leftrightarrow\qquad \mathcal{L}_\text{sQED$_3$}[\phi,a]+\frac{ada}{4\pi}+\frac{adB}{2\pi}\,,
}
where $B$ is a background field, $a$ is a $U(1)$ gauge field, $\phi$ is a complex scalar field, and $\psi$ is a complex 2-component fermion. When the gauge fields $B$ in $\mathcal{L}_\text{QED$_3$}[\psi,B]$ and $\mathcal{L}_\text{sQED$_3$}[\phi,B]$ are non-dynamical, these Lagrangians correspond to the free fermion and critical $O(2)$ theories, respectively. The LHS of \eqref{seed} is thus a single free fermion, while the RHS is sQED$^{N,k}_3$ with Chern-Simons (CS) level $k=1$ coupled to $N=1$ complex scalar.

Starting from this seed duality, one can ``derive''\footnote{These derivations only apply to the UV theory, so they do not rigorously imply dualities for the IR theory, but they are nonetheless suggestive.} other dualities by gauging symmetries and adding CS terms to both sides \cite{Seiberg:2016gmd,Karch:2016sxi}. For instance, we can add a $-m$ CS term to each side of \eqref{seed}, promote $B$ to a dynamical gauge field $b$, and then introduce a new background field $C$ via a new BF term, to get \cite{Karch:2016aux}:
\es{mdual1}{
\mathcal{L}_\text{QED$_3$}[\psi,b]-\Big(m+\frac12\Big)\frac{bdb}{4\pi} +\frac{bdC}{2\pi}\quad\Leftrightarrow\quad \mathcal{L}_\text{sQED$_3$}[\phi,a]+\frac{ada}{4\pi}-m\frac{bdb}{4\pi}+\frac{adb}{2\pi}+\frac{bdC}{2\pi}\,.
}
The LHS is now QED$^{1,-m-1/2}_3$,\footnote{In the main text, we will define more carefully what we mean by the CS level for QED$_3$.} while the RHS is an Abelian quiver gauge theory with a nontrivial CS matrix coupled to one scalar. The equation of motion of the RHS for $b$ is
\es{EOM}{
da-mdb+dC=0\,.
}
For $m=0$, this makes $a$ non-dynamical, so we get the duality between QED$^{1,-1/2}_3$ and the critical $O(2)$ model \cite{PhysRevB.89.235116,PhysRevB.48.13749}:
\es{m0intro}{
\mathcal{L}_\text{QED$_3$}[\psi,b]-\frac12\frac{bdb}{4\pi} +\frac{bdC}{2\pi}\quad\Leftrightarrow\quad \mathcal{L}_\text{sQED$_3$}[\phi,C]+\frac{CdC}{4\pi}\,.
}
 For $m>0$, we can naively integrate out $b$ in the RHS using its equation of motion to get
\es{local}{
\mathcal{L}_\text{sQED$_3$}[\phi,a]+\frac{m+1}{m}\frac{ada}{4\pi}+\frac{1}{m}\frac{adC}{2\pi}+\frac{1}{2m}\frac{CdC}{2\pi}\,,
}
which is sQED$^{1,\frac{m+1}{m}}_3$. From \eqref{EOM} and the definition of the monopole charge in \eqref{topDef}, we see that the fermionic charge $q_f$ is related to the scalar charge $q_s$ here as $q_s=mq_f$.

Since $k$ must be an integer for sQED$_3$, for $m>1$ we should view \eqref{local} as an effective description that may be useful for computing certain local data such as monopole operator scaling dimensions, but to properly treat non-local operators one requires the original quiver formulation in \eqref{mdual1}. For $m=1$, however, \eqref{local} is perfectly well defined, and is in fact one of the original dualities proposed in \cite{Aharony:2016jvv,Benini:2017dus}. This theory was conjectured to furthermore be dual to an $SU(2)$ gauge theory with $k=1$ coupled to one scalar, and an $SU(2)$ gauge theory with $k=-1/2$ coupled to one fermion. These QCD$_3$ descriptions have an explicit $SO(3)$ symmetry, which implies that the $U(1)$ topological symmetry in QED$^{1,-3/2}_3$ and sQED$^{1,2}_3$ must enhance to $SO(3)$ in the IR, such that the lowest $q=1/2$ monopole becomes a conserved current with $\Delta=2$. This theory has been intensely studied in the condensed matter community. Notably, \cite{Zhou:2025rmv} applied the fuzzy sphere method to the transition between a $\nu=2$ fermionic integer quantum Hall state and a $\nu=1/2$ bosonic fractional quantum Hall state, which has an explicit $SO(3)$ symmetry, and which they conjecture is dual to this $m=1$ theory. They computed the low lying spectrum for $q=0,1/2,1$, and found the lowest two $q=1$ states to be\footnote{The lowest $q=1/2$ state is just the aforementioned $SO(3)$ current.}
\es{fuzzy}{
\Delta_{q=1}^\text{spin $0$}=5.1079\,,\qquad \Delta_{q=1}^\text{spin $2$}=5.3435\,,
}
where the spacetime spins $0,2$ are the values expected from the explicit $q=1$ monopole construction in \cite{Chester:2017vdh}. 

We can obtain another duality by promoting $C$ in the $m=0$ duality \eqref{m0intro} to a dynamical gauge field $a$ and adding $-\frac{ada}{2\pi}+\frac{adB}{2\pi}$ to both sides, so that the sQED$_3$ side is now the time reversed LHS of the seed duality \eqref{seed}, and we thus find \cite{Seiberg:2016gmd}
\es{son}{
\mathcal{L}_\text{QED$_3$}[\psi,b]-\frac12\frac{bdb}{4\pi} +\frac{bda}{2\pi}-\frac{ada}{2\pi}+\frac{adB}{2\pi}\quad\Leftrightarrow\quad \mathcal{L}_\text{QED$_3$}[\psi,B]+\frac{BdB}{8\pi}\,.
}
We can naively integrate out $a$ on the LHS to get QED$^{1,0}_3$, where the EOM of $a$ restricts $q$ to be an integer, and thereby avoids the parity anomaly for half-integer $q$. This is the correct formulation of Son's original duality between QED$^{1,0}_3$ and a free fermion \cite{Son:2015xqa}. As with the $m>1$ dualities, we should view  QED$^{1,0}_3$ as an effective description that is appropriate for computing monopole scaling dimensions. 

It is hard to find dynamical evidence for these non-supersymmetric 3d bosonization dualities, such as direct comparison of CFT data on both sides, since at least one side is strongly coupled. While particle-vortex could be verified with a lattice simulation, that is no longer available here due to CS terms that cause a sign problem. The aforementioned fuzzy sphere calculation \cite{Zhou:2025rmv} may provide a non-perturbative window on QED$^{1,-3/2}_3$, but that assumes that the model they study (which is not a gauge theory) happens to be in the same universality class as QED$_3$, and so cannot test the duality. Instead, these dualities were originally motivated by comparing kinematic consistency checks such as 't Hooft anomalies \cite{Seiberg:2016gmd,Karch:2016sxi,Karch:2016aux}. 

Recently, a new method for dynamically checking dualities of 3d gauge theories was proposed in \cite{Chester:2022wur}. Consider QED$_3$ with $N$ flavors of scalars or fermions and CS level $k$ in the limit $N,k$ large with $\kappa=k/N$ fixed. The scaling dimension of monopoles can be computed in this limit via a saddle point analysis. This calculation was first carried out at leading order for $k=0$ for QED$_3$ \cite{Borokhov:2002ib} and sQED$_3$ \cite{Murthy:1989ps}, then generalized to subleading order at $k=0$ for QED$_3$ \cite{Pufu:2013vpa} and sQED$_3$ \cite{Dyer:2015zha}, then at leading order for both QED$_3$ and sQED$_3$ with $k\neq0$ \cite{Chester:2017vdh}, and finally at subleading order for $k\neq0$ sQED$_3$ \cite{Chester:2021drl,Chester:2022wur}.\footnote{Related studies were also carried out for supersymmetric monopoles \cite{Borokhov:2002cg,Klebanov:2011td}, monopoles in $k=0$ QCD$_3$ \cite{Dyer:2013fja}, monopoles in QED$_3$ with a four-fermion potential \cite{Dupuis:2021flq,Dupuis:2019uhs,Dupuis:2019xdo,Dupuis:2021yej}, monopoles in QED$_3$ in a $4-\epsilon$ expansion \cite{Chester:2015wao}, and monopoles in the large $q$ limit \cite{Dumitrescu:2025vfp,DeLaFuente:2018uee}.} Note that the subleading calculation in \cite{Chester:2021drl,Chester:2022wur} has a relative sign error between the CS and matter kernel contributions at subleading order, which we correct in the results reported here.

 It was first observed in \cite{Dyer:2015zha} that the subleading results for sQED$^{N,0}_3$ monopole scaling dimensions matched lattice simulations \cite{2013PhRvB..88v0408H,2012PhRvL.108m7201K} quite accurately even for small $N=2,3,4,5,6$. Then \cite{Chester:2022wur} took this a step further, by showing that the sQED$^{N,0}_3$ results even for $N=1$ matched the dual critical $O(2)$ scaling dimensions, as computed previously from lattice \cite{Banerjee:2017fcx} and conformal bootstrap \cite{Chester:2019ifh}. Further evidence was found for this claim by comparing to bootstrap results for sQED$^{N,0}_3$ for $N=2$ \cite{Chester:2023njo} and $N=3$ \cite{Chester:2025uxb}, bootstrap results for QED$^{N,0}_3$ for $N=4$ \cite{Chester:2016wrc,Albayrak:2021xtd}, and a new conjecture for the QED$^{2,0}_3$ phase diagram \cite{Chester:2024waw}.

In this work, we generalize the subleading large $N,k$ calculation from sQED$_3$ to QED$_3$. In particular, we compute the free energy on $S^2\times \mathbb{R}$ with $4\pi q$ magnetic flux threading the $S^2$ to subleading order at large $N,k$ by expanding around the saddle point configuration. The state-operator correspondence then identifies this energy with the scaling dimension of a charge $q$ monopole on $\mathbb{R}^3$. The fermion zero modes and the CS term each induce a gauge charge proportional to $q$, so that the naive $S^2 \times \mathbb{R}$ vacuum must be dressed by charged matter modes for $\kappa\neq0$ sQED$_3$ and $|\kappa|\neq1/2$ QED$_3$. As in \cite{Chester:2017vdh}, we can enforce this dressing by computing the small temperature $T =\beta^{-1}$ limit of the thermal free energy on $S^2\times S^1_\beta$, where the saddle point value of the holonomy of the gauge field on $S^1_\beta$ acts like a chemical potential for the matter fields. For sQED$_3$, one can always dress with the lowest energy mode for any $\kappa$, but for QED$_3$ one must use modes from progressively bigger Landau levels as $\kappa$ increases. Nevertheless, we find a compact formula for the subleading scaling dimension for all $\kappa$, which depends on the number of filled Landau levels, as well as the filling fraction of the valence level. The result is written as an infinite sum and integral, which can be computed numerically with an appropriate cutoff to achieve a desired accuracy.

To subleading order the above calculation applies to any monopole formed by dressing the vacuum with the necessary charged modes, and the number of such degenerate states generally grows exponentially with $N$. These monopoles transform with different spins and $SU(N)$ flavor symmetry irreps, which correspond to how one contracts the indices of the matter dressing. This large $N$ degeneracy can be seen from a $\beta$-dependent entropy term at leading order, as well as $\log\beta$ terms at subleading order \cite{Chester:2017vdh}. These subleading $\log\beta$ terms are expected to be related to a state-dependent contribution to the scaling dimension, which was worked out for the simplest case where degenerate monopoles are uniquely labeled by their spacetime spin. In this paper, we show how to compute the energy splitting for the most general case, where degenerate monopoles of a given spin can have different $SU(N)$ flavor irreps, and a given spin and flavor irrep can occur more than once. The energy splitting term is obtained by diagonalizing a Hamiltonian formed by oscillators corresponding to the matter modes used for dressing, where the size of the Hamiltonian grows with $q$.

The saddle point equation described above for both QED$_3$ and sQED$_3$ involved a saddle that is symmetric under the $S^2$ isometry $SO(3)$. For QED$_3$ with $|k|>1/2$, which is when one must start dressing with excited modes, this spherically symmetric saddle is no longer stable at zero temperature unless the excited Landau level is completely filled. In particular, one can find an imaginary term at subleading order in $N$ for sufficiently large $\beta$. In this case, we instead consider non-spherically symmetric saddles, which only preserve a subgroup of $SO(3)$, and turn out to be stable and give a lower energy at large $N$ than the spherical saddle. While a given non-spherical saddle only corresponds to one state, there are several non-spherical saddles that give degenerate energies at large $N$, and this degeneracy is encoded in $\log\beta$ terms just like with the spherical saddle. We can then compute an energy splitting term by again diagonalizing a certain finite dimensional Hamiltonian that is related to $SO(3)$ rotation generators.

We then use these results for the scaling dimensions to subleading order plus energy splitting to check the various dualities discussed above. For the $m=1$ duality, i.e. sQED$^{1,2}_3$ dual to QED$^{1,-3/2}_3$, we find the following values
\es{m1}{
\text{sQED}_{3}^{1,2}&:\qquad  \Delta_{q=1/2}^\text{spin $1$}=1.9995\,,\qquad \Delta_{q=1}^\text{spin $0$}=4.8384\,,\qquad \Delta_{q=1}^\text{spin $2$}=5.2476\,,\\
\text{QED}_{3}^{1,-3/2}&:\qquad  \Delta_{q=1/2}^\text{spin $1$}=2.0435\,,\qquad \Delta_{q=1}^\text{spin $0$}=4.9501\,,\qquad \Delta_{q=1}^\text{spin $2$}=5.3037\,.
}
We see that for both theories the unique $q=1/2$ spin 1 monopole is very close to the $\Delta=2$ value predicted for the extra currents\footnote{One for $q=1/2$ the other for $q=-1/2$, both of which have the same scaling dimension.} needed to enhance the topological $U(1)$ symmetry to $SO(3)$. For $q=1$, there are now two monopoles of spins zero and two, which receive contributions from the energy splitting term. Remarkably, their values are very close to the fuzzy sphere prediction \eqref{fuzzy}, which is a strong check of the energy splitting. For higher $q>1$, we list our results in the main text in Table \ref{tab:duality-m1} for each theory, where for each higher $q$ there are many monopoles distinguished by energy splitting. We find that the dual predictions match with an average relative error of 1.0 percent for the available ground-band comparisons, which is small enough that we are sensitive to the energy splitting. 

For the $m=2,3$ dualities, in Tables \ref{tab:duality-m2} and \ref{tab:duality-m3} we compare our QED$_3$ calculation extrapolated to $N=1$ and $|k|=5/2,7/2$ to  sQED$_3$ results extrapolated to $N=1$ and $k=3/2,4/3$, as in the effective theory in \eqref{local}. For the available ground-band comparisons, we find an average relative error of 3.0 percent for $m=2$ and 2.0 percent for $m=3$, which again takes into account the energy splitting for monopoles in different irreps. 

For Son's duality in Table \ref{tab:son} we compare our QED$_3$ calculation extrapolated to $N=1$ and $k=0$ to the dual operators in a free fermion theory, where we restrict to integer $q$ required for this duality. We find a match for all $q$ with an average relative error of 5 percent, which crucially relies on the contribution from the degeneracy breaking term. We also compare the free fermion values to sQED$_3^{1,1}$ in Table \ref{tab:seed}, as predicted by the seed duality. Relative to the previous comparison in \cite{Chester:2022wur}, we now include the energy splitting term, and also fix a relative sign between the CS and matter kernels. We now find an average relative error of 12 percent, which is less accurate than the other dualities. Finally, for $m=0$, which is the duality between QED$^{1,-1/2}_3$ and the critical $O(2)$ model, our extrapolated results do not give a good match for any $q$, as shown in Table \ref{tab:duality-m0}. Note that here all the monopole operators are unique, so there is no energy splitting term.

The rest of this paper is organized as follows. In Section \ref{review} we review the large $N,k$ calculation of the free energy for QED$_3$ and the leading order results, following \cite{Chester:2017vdh}, which is degenerate for many monopoles. In Section \ref{subleading}, we then derive the subleading temperature and state-independent term, which is given as a sum and integral for all $\kappa$. In Section \ref{enSplit}, we compute the energy splitting term, which is related to the temperature-dependent terms at subleading order. We do this for both QED$_3$ and sQED$_3$. In Section \ref{nonS}, we consider the non-spherical saddle, including its leading, subleading, and energy splitting calculations. In Section \ref{num}, we numerically compute the leading, subleading, and energy splitting results for various $q$ and $\kappa$ to compare to the various dualities. We conclude in Section \ref{sec:conclusion} with a review of our results and a discussion of future directions. Technical details of the calculations are given in the appendix and the attached \texttt{Mathematica} notebook.

\section{Free energy}\label{review}

We will start by reviewing known results from \cite{Chester:2017vdh} about the large $N$ expansion of the thermal free energy, which can be used to extract the scaling dimension. We first set up the large $N$ expansion, and then compute the leading order free energy.  

\subsection{Setup}
\label{setup}

The Euclidean action for QED$_3$ with $N$ two component complex fermions and bare CS level $\hat k$ is 
 \es{QEDAction}{
  \mathcal{S} =- \int d^3 x \, \sqrt{g}   \sum_{i=1}^{N} \psi_i^\dagger(i \slashed \nabla +\slashed A) \psi^i 
   - \int d^3x\, \frac{i \hat k}{4 \pi}  \epsilon^{\mu\nu\rho} A_\mu\partial_\nu A_\rho \,,
 }
 where $A_\mu$ is a dynamical $U(1)$ gauge field. Following \cite{Seiberg:2016gmd}, we define the measure of the fermion path integral such that $N$ free fermions in a background gauge field have partition function
\es{FreeFerm}{
Z[A]_\text{free fermions}=\abs{\Det(i \slashed \nabla +\slashed A)}^N\, e^{-{i\pi N\ov 2}\, \eta(A)}\,,
}
where the absolute value of the determinant is the regularized product of the absolute values of the eigenvalues, and the $ \eta(A)$ is the Atiyah-Patodi-Singer eta-invariant \cite{Witten:2015aba}.\footnote{Here we assumed $A$ is real, while later on we will allow for complex $A$, so that \eqref{FreeFerm} should be extended to a holomorphic function of $A$.} The phase in \eqref{FreeFerm} is a $-N/2$ CS term, which we combine with the bare CS level to define $k\equiv \hat k-N/2$. Gauge invariance of the path integral requires that $\hat k$ be an integer, so that $k$ is an integer for even $N$, and a half-integer for odd $N$.

We are interested in computing the thermal free energy $F_q$ in the presence of magnetic flux $\int dA=4\pi q$ through $S^2$, with $q\in \mathbb{Z}/2$. We parameterize the space $S^2\times S^1_\beta$ by $x=(\theta,\phi,\tau)$, with $\tau\in[-\beta/2,\beta/2)$ and metric
\es{metric}{
ds^2=d\theta^2+\sin^2\theta d\phi^2+d\tau^2 \,.
}
We can then integrate out the fermion fields in \eqref{QEDAction} to get the partition function
\es{FQEDdefine}{
Z_q = e^{-\beta F_q} 
=\int_{\int_{S^2} F = 4 \pi q} DA \exp\left[ N\,\tr\log |i\slashed \nabla+\slashed{A}|+N\frac{i\kappa}{4\pi}\int d^3x\epsilon^{\mu\nu\rho} A_\mu \partial_\nu A_\rho \right]\,,
}
where we defined $\kappa=k/N$. Since the action is now proportional to $N$, it can be evaluated in a saddle point approximation at large $N$ with $\kappa$ fixed. We expand $A_\mu$ around a saddle point by taking 
\es{saddle}{
A_\mu=\mathcal{A}_\mu+a_\mu\,,
}
where $a_\mu$ is the fluctuation around a background $A_\mu=\mathcal{A}_\mu$. On $S^2\times S^1_\beta$ with magnetic flux $4\pi q$, the most general spherically symmetric background is
\es{background}{
\mathcal{A}_\tau=-i\alpha\,,\qquad \mathcal{F}_{\theta\phi}d\theta\wedge d\phi=q\sin\theta d\theta\wedge d\phi\,, 
}
where $\alpha=i\beta^{-1}\int_{S^1_\beta}A$ is a real constant called the holonomy of the gauge field. Physically, $\alpha$ corresponds to a chemical potential for the matter fields, and is fixed by the saddle point equation
\es{saddleCond}{
\frac{\partial F_q}{\partial \alpha}=0\,.
}

Since the integrand in \eqref{FQEDdefine} is proportional to $N$, the fluctuation $a_\mu$ has typical size $1/\sqrt{N}$, and so is suppressed at large $N$. The thermal free energy $F_q$ can then be expanded at large $N$ as
\es{largeN}{
F_q=NF_q^{(0)}+F_q^{(1)}+\frac1NF_q^{(2)}+\dots\,,
}
 where $ F_q^{(0)}$ comes from evaluating $F_q$ at the saddle point and $ F_q^{(1)}$  comes from the functional determinant of the fluctuations around the saddle point. These terms can furthermore be expanded at large $\beta$ to get
\es{largeB}{
F_q^{(0)}&=\Delta_q^{(0)}-\frac{1}{\beta}S^{(0)}_q+O(e^{-c\beta})\,,\\
F_q^{(1)}&=\Delta_q^{(1)}+F_{q,\beta}^{(1)}+O(e^{-c\beta})\,,\\
}
where the temperature independent terms are identified with the scaling dimension, the leading-order $\beta^{-1}$ terms give the entropy of the degenerate monopole states, and $\beta$-dependent terms $F_{q,\beta}^{(1)}$ at subleading order can be used to compute the energy splitting of the degenerate monopoles. In the remainder of this section, we will review results for the leading order terms.

\subsection{Leading order free energy}
\label{leading}

After setting $A_\mu$ to its saddle point value \eqref{background} in the free energy \eqref{FQEDdefine} we find 
\es{FQED}{
F_q^{(0)}(\alpha)=-{1\ov \beta}\, \tr\log|i\slashed \nabla+\slashed{\mathcal{A}}|-2\kappa q\alpha\,,
}
The eigenvalues of the operator $i\slashed \nabla+\slashed{\mathcal{A}}$ on $S^2\times S_\beta^1$ with magnetic flux $4\pi q$ are 
\es{QEDeigs}{
&j=q-1/2:\hspace{3.9cm} (\omega_n-i\alpha)\,,\\
&j\in \{q+1/2,\,q+3/2,\,\dots\}:\qquad \pm\sqrt{(\omega_n-i\alpha)^2+\lambda_j^2}\,,
}
and have degeneracy  $d_j=2j+1$ for each distinct eigenvalue.  Here, $\omega_n$ and $\lambda_j$ are the fermionic Matsubara frequencies and the energies of modes of the theory quantized on $S^2\times \R$:
\es{QEDeigs2}{
\omega_n&=\frac{(2 n+1)\pi}{\beta}\,, \qquad n\in \Z\,,\\
\lambda_j&=\sqrt{(j+1/2)^2-q^2}\,.
}
Using this spectrum, we find\footnote{In going from the first to the second line in \eqref{FQED2} we performed the Matsubara sum assuming that $i\alpha$ is real, and then extended the result holomorphically to complex $i\alpha$.}
\es{FQED2}{
   F_q^{(0)}(\alpha) &=-2 \kappa q \alpha - \beta^{-1}\,\sum_{n\in \Z}\left[\sum_{j \geq   q+1/2} {d_j}  \log \left| (\omega_n  - i \alpha)^2 + \lambda_j^2 \right|+d_{q-1/2}\log\left|\omega_n-i\alpha\right|\right]\\
   &= -2  \kappa q \alpha - \beta^{-1} \left[\sum_{j \geq  q+1/2} {d_j}\log\left[ 2\left(\cosh(\beta\lambda_j)+\cosh(\beta\alpha)\right)\right]+d_{q-1/2}\log\left[2\cosh\left(\beta\alpha/2\right)\right]\right] \,.
 }

Lastly, we should solve \eqref{saddleCond} to find the saddle point value for $\alpha$. When $q=0$ or for $k=0$, the physical saddle point is $\alpha=0$, because the sum \eqref{FQED2} is an even function of $\alpha$.  Otherwise, the $\alpha$ saddle that gives the lowest real free energy is
\es{QEDalph}{
\alpha_{\tilde j}(\kappa)&=-\sgn(\kappa-1/2)\left(\lambda_{\tilde j}+\beta^{-1}\log\frac{\xi_{\tilde j}}{1-\xi_{\tilde j}}\right)+O(e^{-(\lambda_{\tilde j+1}-\lambda_{\tilde j})\beta})\,,\\
\xi_{\tilde j}&=\frac{1}{d_{ \tilde j}} \left(2q \le(|\kappa-
1/2 |- \Theta(-1/2-\kappa)\ri)-  \sum_{q-1/2<j'<  \tilde j} d_{j'}  \right)\,,
} 
where $\Theta(x)=1$ for $x>0$ and zero otherwise. The saddle is labeled by the Landau level $\tilde j$, which changes with $\kappa$ and $q$ as
\es{QEDexplicit}{
\tilde j(\kappa)=q+\frac12+\left\lfloor-\le(q+\frac12\ri)+\sqrt{\frac14+2q|\kappa|+q^2}\right\rfloor\,,
}
while $\xi_{\tilde j}$ obeys $0\leq\xi_{\tilde j}(\kappa)\leq1$ and corresponds to the filling fraction for $\tilde j$. 

We can then plug the saddle \eqref{QEDalph} back into \eqref{FQED2}, take the large $\beta$ limit to find the leading order scaling dimension and entropy
\es{QEDFinal}{
\Delta_{q}^{(0)}&=-\Big[\frac{q}{6}(q+2)(2q-1)+\sum_{j \geq q-1/2} \big(d_j \lambda_j-\frac{d_j^2}{2}+q^2\big)\Big]+\sum_{ q-1/2 \leq j < \tilde j} d_j\lambda_j +\xi_{\tilde j} d_{\tilde j}\lambda_{\tilde j}\,,\\
S_{q}^{(0)}&=-d_{\tilde j}\left(\xi_{\tilde j}\log\xi_{\tilde j}+(1-\xi_{\tilde j})\log[1-\xi_{\tilde j}]\right)\,,
}
where for the scaling dimension we applied zeta regularization. For the scaling dimension, the first term corresponds to the zeta regularized Casimir energy, the second term to the filled Landau levels, and the third term to the valence Landau level, with filling fraction $\xi_{\tilde j}$. The entropy $NS_{q}^{(0)}$ corresponds to the logarithm of the number of degenerate monopoles, which as shown in \cite{Chester:2017vdh} correspond to all operators in $SU(N)\times SU(d_{\tilde j})$ irreps
\es{antisum}{
\bigoplus_\nu(\Upsilon_\nu,\tilde\Upsilon_\nu)
}
over all possible irreps $\nu$ with Young diagrams $\Upsilon_\nu$ with a total of $d_{\tilde j}N\xi_{\tilde j}$ boxes (whose conjugates are denoted by $\tilde\Upsilon_\nu$) such that $\Upsilon_\nu$ has maximum width $d_{\tilde j}$ and height $N$. The spacetime spins of the degenerate monopoles are then found by decomposing $SU(d_{\tilde j})$ to $SU(2)$.

		\section{Subleading order free energy}\label{subleading}
	
	We will now discuss the calculation of the subleading free energy. We will start by setting up the calculation, following \cite{Chester:2017vdh}, and review the calculation of the CS and $q=0$ kernels. We will then compute the thermal Green's function and the matter kernels for $q>0$. The final answer is written as a sum/integral, which can be computed numerically as in \cite{Dyer:2015zha,Chester:2022wur,Dupuis:2021flq}. Since the final results are identical for positive and negative $\kappa$, for simplicity we will assume $\kappa>0$.
	
\subsection{Setup}
\label{setup2}
	
We start by reviewing some basic results from \cite{Chester:2017vdh}, which should be consulted for further details. We can expand \eqref{FQEDdefine} to quadratic order in the gauge fluctuations $a_\mu$ to get\footnote{The linear term vanishes because $\mathcal{A}_\mu$ is a saddle.}
\es{subQEDF}{
\exp(-\beta F_q^{(1)})&=\int Da \, \exp\left[-\frac N2 \int d^3xd^3x'\sqrt{g}\sqrt{g'}\, a_\mu(x)K_q^{\mu\nu}(x,x') a_\nu(x')\right]\,,
}
where the kernel $K_q^{\mu\nu}\equiv K_{q,\text{mat}}^{\mu\nu}+K_{q,\text{CS}}^{\mu\nu}$ receives contributions from both the matter and CS terms in \eqref{FQEDdefine}:
\es{Ks}{
K_{q,\text{mat}}^{\mu\nu}(x,x')=\Tr(\gamma^\mu G_q(x,x')\gamma^\nu G_q(x',x))\,,\qquad K_{\text{CS}}^{\mu\nu}=-\frac{i\kappa}{2\pi}\delta(x,x')\epsilon^{\mu\nu\rho}\partial'_\rho\,.
}
The matter sign follows from \eqref{GreenDefF}: the two inverse Dirac operators in the second variation of $-\Tr\log(i\slashed\nabla+\slashed{\mathcal A})$ each equal $-G_q$, so
\es{matterSign}{
 \delta^2[-\Tr\log(i\slashed\nabla+\slashed{\mathcal A})]
 =\Tr(G_q\,\delta\slashed A\,G_q\,\delta\slashed A)\,.
}
Thus the minus-trace convention would require the phase-stripped propagator $G_q/i$, rather than the $G_q$ used here. This corrects the position-space sign in \cite{Chester:2017vdh}; the positive $q=0$ kernels below use the convention in \eqref{Ks}.

The $\gamma^\mu$ are gamma matrices on $S^2\times \mathbb{R}$. We can define them as $\gamma^\mu=e^\mu_a\sigma^a$ where $\sigma^a$ are the usual Pauli matrices, and the frame obtained from the conformal transformation of $S^2\times \mathbb{R}$ to $\mathbb{R}^3$ is\footnote{There is no difference between upper and lower frame indices in Euclidean signature.}
\es{viel}{
e^a=e^{-\tau}dx^a\,,\qquad \vec x=e^\tau \hat x=e^\tau\begin{pmatrix} \sin\theta\cos\phi&\sin\theta\sin\phi&\cos\theta\end{pmatrix}\,.
}
The single-fermion thermal Green's function is $G_q(x,x')=\langle\psi(x)\psi^\dagger(x')\rangle$. 

We can compute the integral over the gauge fluctuations in \eqref{subQEDF} by expanding the fluctuations in Fourier space as 
 \es{aFourier}{
  a(x) = \mathfrak{a}_{00}^E(0) \frac{d\tau}{\sqrt{4 \pi \beta}}  + 
  \sum_{n=-\infty}^\infty \sum_{\ell=1}^\infty \sum_{m=-\ell}^\ell 
   \left[  \mathfrak{a}_{\ell m}^E(\omega_n) {\cal E}_{n \ell m} (x) 
    + \mathfrak{a}_{\ell m}^B(\omega_n) {\cal B}_{n \ell m} (x) \right]  \frac{e^{-i \omega_n \tau}}{\sqrt{\beta}}+ d \lambda(x)\,,
 }
where $d\lambda$ are pure gauge modes and ${\cal E}_{n \ell m} (\theta, \phi, \tau)$ and ${\cal B}_{n \ell m} (\theta, \phi, \tau)$, together with $d\tau/\sqrt{4\pi\beta}$, form an orthonormal basis of polarizations for the one-form $a(x)$:
 \es{EBDef}{
    {\cal E}_{n \ell m}(x) = \frac{ \ell(\ell+1)  Y_{\ell m}d\tau  -i \omega_n dY_{\ell m}}{\sqrt{\ell(\ell+1)} \sqrt{\omega_n^2 + \ell(\ell+1)}}  \,, \qquad
     {\cal B}_{n \ell m}(x) = \frac{*_2 dY_{\ell m}(\theta, \phi)}{\sqrt{\ell (\ell+1)}}  \,.
 }
Here $\omega_n = 2\pi n / \beta$ are the bosonic Matsubara frequencies, and $*_2$ is the Hodge dual on $S^2$. Ignoring the pure gauge modes, the exponent in \eqref{subQEDF} can then be written in Fourier space as
 \es{NumExponent}{
  -\frac{N}{2} \abs{\mathfrak{a}_{00}^E(0)}^2 K^{EE}_{q, 0}(0)  -\frac{N}{2} \sum_{n=-\infty}^\infty \sum_{\ell=1}^\infty \sum_{m=-\ell}^\ell
   \begin{pmatrix} \mathfrak{a}_{\ell m}^E(\omega_n) \\ \mathfrak{a}_{\ell m}^B(\omega_n)
   \end{pmatrix}^\dagger
   {\bf K}_{q, \ell}(\omega_n) \begin{pmatrix} \mathfrak{a}_{\ell m}^E(\omega_n) \\ \mathfrak{a}_{\ell m}^B(\omega_n)
   \end{pmatrix} \,,
 }
The kernel ${\bf K}_{q, \ell}(\omega_n)$ is a $2\times 2$ matrix whose entries we denote with doubled superscripts $E$ or $B$, which is defined by
\es{fourierKernscal}{
\bold{K}_{q,\ell}(\omega_n)=&\frac{4\pi}{2\ell+1}\int d^3x \sqrt{g}e^{i\omega_n \tau} \sum_{m=-\ell}^{\ell}\\
&\quad\times\begin{pmatrix} \mathcal{E}_{n\ell m,\mu}(x)^* \\ \mathcal{B}_{n\ell m,\mu}(x)^* \end{pmatrix}
 K_q^{\mu\nu}(x,0)
 \begin{pmatrix} \mathcal{E}_{n\ell m,\nu}(0) & \mathcal{B}_{n\ell m,\nu}(0) \end{pmatrix}\,,
}
where we used rotational symmetry to remove the $x'$ integral. We can take care of the pure gauge modes by considering $e^{-\beta F_q^{(1)}}/e^{-\beta F_0^{(1)}}$ and using the fact that $F_0^{(1)}=0$. We can then integrate over the gauge modes to get at large $\beta$
\es{F1true}{
 \beta F_q^{(1)} \approx \frac{1}{2} \le[\sum_{n=-\infty}^\infty \sum_{\ell=1}^\infty  (2\ell+1)\log \le({\det{}'{\bf K}_{q, \ell}(\omega_n) \ov \det{\bf K}_{0, \ell}(\omega_n) }\ri)
  + \begin{cases}
   \log \le(N\, 2\pi d_{\tilde j} \,\xi_{\tilde j}(1-\xi_{\tilde j})  \ri) & \text{if $\xi_{\tilde j} \neq 0$} \,,\\
   0 & \text{if $\xi_{\tilde j} = 0$}
  \end{cases} \ri]\,,
}
where $\det'$ denotes that we ignore the zero eigenvalue for $n=0$ coming from the integral over $\mathfrak{a}_{00}(0)$, which has been separately considered in the second term \cite{Chester:2017vdh}. We can then take the large $\beta$ limit of ${\bf K}_{q, \ell}(\omega_n)$ to get
 \es{SmoothApprox}{
  {{\bf K}}_{q, \ell}(\omega_n) = \beta\overline{\bf K}_{q,\ell} \delta_{n0}+\widetilde{{\bf K}}_{q, \ell}(\omega)\big\vert_{\omega=\omega_n}+ O(e^{- \beta}) \,.
}
The $\beta$-independent term then leads to the scaling dimension
\es{freeFinal}{
&\Delta_{q,\kappa}^{(1)}=\frac12\int {d\omega\ov 2\pi } \sum_{\ell=1}^\infty  (2\ell+1)\log \det\le(\frac{\widetilde{\bf K}_{q, \ell,\text{mat}}(\omega)+\widetilde{\bf K}_{\ell,\text{CS}}(\omega) }{\widetilde{\bf K}_{0, \ell,\text{mat}}(\omega)+\widetilde{\bf K}_{\ell,\text{CS}}(\omega) }\ri)\,,
}
where the sum over $\omega_n$ in \eqref{F1true} has been converted into an integral to exponential precision in $\beta$, and we divided the kernels into the matter and CS contributions. We will discuss the $\beta$-dependent term in the next section.

We can already compute some of the kernels in this expression.  As we see from \eqref{Ks}, the Chern-Simons kernel $K_{q,\text{CS}}^{\mu\nu}(x,x')$ is local, so we can compute its Fourier transform without doing any integrals to get \cite{Chester:2021drl}
\es{CSfourier2}{
\widetilde{\mathbf{K}}_{\ell,\mathrm{CS}}(\omega)
=
\frac{i\kappa}{2\pi}
\begin{pmatrix}
0 & \sqrt{\omega^2+\ell(\ell+1)} \\
\sqrt{\omega^2+\ell(\ell+1)} & 0
\end{pmatrix}\,.
}
The $q=0$ matter kernels can also be easily computed using the closed form expression \cite{Pufu:2013vpa}
\es{freeG}{
G_0(x,x')=\frac{i}{4\pi}\frac{\sigma\cdot \big(e^{(\tau-\tau')/2}\hat x-e^{(\tau'-\tau)/2}\hat x'\big)}{\left(2\cosh\left(\tau-\tau'\right)-2\cos\gamma\right)^{3/2}}\,,
}
for the $q=0$ Green's function, where $\gamma$ is the angle between the two points on $S^2$
\es{gamma}{
\cos\gamma=\cos\theta\cos\theta'+\sin\theta\sin\theta'\cos\left(\phi-\phi'\right)\,.
}
We then plug this into \eqref{Ks}, take the Fourier transform \eqref{fourierKernscal} to compute ${\bf K}_{0,\ell,\text{mat}}(\omega_n)$, and send $\beta\to\infty$ to get  \cite{Pufu:2013vpa,Dupuis:2021flq}:
\es{q0}{
\widetilde{\bf K}_{0,\ell,\text{mat}}(\omega)&=
\frac{1}{2}
\begin{pmatrix}
\bigl[\omega^2+\ell(\ell+1)\bigr]D_\ell(\omega) & 0 \\
0 & \bigl(\omega^2+\ell^2\bigr)D_{\ell-1}(\omega)
\end{pmatrix}
\,,\\
}
where we define
\es{D}{
D_\ell(\omega)&=\left|\frac{\Gamma\left((\ell+1+i\omega)/2\right)}{4\Gamma\left((\ell+2+i\omega)/2\right)}\right|^2\,.\\
}
Our next task is to compute the matter kernels for $q>0$.
	
	\subsection{Green's function}
	\label{green}
	
	In order to compute the matter kernels, we must first discuss the thermal Green's function, which is defined as
	\es{GreenDefF}{
\left(i\slashed{D}+\slashed{\mathcal{A}}\right)G_q(x,x')=-\delta(x-x')\,.
}
This function was derived in \cite{Chester:2017vdh}, and given the form\footnote{Note that we corrected the sign for $\bold{N}_{q,j}$ relative to \cite{Chester:2017vdh}.} 
\es{TempGreen}{
&  G_q(x, x')  =e^{\alpha(\tau-\tau')}(G(x,x')+\hat G(x,x'))\,,\\
&G(x,x')=\frac{i}{2}\sum_{j, m} \begin{pmatrix}T_{qjm}(\theta,\phi)&S_{qj m}(\theta,\phi)\end{pmatrix}  e^{-\lambda_{j}|\tau-\tau'|}\left(i\sigma_2+\bold{N}_{q,j}\sgn(\tau-\tau')\right) \begin{pmatrix}T^\dagger_{qjm}(\theta',\phi')\\S^\dagger_{qj m}(\theta',\phi')\end{pmatrix}\,,\\
&\hat G_{\tilde j}(x,x')=-\frac{i}{2}\sum_{j, m} \begin{pmatrix}T_{qjm}(\theta,\phi)&S_{qj m}(\theta,\phi)\end{pmatrix}  \\
&\left[\frac{e^{-\lambda_{j}(\tau-\tau')}}{1+e^{\beta(-\alpha_{\tilde j}+\lambda_{j})}}\left(i\sigma_2+\bold{N}_{q,j}\right)+\frac{e^{\lambda_{j}(\tau-\tau')}}{1+e^{\beta(\alpha_{\tilde j}+\lambda_{j})}}\left(i\sigma_2-\bold{N}_{q,j}\right)\right] \begin{pmatrix}T^\dagger_{qjm}(\theta',\phi')\\S^\dagger_{qj m}(\theta',\phi')\end{pmatrix}\,,
 } 
 where $\sigma_2$ is a Pauli matrix and 
 \es{N}{
 \bold{N}_{q,j}=\begin{pmatrix}
-\frac{q}{j+1/2}&-\frac{\lambda_j}{j+1/2} \\
-\frac{\lambda_j}{j+1/2} &\frac{q}{j+1/2}\\
\end{pmatrix}\,.
 }
 The spinor monopole harmonics $T_{qjm}(\theta,\phi)$ and $S_{qjm}(\theta,\phi)$ are defined in \cite{Pufu:2013vpa,Borokhov:2002ib} as
 \es{spinorHarm}{
 T_{qjm}=\begin{pmatrix}\sqrt{\frac{j+m}{2j}}Y_{q,j-1/2,m-1/2}(\theta,\phi)\\
\sqrt{\frac{j-m}{2j}}Y_{q,j-1/2,m+1/2}(\theta,\phi)\end{pmatrix},\quad
S_{qjm}=\begin{pmatrix}-\sqrt{\frac{j-m+1}{2j+2}}Y_{q,j+1/2,m-1/2}(\theta,\phi)\\
\sqrt{\frac{j+m+1}{2j+2}}Y_{q,j+1/2,m+1/2}(\theta,\phi)\end{pmatrix}\,,
 }
 where $Y_{qjm}(\theta,\phi)$ are standard monopole harmonics. For the lowest Landau level $j=q-1/2$ there is no $T_{qjm}$, so we should set $\sigma_2$ to zero, $\bold{N}_{q,q-1/2}=1$, and then the reduced matrix in harmonic space is one-dimensional. 
 
 The function $G(x,x')$ is the same as the CP preserving Green's function computed for $k=0$ in \cite{Pufu:2013vpa}.\footnote{Up to a sign we corrected for the $i\sigma_2$ term.} For the CP breaking term $\hat G(x,x')$, we should input the saddle point values for $\alpha_{\tilde j}$ in \eqref{QEDalph} for each Landau level $\tilde j$. For the lowest Landau level $\tilde j=q-1/2$ both terms in \eqref{TempGreen} contribute to give
 \es{zeroGreen}{
\hat G_{q-1/2}(x,x')=\frac{i(1-2\xi_{q-1/2})}{2}\sum_{m}S_{q\,q-1/2\, m}(\theta,\phi)S^\dagger_{q\,q-1/2\, m}(\theta',\phi')\,,
 }
where note that unlike the zero mode of $G(x,x')$, there is no $\text{sign}(\tau-\tau')$ term here, so this term violates CP unless $\xi_{q-1/2}=1/2$, i.e. $k=0$. For other $\tilde j$ we have
 \es{nozeroGreen}{
&\hat G_{\tilde j>q-1/2}(x,x')=-\frac{i}{2}\sum_{m,j\leq \tilde j} \begin{pmatrix}T_{qjm}(\theta,\phi)&S_{qj m}(\theta,\phi)\end{pmatrix}  \\
&\qquad(1+\delta_{j,\tilde j}(\xi_{\tilde j}-1)){e^{\text{sgn}(\kappa-1/2)\lambda_{j}(\tau-\tau')}}   \left(i\sigma_2-\text{sign}(\kappa-1/2)\bold{N}_{q,j}\right) \begin{pmatrix}T^\dagger_{qjm}(\theta',\phi')\\S^\dagger_{qj m}(\theta',\phi')\end{pmatrix}\,,
 }
 where as usual for $j=q-1/2$ we set $\sigma_2=0$ and $\bold{N}_{q,q-1/2}=1$ so the reduced matrix in harmonic space is one-dimensional.
	
	\subsection{Matter kernels }
	\label{qkern}

We now consider the matter kernels for $q>0$. In this section, we will find it convenient to write the kernels in the Fourier basis defined by the vector spherical harmonics
\es{UVWdef}{
U_{(\ell+1)m}&=-\sqrt{\frac{{\ell+1}}{{2\ell+1}}}Y_{\ell m}d\tau+\frac{1}{\sqrt{(2\ell+1)(\ell+1)}}dY_{\ell m}\\
V_{\ell m}&=i\frac{*_2dY_{\ell m}}{\sqrt{\ell(\ell+1)}}\\
W_{(\ell-1)m}&=\sqrt{\frac{{\ell}}{{2\ell+1}}}Y_{\ell m}d\tau+\sqrt{\frac{(\ell+1)}{\ell(2\ell+1)(\ell+1)}}dY_{\ell m}\,,\\
}
such that the kernels are related to the $\mathcal{E},\mathcal{B}$ basis of \eqref{EBDef} as
\begin{equation}
\mathbf{K}^{(E,B)}_{q,\ell}(\omega)=
\begin{pmatrix}
0 & -i \\[4pt]
-\sqrt{\frac{\ell}{2\ell+1}}\,
 \frac{\ell+1+i\omega}{\sqrt{\omega^2+\ell(\ell+1)}} & 0 \\[6pt]
\sqrt{\frac{\ell+1}{2\ell+1}}\,
 \frac{\ell-i\omega}{\sqrt{\omega^2+\ell(\ell+1)}} & 0
\end{pmatrix}^\dagger
\mathbf{K}^{(V,U,W)}_{q,\ell}(\omega)
\begin{pmatrix}
0 & -i \\[4pt]
-\sqrt{\frac{\ell}{2\ell+1}}\,
 \frac{\ell+1+i\omega}{\sqrt{\omega^2+\ell(\ell+1)}} & 0 \\[6pt]
\sqrt{\frac{\ell+1}{2\ell+1}}\,
 \frac{\ell-i\omega}{\sqrt{\omega^2+\ell(\ell+1)}} & 0
\end{pmatrix}.
\end{equation}

These kernels can be computed using the algorithmic method first introduced for sQED$_3$ in \cite{Dyer:2015zha}, and then applied to fermionic QED$_3$ in \cite{Dyer:2013fja}. The Fourier space kernels in \eqref{fourierKernscal} are written as integrals of six monopole spherical harmonics: one each from the vector harmonics in \eqref{aFourier}, and then four more from the position space kernels in \eqref{Ks}, where each Green's function includes a pair of harmonics \eqref{TempGreen}. Three of the harmonics are functions of $x'$, which we can eliminate by using rotational invariance to fix $x'=0$, and using the relation
\es{Y0}{
Y_{q,\ell,m}(0,0) & =\delta_{q,-m}\sqrt{\frac{2\ell+1}{4\pi}}\,.
}
For the remaining three harmonics, we can first relate conjugate harmonics as 
\es{Yconj}{
Y_{q,\ell,m}^{*}(\theta,\phi)=\left(-1\right)^{q+m}Y_{-q,\ell,-m}(\theta,\phi)\,,
}
and then use the triple harmonic formula
\es{tripY}{
\int_0^{2\pi} d\phi\int_0^\pi d\theta \sin\theta Y_{q,\ell,m}(\theta,\phi)Y_{q^{\prime},\ell^{\prime},m^{\prime}}(\theta,\phi)Y_{q^{\prime\prime},\ell^{\prime\prime},m^{\prime\prime}}(\theta,\phi)=\left(-1\right)^{\ell+\ell^{\prime}+\ell^{\prime\prime}}\\
\times\sqrt{\frac{\left(2\ell+1\right)\left(2\ell^{\prime}+1\right)\left(2\ell^{\prime\prime}+1\right)}{4\pi}}\begin{pmatrix}\ell & \ell^{\prime} & \ell^{\prime\prime}\\
q & q^{\prime} & q^{\prime\prime}
\end{pmatrix}\begin{pmatrix}\ell & \ell^{\prime} & \ell^{\prime\prime}\\
m & m^{\prime} & m^{\prime\prime}
\end{pmatrix}\,,
}
where the bottom line is written in terms of two $3-j$ symbols. The result for the Fourier space matter kernels is then written in terms of these $3-j$ symbols, while the dependence on $\tau$ is just simple exponentials. Some of these $\tau$ integrals will give linear in $\beta$ terms, which were already computed in \cite{Chester:2017vdh}, and do not contribute to the scaling dimension. For the other terms, we can immediately take $\beta\to\infty$ to get $\widetilde{K}$.

The CP preserving kernels, which arise from CP preserving Green's function $G(x,x')$ in \eqref{TempGreen}, were computed in this way in \cite{Dyer:2013fja}. By looking at many values of $q$, we found a particularly compact formulation that applies for all $q$. First let us define the spherical Pauli matrices
\begin{equation}
 \tilde\sigma_{-1}=\begin{pmatrix}0&-\sqrt2\\0&0\end{pmatrix},\qquad
 \tilde\sigma_0=\begin{pmatrix}1&0\\0&-1\end{pmatrix},\qquad
 \tilde\sigma_{+1}=\begin{pmatrix}0&0\\\sqrt2&0\end{pmatrix}\,,
\end{equation}
and the three vector-harmonic coefficient triples
{\small
\es{vs}{
 \hspace{-.9in}(u_U^{-1},u_U^0,u_U^{+1})={}&\left(
 \sqrt{\frac{(\ell-j+j'+1)(\ell-j+j'+2)}{(2\ell+2)(2\ell+3)}},
 -\sqrt{\frac{(\ell-j+j'+1)(\ell+j-j'+1)}{(\ell+1)(2\ell+3)}},
 \sqrt{\frac{(\ell+j-j'+1)(\ell+j-j'+2)}{(2\ell+2)(2\ell+3)}}\right),\\
 \hspace{-.9in} (u_V^{-1},u_V^0,u_V^{+1})={}&\left(
 -\sqrt{\frac{(\ell-j+j'+1)(\ell+j-j')}{2\ell(\ell+1)}},
 \frac{j-j'}{\sqrt{\ell(\ell+1)}},
 \sqrt{\frac{(\ell-j+j')(\ell+j-j'+1)}{2\ell(\ell+1)}}\right),\\
 \hspace{-.9in} (u_W^{-1},u_W^0,u_W^{+1})={}&\left(
 \sqrt{\frac{(\ell+j-j'-1)(\ell+j-j')}{2\ell(2\ell-1)}},
 \sqrt{\frac{(\ell-j+j')(\ell+j-j')}{\ell(2\ell-1)}},
 \sqrt{\frac{(\ell-j+j'-1)(\ell-j+j')}{2\ell(2\ell-1)}}\right).
}
}
The kernels can then all be written in terms of the following $2\times 2$ matrix:
\begingroup
\small
\begin{align}
 {\bold{R}^{X,(j,j')}_\ell}_{\alpha\beta}
={}&\left[
 \begin{pmatrix}j&\ell&j'\\-j&j-j'&j'\end{pmatrix}\right]^{-1}
 \sum_{a,b=1}^{2}\sum_{\nu=-1}^{1}
 \left[\delta_{\alpha,T}\delta_{a,1}+\delta_{\alpha,S}
 \left(-\frac{\delta_{a,1}}{\sqrt{2j+2}}
 +\sqrt{\frac{2j+1}{2j+2}}\delta_{a,2}\right)\right]\nonumber\\[-2pt]
&\times\left[\delta_{\beta,T}\delta_{b,1}+\delta_{\beta,S}
 \left(-\frac{\delta_{b,1}}{\sqrt{2j'+2}}
 +\sqrt{\frac{2j'+1}{2j'+2}}\delta_{b,2}\right)\right]
 (\tilde\sigma_\nu)_{ab}\,u_X^\nu(\ell,j-j')\nonumber\\[-2pt]
&\times(-1)^{q+2j+j'+a+\ell_X-\frac12
 -\delta_{\alpha,T}-\delta_{\beta,T}}
 \sqrt{\frac{[2(j+\frac12-\delta_{\alpha,T})+1](2\ell_X+1)
 [2(j'+\frac12-\delta_{\beta,T})+1]}{4\pi}}\nonumber\\[-2pt]
&\times
 \begin{pmatrix}j+\frac12-\delta_{\alpha,T}&\ell_X&
 j'+\frac12-\delta_{\beta,T}\\-q&0&q\end{pmatrix}
 \begin{pmatrix}j+\frac12-\delta_{\alpha,T}&\ell_X&
 j'+\frac12-\delta_{\beta,T}\\
 -(j-\frac32+a)&j-j'+\nu&j'-\frac32+b\end{pmatrix}\,, \label{eq:reduced-current}
\end{align}
\endgroup
where $\alpha,\beta$ run over the spinor harmonic basis $T,S$, we define $\ell_U=\ell+1,\ell_V=\ell,\ell_W=\ell-1$, and $X=U,V,W$. The CP preserving kernels for each fermion level $(j,j')$ are then
\es{Kjjp}{
 \widetilde{K}^{XX',(j,j')}_{q,\ell,CP}(\omega)=&\frac{\eta_X}{2\ell+1}\frac{(-1)^{j-j'+1}(\lambda_j+\lambda_{j'})}{2(\lambda_j+\lambda_{j'})^2+2\omega^2}\Tr\left[\sigma_2\bold{R}^{X',(j,j')}_\ell \sigma_2\bold{R}^{X,(j',j)}_\ell+\bold{N}_{q,j}\bold{R}^{X',(j,j')}_\ell\bold{N}_{q,j'}\bold{R}^{X,(j',j)}_\ell\right]\\
 &+\frac{\eta_X}{2\ell+1}\frac{(-1)^{j-j'+1}\omega}{2(\lambda_j+\lambda_{j'})^2+2\omega^2}\Tr\left[\bold{N}_{q,j}\bold{R}^{X',(j,j')}_\ell \sigma_2\bold{R}^{X,(j',j)}_\ell-\sigma_2\bold{R}^{X',(j,j')}_\ell\bold{N}_{q,j'}\bold{R}^{X,(j',j)}_\ell\right]\,,
}
where $\eta_V=1$, $\eta_U=\eta_W=-1$, and as usual for the lowest level $j,j'=q-1/2$, we restrict to the $S$ entry in the matrix in the trace. We can then use these to construct the full CP preserving kernel
\es{CPfin}{
 \widetilde{K}^{XX'}_{q,\ell,CP}(\omega)
 =&\sum_{j=q+1/2}^\infty\left[ \widetilde{K}^{XX',(q-1/2,j)}_{q,\ell,CP}(\omega)+\widetilde{K}^{XX',(j,q-1/2)}_{q,\ell,CP}(\omega)\right]\\
 &+\sum_{j=q+1/2}^\infty\left[\sum_{j'=q+1/2}^\infty \widetilde{K}^{XX',(j,j')}_{q,\ell,CP}(\omega)+\frac{f^{XX'}}{4\pi}\right]-\frac{f^{XX'}}{4\pi}\zeta(0,q+1)\,,
}
where the first line corresponds to cross terms between the lowest level and higher levels, while in the second line we only have higher levels. The term with only the lowest level Green's functions only contributes to linear in $\beta$ terms, and so does not appear here. Following \cite{Pufu:2013vpa}, in the second line we used zeta regularization to make the sum over the  levels finite. The coefficients of this regularization $f^{XX'}$ for each kernel are
\es{FXX}{
f^{VV}=1\,,\qquad f^{UU}=\frac{\ell}{2\ell+1}\,,\qquad f^{WW}=\frac{\ell+1}{2\ell+1}\,,\qquad f^{UW}=\frac{\sqrt{\ell(\ell+1)}}{2\ell+1}\,,\\
}
while the remaining kernels are finite so have  $f^{XX'}=0$. 

The CP violating kernels arise from including the CP violating Green's function in \eqref{zeroGreen} and \eqref{nozeroGreen}. We again ignore linear in $\beta$ terms, that were already considered in \cite{Chester:2017vdh}. The result for the kernel now depends on the highest Landau level $\tilde j$ that we are filling. By computing the kernel for many values of $q$ and $\tilde j$, we find they can be written using the same ingredient \eqref{eq:reduced-current} as the CP preserving kernels. We first consider the cross term between the lowest level CP violating Green's function and that of all other levels, which is only nonzero for $VU$ and $UV$: 
\es{NCPfin1}{
  \widetilde{K}^{VU,0}_{q,\ell,NCP}(\omega)=&
 -\frac{(\ell+1-i\omega)}
 {2\pi(\ell+1)\sqrt{\ell(2\ell+1)}}
 \sum_{j=q+1/2}^{q+\ell-1/2}
 \frac{j+\frac12}{\lambda_j^2+\omega^2}\\[-2pt]
 &\quad\times
 \frac{\Gamma(\ell+j-q+\frac32)\Gamma(j+q+\frac32)\Gamma(2q+1)}
 {\Gamma(j-q+\frac12)\Gamma(\ell-j+q+\frac12)
 \Gamma(j+q-\ell+\frac12)\Gamma(\ell+j+q+\frac32)}\,,
}
We next consider the cross term between higher level CP violating Green's function and the lowest level of the CP preserving Green's function:
\es{NCPfin2}{
 \widetilde{K}^{VU,j}_{q,\ell,NCP}(\omega)=&\frac{\sqrt{\ell}(\ell+1-i\omega)\lambda_j^2}{\sqrt{2\ell+1}} T_{j,\ell}(\omega)\,, \\
 \widetilde{K}^{VV,j}_{q,\ell,NCP}(\omega)=&\lambda^3_jT_{j,\ell}(\omega)\,,\\
  \widetilde{K}^{UU,j}_{q,\ell,NCP}(\omega)+ \widetilde{K}^{WW,j}_{q,\ell,NCP}(\omega)=&\lambda_j(\lambda_j^2-\ell(\ell+1))T_{j,\ell}(\omega)\,,
}
where we define
\es{Tj}{
T_{j,\ell}(\omega)= \frac{( j+\frac12)
 (2q+1)_{ j-q-1/2}
 (\ell- j+q+\frac12)_{ j-q-1/2}
 (\ell+2)_{ j-q-1/2}\Gamma(2q+1)^2}
 {2\pi\,\Gamma( j-q+\frac32)
 (\omega^2+\lambda_{ j}^2)
 \Gamma(q+ j-\ell+\frac12)
 \Gamma(q+ j+\ell+\frac32)}.
}
Finally, there is the cross term between a nonzero CP propagator and an excited-level contribution:
\es{NCPfin3}{
\widetilde{K}^{XX',(j,j')}_{q,\ell,NCP}(\omega)=&\frac{(-1)^{j-j'}\eta_X\lambda_j}{2(2\ell+1)(\lambda^2_j+(\omega+i\lambda_{j'})^2)}\Tr\left[\sigma_2\bold{R}_\ell^{X',(j,j')}(i\bold{N}_{q,j'}+\sigma_2)\bold{R}_\ell^{X,(j',j)}\right]\\
&+\frac{(-1)^{j-j'}\eta_X(\omega+i\lambda_{j'})}{2(2\ell+1)(\lambda^2_j+(\omega+i\lambda_{j'})^2)}\Tr\left[\bold{N}_{q,j}\bold{R}_\ell^{X',(j,j')}(i\bold{N}_{q,j'}+\sigma_2)\bold{R}_\ell^{X,(j',j)}\right]\\
&+\frac{(-1)^{j-j'}\eta_X\lambda_j}{2(2\ell+1)(\lambda^2_j+(\omega-i\lambda_{j'})^2)}\Tr\left[(i\bold{N}_{q,j'}+\sigma_2)\bold{R}_\ell^{X',(j',j)}\sigma_2\bold{R}_\ell^{X,(j,j')}\right]\\
&-\frac{(-1)^{j-j'}\eta_X(\omega-i\lambda_{j'})}{2(2\ell+1)(\lambda^2_j+(\omega-i\lambda_{j'})^2)}\Tr\left[(\sigma_2+i\bold{N}_{q,j'})\bold{R}_\ell^{X',(j',j)}\bold{N}_{q,j}\bold{R}_\ell^{X,(j,j')}\right]\,,
}
where as before $\eta_V=1$, $\eta_U=\eta_W=-1$. We can now assemble the full CP violating kernel
\es{NCPff}{
\widetilde{K}^{XX'}_{q,\ell,NCP}(\omega)=(1-2\xi_{q-1/2})\widetilde{K}^{XX',0}_{q,\ell,NCP}(\omega)+\sum_{j'=q+1/2}^{\tilde j} \xi_{j'} \Big[\widetilde{K}^{XX',j'}_{q,\ell,NCP}(\omega)+\sum_{j=q+1/2}^{j'+\ell+1}\widetilde{K}^{XX',(j,j')}_{q,\ell,NCP}(\omega)\Big]\,,
}
where the excited filling fractions $\xi_j$ are one for $q+1/2\leq j<\tilde j$, while $\xi_{q-1/2}=0$ for $\tilde j>q-1/2$. Note that CP is preserved when the lowest level is half-filled, which is why we have a factor $1-2\xi_{q-1/2}$. The sums here are all finite, so there is no need for the zeta regularization used for the CP preserving kernels.

If we now plug all the explicit $q>0$ matter kernels into the subleading scaling dimension formula in \eqref{freeFinal}, we find that the integral/sum diverges. To see this divergence, let us write \eqref{freeFinal} as 
\es{freeFinalReg2}{
&\Delta_{q,\kappa}^{(1)}=\frac12\int_{-\infty}^{\infty} {d\omega\ov 2\pi } \sum_{\ell=1}^{\infty}  (2\ell+1)L_\ell^{q,\kappa}(\omega)\,.
}
In Appendix \ref{asymp}, we expand $L_\ell^{q,\kappa}(\omega)$ at large $\ell,\omega$ for all $\kappa>0$ and find
\es{asympLead}{
L_\ell^{q,\kappa}(\omega)=\frac{128\kappa q(2\xi_{q-1/2}-1)}{\pi^2+64\kappa^2}\frac{1}{(\ell+1/2)^2+\omega^2}+O\Bigg(\frac{1}{((\ell+1/2)^2+\omega^2)^{3/2}}\Bigg)\,,
}
where as usual when $\kappa>1/2$ we set $\xi_{q-1/2}=0$. Note there is no other dependence on the filling fractions of the higher levels. The leading term gives a linear divergence after we plug it into the scaling dimension \eqref{freeFinalReg2}. The renormalized scaling dimension is obtained as in the sQED$_3$ case \cite{Dyer:2015zha,Chester:2022wur} by subtracting off the divergent term then adding back its zeta regularized version
\es{zetaFin}{
\int_{-\infty}^{\infty} {d\omega\ov 2\pi } \sum_{\ell=1}^{\infty}\frac{2\ell+1}{(\ell+1/2)^2+\omega^2}=\sum_{\ell=1}^\infty\frac{2\ell+1}{2\ell+1}={\zeta(0,3/2)}=-1\,,
}
to get the final answer
\es{freeFinalReg3}{
&\Delta_{q,\kappa}^{(1)}=\frac12\int_{-\infty}^{\infty} {d\omega\ov 2\pi } \sum_{\ell=1}^{\infty}  (2\ell+1)\Big[L_\ell^{q,\kappa}(\omega)-\frac{128\kappa q(2\xi_{q-1/2}-1)}{\pi^2+64\kappa^2}\frac{1}{(\ell+1/2)^2+\omega^2}\Big]
-\frac{64\kappa q(2\xi_{q-1/2}-1)}{\pi^2+64\kappa^2}\,,
}	
where as before when $\kappa>1/2$ we set $\xi_{q-1/2}=0$. This is the state-independent contribution to the scaling dimension. To compute the scaling dimension in practice, we require two cutoffs. The first cutoff relates to the infinite sum over $j'$ used to define the CP preserving kernels in \eqref{CPfin}. As described in \cite{Dyer:2013fja}, an efficient method of computing this sum is to subtract its asymptotic tail as expanded to order $j'^{-a_\text{cut}}$, then add it back analytically summed using zeta functions. The sum minus its asymptotic tail is then performed up to some cutoff $j'_\text{max}$. 

The second cutoff concerns the final sum and integral in \eqref{freeFinalReg3}. As in \cite{Pufu:2013vpa}, we impose a relativistic cutoff
\es{cutoff}{
\omega^2+\ell(\ell+1)\leq(\ell_\text{max}+1)(\ell_\text{max}+2)\,,
}
which ensures the vanishing of the potential logarithmic divergence, given by the $((\ell+1/2)^2+\omega^2)^{-3/2}$ terms in the asymptotic \eqref{LasApp}. We then subtract the asymptotic, which is computed up to order $((\ell+1/2)^2+\omega^2)^{-b_\text{cut}/2}$, and add back its exactly analytically summed and integrated expression. Finally, we perform the sum/integral for a given $\ell_\text{max}$, and then increase $\ell_\text{max}$ until we get the desired precision. 

\section{Energy splitting}
\label{enSplit}

We will now discuss how to compute the energy splitting term. We will do this first for the QED$_3$ case discussed above, and then for sQED$_3$ following the discussion in \cite{Chester:2017vdh}.

\subsection{QED$_3$}
\label{splitQED}

We will start by reviewing the $\beta$ dependent terms at order $N^0$, following \cite{Chester:2017vdh}. The $\beta$-dependent kernel $\overline{\bf K}_{q,\ell} $ that comes from the large $\beta$ limit of the full kernel is only nonzero for $\omega_n=0$ and $\ell\leq d_{\tilde j}-1$, and takes the form
\es{overK}{
 \overline{\bf K}_{q,\ell} = \xi_{\tilde j}(1-\xi_{\tilde j})  \begin{pmatrix}
    v^E_{q, \ell} \\ v^B_{q, \ell} 
   \end{pmatrix} \begin{pmatrix}
    v^E_{q, \ell} & v^B_{q, \ell} 
   \end{pmatrix} \,,
 } 
where for $\tilde j>q-1/2$ we define
\es{vsSplitF}{
v^E_{\tilde j,q,\ell}
&=
\frac{\operatorname{sgn}(\kappa-\frac12)
      (2\tilde j+1)(-1)^{2\tilde j+\ell}}
     {2\sqrt{4\pi}}
\left[
\begin{pmatrix}
\tilde j & \ell & \tilde j \\
q-\frac12 & 0 & -q+\frac12
\end{pmatrix}
-
\begin{pmatrix}
\tilde j & \ell & \tilde j \\
q+\frac12 & 0 & -q-\frac12
\end{pmatrix}
\right],
\\
v^B_{\tilde j,q,\ell}
&=
-\frac{i(2\tilde j+1)(-1)^{2\tilde j+\ell}
        \lambda_{\tilde j}}
       {\sqrt{4\pi\ell(\ell+1)}}
\left[
\begin{pmatrix}
\tilde j & \ell & \tilde j \\
q-\frac12 & 0 & -q+\frac12
\end{pmatrix}
+
\begin{pmatrix}
\tilde j & \ell & \tilde j \\
q+\frac12 & 0 & -q-\frac12
\end{pmatrix}
\right]\,,
}
while for the zero mode we have
\es{vszero}{
v^E_{q-\frac12,q,\ell}
=
\frac{\operatorname{sgn}(\kappa-\frac12)(2q)!}
     {\sqrt{4\pi(2q-\ell-1)!(2q+\ell)!}},
\qquad
v^B_{q-\frac12,q,\ell}=0\,.
}
We can then plug this into \eqref{F1true} to get the temperature dependent subleading free energy
\es{F1smooth}{
 F_{q,\beta}^{(1)}=& \frac{\log \le(N\, 2\pi d_{\tilde j} \,\xi_{\tilde j} (1-\xi_{\tilde j}) \ri)  +(d_{\tilde j}^2-1){\log\beta} + \sum_{\ell=1}^{d_{\tilde j}-1} (2\ell+1)\log\left( \xi_{\tilde j}(1-\xi_{\tilde j}) C_{\tilde j, \ell}+\beta^{-1}\right)}{2 \beta} \,,
}
where note that the first term came from the second term in \eqref{F1true} and we defined 
 \es{CDef}{
  C_{\tilde j, \ell} \equiv  \begin{pmatrix}
    v^E_{q, \ell} & v^B_{q, \ell} 
   \end{pmatrix}  \widetilde{{\bf K}}^{-1}_{q, \ell}(0) \begin{pmatrix}
    v^E_{q, \ell} \\ v^B_{q, \ell} 
   \end{pmatrix} \,.
 }

These temperature-dependent terms in the thermal free energy arise from a trace over the dressing modes in the thermal path integral. To determine the energy splitting, we need to find the Hamiltonian formed from oscillators on $S^2\times \mathbb{R}$ whose trace recovers these terms. These oscillators are defined as \cite{Chester:2017vdh}
\es{psiDecomp}{
\psi^i(x)={}&\sum_{m=1/2-q}^{q-1/2}c^{i,\dagger}_{q-1/2,m}\Psi_{q-1/2,m}(x)
 +\sum_{j>q-1/2,m}\left(a^{i,\dagger}_{jm}\Psi_{jm,-}^{\rm sph}(x)e^{\lambda_j\tau}
 +b^i_{jm}\Psi_{jm,+}^{\rm sph}(x)e^{-\lambda_j\tau}\right),\\
\psi_i^\dagger(x)={}&i\Bigg[\sum_{m=1/2-q}^{q-1/2}c_{q-1/2,m,i}\Psi_{q-1/2,m}^\dagger(x)\\
 &\qquad+\sum_{j>q-1/2,m}\left(a_{jm,i}\Psi_{jm,-}^{\rm sph\dagger}(x)e^{-\lambda_j\tau}
 +b^\dagger_{jm,i}\Psi_{jm,+}^{\rm sph\dagger}(x)e^{\lambda_j\tau}\right)\Bigg]\gamma^\tau(x)\,.
}
where $t=-i\tau$, so creation operators have $e^{+\lambda_j\tau}$ and annihilation operators have $e^{-\lambda_j\tau}$. The spinors $\Psi_{jm,\pm}^{\rm sph}$ have static energies $\pm\lambda_j$; hence creation operators in $\psi$ multiply the negative-energy branch. The second line is the Euclidean conjugate appearing in \eqref{QEDAction}: it equals $i$ times the canonical oscillator adjoint times $\gamma^\tau$. This converts the temporal kinetic term to the usual canonical one and reproduces the propagator convention \eqref{GreenDefF}. We define
\es{ABC}{
 \Psi_{j m,+}^\text{sph}(x)&=\frac{(qT_{qjm}(x)+(\lambda_j+j+1/2)S_{qjm}(x))}{\sqrt{(2j+1)(j+1/2+\lambda_j)}}\,,\\
  \Psi_{j m,-}^\text{sph}(x)&=-\frac{(qT_{qjm}(x)+(\lambda_j-j-1/2)S_{qjm}(x))}{\sqrt{(2j+1)(j+1/2-\lambda_j)}}\,, \quad \Psi_{q-1/2,m}(x)=S_{q,q-1/2,m}(x)\,.
}
The coefficients $c^{i,\dagger}_{q-1/2, m}$, $a^{i,\dagger}_{j m}$, $b^\dagger_{j m, i}$ can be interpreted as fermionic creation operators, and $c_{q-1/2, m, i}$, $a_{j m, i}$, $b_{j m}^i$ are the corresponding annihilation operators. In the generalized free field theory defined by the large $N$ saddle, the nonzero Wick contractions are
\es{wickFerm}{
 \left\langle a^{i,\dagger}_{\tilde j m_1}
                    a_{\tilde j m_2,i'}\right\rangle_q
 &=\xi_{\tilde j}\,\delta^i_{i'}\delta_{m_1m_2},\qquad \left\langle a_{\tilde j m_1,i}
                    a^{i',\dagger}_{\tilde j m_2}\right\rangle_q
 =(1-\xi_{\tilde j})\,\delta_i^{i'}\delta_{m_1m_2}\,,
}
These contractions describe holes in the negative-energy valence shell for $\kappa>1/2$; for $\kappa<-1/2$ the positive-energy $b^\dagger$ modes have the analogous contractions. With the zero-mode convention in \eqref{psiDecomp}, $\langle c^\dagger c\rangle_q=1-\xi_{q-1/2}$ and $\langle c c^\dagger\rangle_q=\xi_{q-1/2}$, with the same flavor and magnetic Kronecker deltas. In particular both equal $1/2$ at $\kappa=0$.

We then plug this oscillator expansion into the action \eqref{QEDAction}, integrate the triple product of monopole harmonics using \eqref{tripY}, and find that the modes corresponding to the dressing $\tilde j$ appear (for $\kappa>\frac12$) as
\es{shellOsc}{
 &-i\sqrt{\beta}
 \sum_{\ell=1}^{2\tilde j}\sum_{m=-\ell}^{\ell}
\sqrt{({2\ell+1})} \left[\mathfrak a^E_{\ell m}(0)^*v^E_{q,\ell}
      +\mathfrak a^B_{\ell m}(0)^*v^B_{q,\ell}\right]\\
 &\qquad\times\sum_{i=1}^N\sum_{m_1,m_2=-\tilde j}^{\tilde j}
 a^{i,\dagger}_{\tilde j m_1}
 a_{\tilde j m_2,i}\,(-1)^{\tilde j-q-1/2+\ell}(-1)^{\tilde j-m_1}
 \begin{pmatrix}\tilde j&\ell&\tilde j\\-m_1&m&m_2\end{pmatrix}\,,
}
while for other $\kappa$ the corresponding modes appear. The phase $(-1)^{\tilde j-q-1/2+\ell}$ follows from the spinors in \eqref{ABC} and the harmonic phases in \eqref{vsSplitF}; it squares to one and drops out of the splitting Hamiltonian. The full action at this order is this term plus the usual $\beta$-independent term in \eqref{NumExponent}. We can then integrate out the eigenmodes $\mathfrak{a}$ of the full action to get the usual $\beta$-independent, state-independent scaling dimension \eqref{freeFinal}, as well as the $\beta$-dependent term 
\es{therm}{
F_{q,\beta}^{(1)}=&\frac{1}{2\beta}\log \le(N\, 2\pi d_{\tilde j} \,\xi_{\tilde j} (1-\xi_{\tilde j}) \ri)
-\frac{1}{\beta}\log\left(\frac{\Tr e^{-\beta H^{N,\tilde j,q}_\text{split}}}{\binom{Nd_{\tilde j}}{Nd_{\tilde j}\xi_{\tilde j}}}\right)\,.
}
Here the trace is restricted to $Nd_{\tilde j}\xi_{\tilde j}$ dressing excitations in a nonzero-energy shell. In the zero-mode hole convention of \eqref{psiDecomp}, the occupation is instead $Nd_{q-1/2}(1-\xi_{q-1/2})$; the binomial coefficient is unchanged under $\xi\leftrightarrow1-\xi$. The binomial coefficient in the denominator is the size of this fixed-occupation Hilbert space. The energy-splitting Hamiltonian (for $\kappa>\frac12$) is 
\es{HsplitF}{
H^{N,\tilde j,q}_\text{split}=&\frac1N\sum_{\ell=1}^{2\tilde j}C_{\tilde j,\ell}
 \sum_{m=-\ell}^{\ell}\sum_{i,i'=1}^{N}
 \sum_{m_1,m_2,m_3,m_4=-\tilde j}^{\tilde j}
 \frac{2\ell+1}{2}(-1)^{2\tilde j-m_1-m_3}\\
& \times\begin{pmatrix}\tilde j&\ell&\tilde j\\-m_1&m&m_2\end{pmatrix}
 \begin{pmatrix}\tilde j&\ell&\tilde j\\-m_3&m&m_4\end{pmatrix}a^{i,\dagger}_{\tilde j m_2}a_{\tilde j m_1,i}
 a^{i',\dagger}_{\tilde j m_3}a_{\tilde j m_4,i'}\,,
}
while for other $\kappa$ the corresponding modes appear. As a consistency check, one can perform the trace over the oscillators in \eqref{therm} using the Wick contractions for the oscillators in \eqref{wickFerm} to recover the usual $\beta$-dependent term in \eqref{F1smooth}. The advantage of the oscillator formulation in \eqref{therm} is that one can diagonalize it in terms of the oscillator basis to find the energy splitting $\Delta_{q,\ell,R}$ corresponding to a monopole with spin $\ell$ and flavor irrep $R$ of $SU(N)$, with a further multiplicity label understood when the same pair $(\ell,R)$ occurs more than once. The Hamiltonian must also be diagonalized within this multiplicity space. In the tables, an asterisk distinguishes the second operator with the same charge and spin. This splitting is of order $1/N$ when the representation of the traceless orbital bilinears under $SU(d_{\tilde j})$ is held fixed as $N$ increases, even though it arose from an $N^0$ term in the original action.

For instance, consider $N=2$, $q=1$, $k=0$, and $\tilde j=1/2$, then the energy-splitting Hamiltonian is
\es{Hexp}{
H^{2,1/2,1}_\text{split}=\frac{C_{\tilde j,1}}N L^2\,,\qquad  L^a&=\sum_{i=1}^N\sum_{m,m'=-1/2}^{1/2}
 c^{i,\dagger}_{1/2,m}\left(\frac{\sigma_a}{2}\right)_{mm'}
 c_{1/2,m',i}\,,
}
where $L^a$ is a generator of the spacetime rotation $SU(2)$. Two representative eigenstates of this Hamiltonian are $c^{1,\dagger}_{1/2,1/2}c^{1,\dagger}_{1/2,-1/2}
 |M_{\rm bare}\rangle$ and $c^{1,\dagger}_{1/2,1/2}c^{2,\dagger}_{1/2,1/2}
 |M_{\rm bare}\rangle$. Under $SU(2)_{\rm rot}\times SU(2)_{\rm flavor}$, the first state lies in $(\mathbf{1},\mathbf{3})$ and has splitting energy $\Delta^{\rm split}=0$, while the second lies in $(\mathbf{3},\mathbf{1})$ and has splitting energy $\Delta^{\rm split}=2C_{1/2,1}/N=C_{1/2,1}$ at $N=2$.

\subsection{sQED$_3$}
\label{splitsQED}

We will now discuss the energy splitting for the scalar case. As these details are described in \cite{Chester:2017vdh,Chester:2022wur}, we will be brief. The action of $N$ complex scalars $\phi_i$ on $S^2\times S^1_\beta$ is
\es{actionS}{
\mathcal{S}&=\int\! d^3x\Big[  { \sqrt{g}}  \left(\abs{(\nabla_\mu-iA_\mu)\phi^i}^2+(\frac{1}{4}+i\lambda)|\phi^i|^2\right)-\frac{ik}{4\pi}\epsilon^{\mu\nu\rho}A_\mu\partial_\nu A_\rho\Big]\,,
}
where $g$ is the determinant of the metric, $\lambda$ is a Hubbard-Stratonovich field, and $i=1,\dots, N$. After integrating out the matter, we now expand both the gauge field and $i\lambda$ around saddle point values
\es{saddleS}{
A_\mu=\mathcal{A}_\mu+a_\mu\,,\qquad i\lambda=\mu+i\sigma\,.
}
The saddle point equations fix the holonomy to be
\es{CPNalph}{
\alpha(\kappa)&=-\sgn(\kappa)\left(\lambda_{q}+\beta^{-1}\log\frac{\xi}{1+\xi}\right)+O(e^{-(\lambda_{q+1}-\lambda_q)\beta})\,,\qquad
\xi \equiv \frac{2q |\kappa |}{d_q} \,,
} 
while $\mu$ is given by the zeta regularized expression
\es{muSad}{
\sum_{j\geq q} \left(\frac{d_j}{ 2\lambda_j(\mu)}-1\right)-q+\frac{\xi d_q}{2 \lambda_q(\mu)}=0\,.
}
Here $\lambda_j=\sqrt{(j+1/2)^2-q^2+\mu}$ is the eigenvalue of the Klein-Gordon operator on $S^2\times \mathbb{R}$ with the saddle-point value $\mu$, with degeneracy $d_j=2j+1$. Note that the holonomy saddle corresponds to the lowest mode $j=q$ for all $\kappa\neq0$, so there are no Landau levels unlike the fermionic case. We can then use these saddle point values to compute the leading order free energy
\es{CPNFinal}{
\Delta_{q}^{(0)}&=\sum_{j \geq q} (d_j\lambda_j-2(j+1/2)^2-\mu+q^2)+\frac{2q^3+q-6\mu q}{6}+\xi d_q\lambda_{q} \,,\\
S_{q}^{(0)}&=-d_q\left(\xi\log\xi-(1+\xi)\log[1+\xi]\right)\,,
}
where we show the zeta regularized scaling dimension, which as in the fermionic case consists of a Casimir term plus the dressing. 
The entropy $NS_{q}^{(0)}$ corresponds to the logarithm of the number of degenerate monopoles, which correspond to all operators in $SU(N)\times SU(d_q)$ irreps
\es{symsum}{
\bigoplus_\nu(\Upsilon_\nu,\Upsilon_\nu)
}
where $\Upsilon_\nu$ is a Young diagram with a total of $d_qN\xi $ boxes such that $\Upsilon_\nu$ has maximum height $\min(d_q,N)$.

The subleading calculation proceeds similarly to the fermionic case, except now we will have modes for $\sigma$ as well. For instance, the scalar version of the subleading Fourier space action \eqref{NumExponent} is 
 \es{NumExponentScal}{
  &{}-\frac{N}{2} \begin{pmatrix} b_{00}(0) \\ \mathfrak{a}_{00}^E(0)
   \end{pmatrix}^\dagger
   {\bf K}_{q, 0}(0) \begin{pmatrix} b_{00}(0) \\ \mathfrak{a}_{00}^E(0)
   \end{pmatrix}
    - \frac{N}{2} \sum_{n \neq 0} \abs{b_{00}(\omega_n)}^2 K_{q, 0}(\omega_n) \\
  &{}-\frac{N}{2} \sum_{n=-\infty}^\infty \sum_{\ell=1}^\infty \sum_{m=-\ell}^\ell
   \begin{pmatrix} b_{\ell m}(\omega_n) \\ \mathfrak{a}_{\ell m}^E(\omega_n) \\ \mathfrak{a}_{\ell m}^B(\omega_n)
   \end{pmatrix}^\dagger
   {\bf K}_{q, \ell}(\omega_n) \begin{pmatrix} b_{\ell m}(\omega_n) \\ \mathfrak{a}_{\ell m}^E(\omega_n) \\ \mathfrak{a}_{\ell m}^B(\omega_n)
   \end{pmatrix} \,,
 }
where $b$ is the eigenmode for $\sigma$, ${\bf K}_{q, \ell}(\omega_n)$ is a $3\times 3$ matrix for $\ell \geq 1$, ${\bf K}_{q, 0}(0)$ is a $2\times 2$ matrix, and $K_{q, 0}(\omega_n)$ for $n \neq 0$ is just a number. The $\beta$-dependent kernel $\overline{\bf K}_{q,\ell} $ for the scalar case is 
\es{overKS}{
 \overline{\bold K}_{q,\ell}=\xi(1+\xi)  \begin{pmatrix}
   v^\sigma_{q, \ell} \\   v^E_{q, \ell} \\ v^B_{q, \ell} 
   \end{pmatrix} 
   \begin{pmatrix}
   v^\sigma_{q, \ell} & v^E_{q, \ell} & v^B_{q, \ell} 
   \end{pmatrix}
 } 
where we now define 
\es{scalQEDKvals}{
v_{q,\ell}^\sigma&=\frac{(2q+1)!}{2\lambda_q\sqrt{4\pi(2q-\ell)!(2q+\ell+1)!}}\,,\\
v_{q,\ell}^E&=\frac{\sgn(\kappa)(2q+1)!}{\sqrt{4\pi(2q-\ell)!(2q+\ell+1)!}}\,,\\
v_{q,\ell}^B&=\frac{-i(2q+1)!\sqrt{\ell(\ell+1)}}{2\lambda_q\sqrt{4\pi(2q-\ell)!(2q+\ell+1)!}}\,.
}
The $\beta$-dependent subleading free energy is
\es{F1smoothS}{
 F_{q,\beta}^{(1)}=& \frac{\log \le(N\, 2\pi d_q \,\xi (1+\xi) \ri)  +(d_q^2-1){\log\beta} + \sum_{\ell=1}^{d_{q}-1} (2\ell+1)\log\left( \xi(1+\xi) C_{q, \ell}+\beta^{-1}\right)}{2 \beta} \,,
}
where we define
 \es{DivInKscal}{
  C_{q,\ell}= \begin{pmatrix}
   v^\sigma_{q, \ell} &   v^E_{q, \ell} & v^B_{q, \ell} 
   \end{pmatrix} \widetilde{\bold K}^{-1}_{q,\ell}(0)
   \begin{pmatrix}
   v^\sigma_{q, \ell} \\  v^E_{q, \ell} \\ v^B_{q, \ell} 
   \end{pmatrix} \,,
 } 
 and the $\beta$-independent kernels $ \widetilde{\bold K}_{q,\ell}(\omega)$ are given explicitly in \cite{Chester:2022wur}.\footnote{As discussed in the introduction, the sign of the CS kernel is the opposite of what is written in \cite{Chester:2022wur}.}
 
 The oscillator expansions analytically continued to Euclidean space are defined as 
 \es{ModeExpansionScal}{
\phi^{i}(x)&=\sum_{j=q}^\infty \sum_{m=-j}^j \frac{1}{\sqrt{ 2 \lambda_j}} \left[ a^{i,\dagger}_{j m}\, Y_{q j m}(\theta, \phi)e^{\lambda_j \tau}+b^{i}_{j m}\,Y_{q j m}(\theta, \phi)e^{-\lambda_j \tau} \right] \,,\\
\phi_i^\dagger(x)&=\sum_{j=q}^\infty \sum_{m=-j}^j  \frac{1}{\sqrt{ 2 \lambda_j}} \left[ a_{j m, i}\, Y^*_{q j m}(\theta, \phi)e^{-\lambda_j \tau}+b^\dagger_{j m, i}\,Y^*_{q j m}(\theta, \phi)e^{\lambda_j \tau} \right] \,.
}
In the generalized free field theory defined by the large $N$ saddle, the nonzero Wick contractions are
\es{wickS}{
 \left\langle a^{i,\dagger}_{q m_1}
                    a_{q m_2,i'}\right\rangle_q
 &=\xi\,\delta^i_{i'}\delta_{m_1m_2},\qquad \left\langle a_{q m_1,i}
                    a^{i',\dagger}_{q m_2}\right\rangle_q
 =(1+\xi)\,\delta_i^{i'}\delta_{m_1m_2}\,.
}
We then plug this oscillator expansion into the action \eqref{actionS}, integrate the triple product of monopole harmonics using \eqref{tripY}, and find that the modes corresponding to the dressing $j=q$ appear (for $\kappa<0$) as
\es{shellOscS}{
 -i\sqrt{\beta}
 \sum_{\ell=1}^{2q}\sum_{m=-\ell}^{\ell}
&\sqrt{({2\ell+1})} \left[\mathfrak b_{\ell m}(0)^*v^\sigma_{q,\ell}+\mathfrak a^E_{\ell m}(0)^*v^E_{q,\ell}
      +\mathfrak a^B_{\ell m}(0)^*v^B_{q,\ell}\right]\\
  &\qquad   \times \sum_{i=1}^N\sum_{m_1,m_2=-q}^{q}
 a^{i,\dagger}_{q m_1}
 a_{qm_2,i}\,(-1)^{ q-m_1}
 \begin{pmatrix} q&\ell&q\\-m_1&m&m_2\end{pmatrix}\,,
}
while for $\kappa>0$ the corresponding mode appears. The full action at this order is this term plus the usual $\beta$-independent term in \eqref{NumExponentScal}. We can then integrate out the eigenmodes of the full action to get the usual $\beta$-independent, state-independent scaling dimension, as well as the $\beta$-dependent term 
\es{thermS}{
F_{q,\beta}^{(1)}=&\frac{1}{2\beta}\log \le(N\, 2\pi (2q+1) \,\xi (1+\xi) \ri)
-\frac{1}{\beta}\log\left(\frac{\Tr e^{-\beta H^{N,q}_\text{split}}}{\binom{N(2q+1)+2q|k|-1}{2q|k|}}\right)\,,
}
Here the trace is restricted to states with $2q|k|$ scalars in the dressing shell. The binomial coefficient in the denominator is the size of this fixed-occupation Hilbert space. The energy-splitting Hamiltonian (for $\kappa<0$) is 
\es{HsplitS}{
H^{N,q}_\text{split}=&\frac1N\sum_{\ell=1}^{2 q}C_{ q,\ell}
 \sum_{m=-\ell}^{\ell}\sum_{i,i'=1}^{N}
 \sum_{m_1,m_2,m_3,m_4=-q}^{q}
 \frac{2\ell+1}{2}(-1)^{2q-m_1-m_3}\\
& \times\begin{pmatrix}q&\ell&q\\-m_1&m&m_2\end{pmatrix}
 \begin{pmatrix}q&\ell&q\\-m_3&m&m_4\end{pmatrix}a^{i,\dagger}_{q m_2}a_{q m_1,i}
 a^{i',\dagger}_{q m_3}a_{ q m_4,i'}\,,
}
while for other $\kappa$ the corresponding modes appear. As a consistency check, one can perform the trace over the oscillators in \eqref{thermS} using the Wick contractions for the oscillators in \eqref{wickS} to recover the usual $\beta$-dependent term in \eqref{F1smoothS}. As in the fermionic case, one can then diagonalize this action to get the energy splitting for each state.

\section{Non-spherical saddles}\label{nonS}

In the previous section, we discussed the $\beta$ dependent terms at subleading order in $N$ for QED$_3$, which we gave in \eqref{F1smooth} in terms of a quantity $ C_{\tilde j, \ell} $ that is written in terms of the matter kernel in \eqref{CDef}. If $C_{\tilde j, \ell}$ is negative, then for some large $\beta$ we will get an imaginary term in \eqref{F1smooth}, which makes the saddle unstable. This occurs for $|k|>1/2$ when the excited Landau levels are partially filled. In these cases, we must instead consider saddles that break the $S^2$ spherical symmetry $SO(3)$ down to one of $G=O(2), T, O, D_n$, where $T$ is the tetrahedral group, $O$ is the octahedral, and $D_n$ is the order $2n$ dihedral group. We compute the leading and subleading in $N$ terms for these non-spherical saddles below. Sphere rotations also relate different non-spherical saddles that give degenerate energies at large $N$, analogous to the degeneracy in the spherical saddle. We compute the energy splitting term that distinguishes these states.

\subsection{Leading order free energy}
\label{leadNonS}

We consider a general saddle $\cA$ that satisfies $\int_{S^2} d\cA=4\pi q$:
\es{genSad}{
\mathcal{A}_\tau=-i\Big(\alpha+\sum_{\ell=1}^{\ell_\text{bg}}\sum_{m=-\ell}^\ell\alpha_{\ell m}Y_{\ell m}(\theta,\phi)\Big)\,,\qquad \cA_a=\cA_a^\text{sph}+\sum_{\ell=1}^{\ell_\text{bg}}\sum_{m=-\ell}^\ell\cA^B_{\ell m}\cB_{\ell m,a}(\theta ,\phi)\,, 
}
Here $\alpha$ is the usual holonomy, $\cA_a^\text{sph}$ is the spherically symmetric saddle in \eqref{background}, $\cB_{\ell m,a}$ is the vector harmonic in \eqref{EBDef}, and $\alpha_{\ell m}, \cA^B_{\ell m}$ are the new saddle variables. At zero frequency, $\mathcal E_{0\ell m}=Y_{\ell m}d\tau$, so $-i\alpha_{\ell m}$ and $\mathcal A^B_{\ell m}$ are precisely the static $E,B$ coefficients. The temporal factor $-i$ is the same convention as for the constant holonomy $\alpha$. When these variables are nonzero, the saddle breaks the spherical symmetry. A generic non-spherical saddle will include all infinite $\ell$ modes, but for practical numerical calculations we put a cutoff $\ell_\text{bg}$, which we choose big enough to get convergence to a certain level of precision. Note that the B-modes integrate to zero over the sphere, which is why we still satisfy $\int_{S^2} d\cA=4\pi q$ via the spherical term $\cA_a^\text{sph}$.

Consider the Dirac equation for this non-spherical saddle:
\es{DiracNonS}{
\left[\gamma^\tau\gamma^a(\nabla_a-i\mathcal A_a)
 -\sum_{\ell=1}^{\ell_\text{bg}}\sum_{m=-\ell}^{\ell}\alpha_{\ell m}Y_{\ell m}(\theta,\phi)\right]\Psi_r
 =\varepsilon_r\Psi_r\,,\qquad a\in\{\theta,\phi\}\,.
}
Here only the spatial components $\mathcal A_a$ enter the covariant derivative, and the constant holonomy is excluded from $\varepsilon_r$. This is the static Hamiltonian obtained by factoring $i\gamma^\tau$ from the three-dimensional Dirac operator. Retaining complete spherical multiplets gives the finite dimension 
\es{rcut}{
D=2q+2\sum_{j=q+1/2}^{j_{\max}}(2j+1),\qquad r=1,\ldots,D\,.
}
The first term counts the zero modes; the factor of two counts both nonzero-energy branches. We use $j_{\max}$ for this internal Dirac cutoff throughout; its numerical value can be chosen separately for the leading energy, Gaussian correction, and rotational splitting. We can expand the non-spherical eigenfunctions in terms of the spherical eigenfunctions in \eqref{ABC} as
\es{nsPsi}{
\Psi_r(x)=\sum_{t=1}^{D} c_{tr}\Psi_t^\text{sph}(x)\,,
}
where the $t$ index here runs over the usual $(j,m,\pm)$ indices for the spherical eigenfunction in \eqref{ABC}. Using this basis, we can then express the non-spherical Dirac equation \eqref{DiracNonS} as the finite algebraic eigenvalue problem
\es{matrixeigen}{
 \sum_{t=1}^{D}\Big[
  \varepsilon_s^{\rm sph}\delta_{st}
  +\sum_{\ell=1}^{\ell_\text{bg}}\sum_{m=-\ell}^\ell\mathcal A^B_{\ell m}(V_{B\ell m}^{\rm sph})_{st}
  +\sum_{\ell=1}^{\ell_\text{bg}}\sum_{m=-\ell}^\ell\alpha_{\ell m}(V_{\alpha\ell m}^{\rm sph})_{st}
 \Big]c_{tr}=\varepsilon_r c_{sr}\,,
}
where the eigenfunction coefficients satisfy $\sum_s c_{sr}^*c_{sr'}=\delta_{rr'}$, the spherical eigenvalues $  \varepsilon_s^{\rm sph}$ are zero for $j=q-1/2$ and $\eta\lambda_j$ for $j>q-1/2$, with $\eta=\pm$, indexed by $t=(j,m,\eta)$, and we use the triple harmonic formula \eqref{tripY} to compute the coefficients
\es{VsNS}{
 &(V_{\alpha\ell M}^{\rm sph})_{jm\eta,j'm'\eta'}
 =-(-1)^{q+m-1/2}
 \sqrt{\frac{(2j+1)(2j'+1)(2\ell+1)}
 {4\pi(2-\delta_{j,q-1/2})(2-\delta_{j',q-1/2})}}\\
 &\quad\times\begin{pmatrix}j&\ell&j'\\-m&M&m'\end{pmatrix}\left[
 \begin{pmatrix}j&\ell&j'\\q-1/2&0&-q+1/2\end{pmatrix}
 -\eta\eta'
 \begin{pmatrix}j&\ell&j'\\q+1/2&0&-q-1/2\end{pmatrix}
 \right]\,,\\
&(V_{B\ell M}^{\rm sph})_{jm\eta,j'm'\eta'}
 =\frac{(-1)^{q+m-1/2}(\eta\lambda_j+\eta'\lambda_{j'})}
 {\sqrt{\ell(\ell+1)}}
 \sqrt{\frac{(2j+1)(2j'+1)(2\ell+1)}
 {4\pi(2-\delta_{j,q-1/2})(2-\delta_{j',q-1/2})}}\\
 &\quad\times\begin{pmatrix}j&\ell&j'\\-m&M&m'\end{pmatrix}\left[
 \begin{pmatrix}j&\ell&j'\\q-1/2&0&-q+1/2\end{pmatrix}
 +\eta\eta'
 \begin{pmatrix}j&\ell&j'\\q+1/2&0&-q-1/2\end{pmatrix}
 \right]\,,
}
where $\eta,\eta'=\pm$ for the nonzero-energy branches. For a zero mode there is only one branch; the second $3j$ symbol in each bracket vanishes, so its arbitrary $\eta$ label drops out. 

Using these eigenvalues, we can then compute the leading free energy as in the spherical case to get
\es{leadingNSF}{
 F_{q}^{(0)}(\alpha,\mathcal A^B_{\ell m},\alpha_{\ell m})
 ={}&-\frac1\beta\sum_{r=1}^{D}
       \log\left[2\cosh\frac{\beta(\varepsilon_r[\mathcal A^B_{\ell m},\alpha_{\ell m}]-\alpha)}2\right]
 -2q\kappa\alpha\\
 &+\frac{\kappa}{2\pi}\sum_{\ell=1}^{\ell_\text{bg}}\sum_{m=-\ell}^\ell\sqrt{\ell(\ell+1)}
                      \alpha_{\ell m}^*\mathcal A^B_{\ell m}\,,
}
where we have imposed the cutoffs $D$ on the eigenvalues and $\ell_\text{bg}$ on the saddles. We can then fix the saddles using the saddle point equations. For the holonomy we get 
\es{saddlesHol}{
0=\frac{\partial F^{(0)}_q}{\partial\alpha}
 =-2q\kappa-\frac12\sum_{r=1}^{D}\tanh\!\left[\frac\beta2(\alpha-\varepsilon_r)\right]\,.
}
For the gapped branches considered here, $D/2-2q\kappa$ is an integer and $\varepsilon_{D/2-2q\kappa}<\varepsilon_{D/2-2q\kappa+1}$. All levels below this gap are filled. In this case the large-$\beta$ solution, up to exponentially small terms, is
\es{alphaSolNS}{
\alpha={}&\frac{\varepsilon_{D/2-2q\kappa}
                    +\varepsilon_{D/2-2q\kappa+1}}2+\frac1{2\beta}\log
 \frac{\#\{r:\varepsilon_r=\varepsilon_{D/2-2q\kappa}\}}
      {\#\{r:\varepsilon_r=\varepsilon_{D/2-2q\kappa+1}\}}\,,
}
where note that $\varepsilon_r$ are ordered in increasing size. The second term comes from the fact that there are degeneracies that can occur when a non-spherical saddle preserves a subgroup $G\subset SO(3)$, which is generically the case. We can plug this holonomy into the free energy to get the leading order scaling dimension
\es{rawenergy}{
& \Delta_{q}^{(0)}[\mathcal A^B_{\ell m},\alpha_{\ell m}]
 =\lim_{\ell_\text{bg}\to\infty}\lim_{\Lambda\to\infty}\lim_{D\to\infty}\Bigg[\sum_{r=1}^{D}(n_r-\tfrac12)
                   \varepsilon_r[\mathcal A^B_{\ell m},\alpha_{\ell m}]\\
 &\quad+\frac12\sum_{r=1}^{D}|\varepsilon_r[\mathcal A^B_{\ell m},0]|
           \left(1-e^{-\varepsilon_r[\mathcal A^B_{\ell m},0]^2/\Lambda^2}\right)+\frac{\sqrt{\pi}}{2}\Lambda^3
           +\frac{\kappa}{2\pi}\sum_{\ell m}\sqrt{\ell(\ell+1)}
                  \alpha_{\ell m}^*\mathcal A^B_{\ell m}\Bigg]\,,
}
where the occupation number $n_r$ is $1$ for $r\leq D/2-2q\kappa$ and $0$ otherwise, and we regularized the numerical sum with a cutoff $\Lambda$. It is important we take the three cutoffs to infinity in the order listed. Note that there is no entropy term, unlike the spherical case, because the non-spherical saddle corresponds to a unique state.

Finally, we consider the saddle point equations for $\alpha_{\ell m}$ and $\cA^B_{\ell m}$:
\es{sadAlphas}{
& 0=\frac{\partial F^{(0)}_q}{\partial \alpha_{\ell m}}=\sum_{r,t,u=1}^D(n_r-\tfrac12)
          c_{tr}^*(V_{\alpha\ell m}^{\rm sph})_{tu}c_{ur}
       +\frac{\kappa\sqrt{\ell(\ell+1)}}{2\pi}
                         \mathcal A^{B*}_{\ell m},\\
&0=\frac{\partial F^{(0)}_q}{\partial \mathcal A^B_{\ell m}}=\frac{\kappa\sqrt{\ell(\ell+1)}}{2\pi}\alpha_{\ell m}^*+\sum_{r,t,u=1}^D(n_r-\tfrac12)
          c_{tr}^*(V_{B\ell m}^{\rm sph})_{tu}c_{ur}\\
 &\quad+\sum_{r,t,u=1}^D \frac{\sgn(\varepsilon_r[\mathcal A^B_{\ell m},0])}{2}\Big[1+(\frac{2\varepsilon_r[\mathcal A^B_{\ell m},0]^2}{\Lambda^2}-1)e^{-\varepsilon_r[\mathcal A^B_{\ell m},0]^2/\Lambda^2}\Big]
\\
 &\qquad\times c_{tr}^*[\mathcal A^B_{\ell m},0](V_{B\ell m}^{\rm sph})_{tu}
          c_{ur}[\mathcal A^B_{\ell m},0]\,,
}
where we used \eqref{matrixeigen}. With complex harmonics, $\alpha_{\ell,-m}=(-1)^m\alpha_{\ell m}^*$ and $\mathcal A^B_{\ell,-m}=(-1)^m\mathcal A^{B*}_{\ell m}$. Consequently the CS sum in \eqref{rawenergy} can equivalently be written as $\sum_{\ell m}\sqrt{\ell(\ell+1)}\alpha_{\ell m}\mathcal A^{B*}_{\ell m}$. The derivatives above are with respect to the unstarred coefficients; varying their independent real and imaginary parts gives the same saddle equations. These are variations of the static $E,B$ components in \eqref{genSad}. To compute the scaling dimension in practice, we start by plugging in trial values of $\alpha_{\ell m}$ and $\cA^B_{\ell m}$ into \eqref{matrixeigen}, and numerically computing the eigenvalues $\varepsilon_r$ and eigenfunction values $c_{tr}$, also for zero $\alpha_{\ell m}$ as needed for the second equation in \eqref{sadAlphas}. We then insert these values into the saddle point equations \eqref{sadAlphas}, and iterate the process using Newton's method until the saddle point equations are numerically close to zero. We do this for increasingly larger $\ell_\text{bg}$ and $D$ until we find the scaling dimension converges to the desired number of digits. The results for the various $\kappa$ we consider will be reported in the next section.

\subsection{Subleading order free energy}
\label{subNonS}

We start by computing the Green's function, by solving \eqref{GreenDefF} for the non-spherical saddle. We get
\es{nonSGreen}{
G_q^{\rm new}(x,x')&=
 \frac{i}{\beta}\sum_{n\in\mathbb Z}\sum_r
 \frac{e^{-\pi i(2n+1)(\tau-\tau')/\beta}
       \Psi_r(\Omega)\Psi_r^\dagger(\Omega')\gamma^\tau(\Omega')}
      {-\pi i(2n+1)/\beta+\varepsilon_r-\alpha}\\
 ={}&i\sum_{r}\Psi_r(\Omega)\Psi_r^\dagger(\Omega')\gamma^\tau(\Omega')
 e^{-(\varepsilon_r-\alpha)(\tau-\tau')}\left[\frac{\Theta(\tau-\tau')}{e^{-\beta(\varepsilon_r-\alpha)}+1}
                 -\frac{\Theta(\tau'-\tau)}{e^{\beta(\varepsilon_r-\alpha)}+1}\right]\,.
}
We can then use this to define the position space matter kernel \eqref{Ks}, and go to momentum space as in the spherical case to get 
\es{KnonS}{
 K_{q;\ell m,\ell'm'}^{XX',\rm new}(\omega_n)
 ={}&\int_{-\beta/2}^{\beta/2}d\tau\,e^{i\omega_n\tau}
 \int_0^{2\pi}d\phi\int_0^\pi d\theta\sin\theta
 \int_0^{2\pi}d\phi'\int_0^\pi d\theta'\sin\theta'\\
 &\times\mathcal X^*_{\omega\ell m,\mu}(\Omega)
 K_q^{\mu\nu,\rm new}(\tau,\Omega;0,\Omega')
 \mathcal X'_{\omega\ell'm',\nu}(\Omega'),\qquad X,X'\in\{E,B\}\,,
 }
 where relative to the spherical case \eqref{fourierKernscal} we can no longer remove one of the spherical integrals using rotational symmetry. We can compute this by plugging the Green's function \eqref{nonSGreen} into the position space kernel \eqref{Ks}, expanding the Dirac eigenfunctions in spherical eigenfunctions as \eqref{nsPsi}, and computing the integrals of monopole spherical harmonics using the triple harmonic formula \eqref{tripY}. At large $\beta$, up to exponentially small terms, we obtain\footnote{For coincident spatial eigenvalues, including $r=s$, the divided difference in the regulating term means the derivative of its displayed numerator function with respect to the common eigenvalue. Its limiting contribution, including the factor $1/2$ outside that sum, is $|\varepsilon_r|\Lambda^{-2}(3-2\varepsilon_r^2/\Lambda^2)e^{-\varepsilon_r^2/\Lambda^2}$, with all eigenvalues here evaluated at $[\mathcal A^B,0]$; this vanishes at $\varepsilon_r=0$. In the matter term, coincident levels on the same side of the gap have equal occupations and contribute zero in the gapped zero-temperature limit.}
 \es{Knew}{
&\widetilde K^{XX',\rm new}_{q;\ell m,\ell'm'}(\omega)
 =\lim_{\Lambda\to\infty}\lim_{j_{\max}\to\infty}\Bigg[
 \sum_{r,s=1}^{D}\frac{n_r-n_s}{\varepsilon_r-\varepsilon_s+i\omega}\\
 &\quad\times\sum_{t,t'=1}^{D}c_{tr}^*
 \left[i\delta_{XE}\sqrt{1+\frac{\omega^2}{\ell(\ell+1)}}(V_{\alpha\ell m}^{\rm sph})^\dagger
       +\delta_{XB}(V_{B\ell m}^{\rm sph})^\dagger\right]_{tt'}c_{t's}\\
 &\quad\times\sum_{u,u'=1}^{D}c_{us}^*
 \left[i\delta_{X'E}\sqrt{1+\frac{\omega^2}{\ell'(\ell'+1)}}V_{\alpha\ell'm'}^{\rm sph}
       +\delta_{X'B}V_{B\ell'm'}^{\rm sph}\right]_{uu'}c_{u'r}\\
 &+\frac{\delta_{XB}\delta_{X'B}}2\sum_{r,s=1}^{D}
 \frac{\sgn(\varepsilon_r[\mathcal A^B_{\ell m},0])
 [1+(2\varepsilon_r[\mathcal A^B_{\ell m},0]^2/\Lambda^2-1)e^{-\varepsilon_r[\mathcal A^B_{\ell m},0]^2/\Lambda^2}]
 -(r\leftrightarrow s)}
 {\varepsilon_r[\mathcal A^B_{\ell m},0]-\varepsilon_s[\mathcal A^B_{\ell m},0]}\\
 &\quad\times\sum_{t,t'=1}^{D}c_{tr}[\mathcal A^B_{\ell m},0]^*
 ((V_{B\ell m}^{\rm sph})^\dagger)_{tt'}c_{t's}[\mathcal A^B_{\ell m},0]
 \sum_{u,u'=1}^{D}c_{us}[\mathcal A^B_{\ell m},0]^*
 (V_{B\ell'm'}^{\rm sph})_{uu'}c_{u'r}[\mathcal A^B_{\ell m},0]\Bigg]\,.
}
Here $X,X'\in\{E,B\}$, while the $V$'s are the static vertices in \eqref{VsNS}. The $BB$ regulating term comes from expanding \eqref{rawenergy} to quadratic order. Note that unlike the spherical case, there is no linear-in-$\beta$ kernel $\overline K$. 

We can then integrate over gauge modes as in the spherical case to obtain
\es{DeltaNS}{
 &\Delta_q^{(1),\rm new}-\Delta_q^{(1),\rm sph}
 =\lim_{\ell_{\max}\to\infty}\Bigg\{\frac12
 \int_{-\infty}^{\infty}\frac{d\omega}{2\pi}
 \log\frac{\det(\widetilde{\mathbf K}^{\rm new}_q(\omega)+\widetilde{\mathbf K}_{\rm CS}(\omega))}
               {\det(\widetilde{\mathbf K}^{\rm sph}_q(\omega)+\widetilde{\mathbf K}_{\rm CS}(\omega))}\\
 &\quad-\frac{64\kappa}{\pi^2+64\kappa^2}
 \left[1+\int_{-\infty}^{\infty}\frac{d\omega}{2\pi}
 \sum_{\ell}\frac{2\ell+1}{\omega^2+(\ell+1/2)^2}\right]\\
 &\quad\times\sum_{r=1}^{D/2-2q\kappa}\left[
 \sum_{m=1/2-q}^{q-1/2}|c_{(q-1/2,m),r}|^2
 +2\operatorname{Re}\sum_{j=q+1/2}^{j_{\max}}\sum_{m=-j}^{j}
 c_{(j,m,+),r}^*c_{(j,m,-),r}\right]\Bigg\}\,.
}
       where the spherical spinor matrix elements have been evaluated using \eqref{N} and \eqref{ABC}. Both integrals carry the same relativistic cutoff \eqref{cutoff} before the limit is taken. The term proportional to the second integral subtracts the linear divergence, while the $1$ in its square bracket adds back the zeta regularized value, as in the spherical calculation. Note that it is numerically efficient to subtract the previously considered spherical result, which also takes into account the pure gauge modes. For the spherically invariant matter kernel and CS kernel, we define
\es{NSblockSums}{
 \widetilde{\mathbf K}^{\rm sph}_q(\omega)
 &=\bigoplus_{\ell\geq1}\bigoplus_{m=-\ell}^{\ell}\widetilde{\mathbf K}_{q,\ell,\rm mat}(\omega),\qquad
 \widetilde{\mathbf K}_{\rm CS}(\omega)
 =\bigoplus_{\ell\geq1}\bigoplus_{m=-\ell}^{\ell}\widetilde{\mathbf K}_{\ell,\rm CS}(\omega),\\
 \log\det(\widetilde{\mathbf K}^{\rm sph}_q+\widetilde{\mathbf K}_{\rm CS})
 &=\sum_{\ell\geq1}(2\ell+1)\log\det\big(\widetilde{\mathbf K}_{q,\ell,\rm mat}+\widetilde{\mathbf K}_{\ell,\rm CS}\big)\,.
}
The cutoff restricts both external indices of the non-spherical matrix, as well as the spherical blocks in this ratio.
       
 For the $\beta$ dependent terms, there are two sources. The first comes from considering the difference between the Matsubara sum and the integral at $\beta\to\infty$:
 \es{beta1}{
 \frac{1}{2\beta}\sum_{n\in\mathbb{Z}}\log\det{}'(\widetilde{\mathbf K}^{\rm new}_q(\omega_n)+\widetilde{\mathbf K}_\text{CS}(\omega_n))- \frac{1}{2}\int_{-\infty}^\infty\frac{d\omega}{2\pi}\log\det(\widetilde{\mathbf K}^{\rm new}_q(\omega)+\widetilde{\mathbf K}_\text{CS}(\omega))\,,
 }
 where $\det'$ means that we consider only the nonzero eigenvalues; we will consider the zero eigenvalues for $n=0$ afterwards. This difference is nontrivial only for terms that give divergences in each separate sum and integral, which come from considering the small $\omega$ expansion of the kernel. We find that 
 \es{zG}{
\det\big(\widetilde{\mathbf K}^{\rm new}_q(\omega)+\widetilde{\mathbf K}_{\rm CS}(\omega)\big)
 \propto\omega^{2z_G}\qquad(\omega\to0)\,.
}
 where each rotational eigenvalue vanishes quadratically, while the complementary block has a nonzero determinant at $\omega=0$. Here $z_G$ is the number of rotational zero eigenvalues for subgroup $G\subset SO(3)$ that the non-spherical saddle preserves, which is 2 for $G=O(2)$, and 3 when $G=T,O,D_n$. We can then evaluate the difference \eqref{beta1} to get
 \es{beta12}{
& z_G\frac{\log\beta}{\beta}-\frac{1}{2\beta}\log\det_{\rm rot}\Big[\frac12(\widetilde{\mathbf K}^{\rm new}_q+\widetilde{\mathbf K}_\text{CS})''(0)\\
 &\qquad-(\widetilde{\mathbf K}^{\rm new}_q+\widetilde{\mathbf K}_\text{CS})'(0)(\widetilde{\mathbf K}^{\rm new}_q+\widetilde{\mathbf K}_\text{CS})^{+}(0)(\widetilde{\mathbf K}^{\rm new}_q+\widetilde{\mathbf K}_\text{CS})'(0)\Big]\,.
 }
 Here $+$ denotes the inverse on the complement of the rotational zero modes, with zero on the null space. The determinant is taken after restricting the bracket to an orthonormal basis of the $z_G$ rotational zero modes, using the same integration coordinates as the static measure below. 
 
 The second contribution to the $\beta$-dependent terms comes from the zero eigenvalues for $n=0$. For these terms, we expand around the saddle points as $\cA_{\ell m}^B+\delta\cA_{\ell m}^B$ and $\alpha_{\ell m}+\delta\alpha_{\ell m}$, and consider fluctuations corresponding to the $z_G$ zero eigenvalues. For these terms the action vanishes, and we integrate over the $z_G$ modes in the path integral and then take the logarithm to get a $\beta$ dependent contribution to the free energy:
 \es{zGmodes}{
- \frac{z_G}{2\beta}\log\left(\frac{N\beta}{2\pi}\right)-\frac{1}{\beta}\log\Vol_{(\cA^B_{\ell m},\alpha_{\ell m})}(SO(3)/G)\,,
 }
 where $\Vol_{(\cA^B_{\ell m},\alpha_{\ell m})}(SO(3)/G)$ is the volume of the rotation orbit of the saddle, which is a surface volume with the metric
 \es{metNS}{
\left.\sum_{\ell m}\left\{
 \left|d(\sqrt N\,i\mathfrak a^E_{\ell m}(0))\right|^2
 +\left|d(\sqrt N\,\mathfrak a^B_{\ell m}(0))\right|^2\right\}\right|_{\rm tangent}
 =N\beta\left.\sum_{\ell m}\left[
 |d\delta\alpha_{\ell m}|^2+|d\delta\mathcal A^B_{\ell m}|^2\right]\right|_{\rm tangent}\,.
}
 For instance, for $G=O(2)$ we get
 \es{VolO2}{
\Vol_{(\cA^B_{\ell m},\alpha_{\ell m})}(SO(3)/O(2))&=\int_0^{2\pi}d\phi\int_0^{\pi/2}d\theta\,\sin\theta\,
       \frac12\sum_\ell\ell(\ell+1)
                    \big[(\mathcal A^B_{\ell0})^2+\alpha_{\ell0}^2\big]\\
 &=\pi\sum_\ell\ell(\ell+1)
                    \big[(\mathcal A^B_{\ell0})^2+\alpha_{\ell0}^2\big]\,,
 }
where note that only the $m=0$ modes are nonzero when we have axial symmetry. For the tetrahedral and octahedral groups we get
\es{TOvol}{
\Vol_{(\cA^B_{\ell m},\alpha_{\ell m})}(SO(3)/G)
 =\frac{8\pi^2}{|G|}\left[\frac13\sum_{\ell m}\ell(\ell+1)
       \big[|\mathcal A^B_{\ell m}|^2+|\alpha_{\ell m}|^2\big]\right]^{3/2}\,,
}
 for the $n>2$ dihedral groups
\es{D2nVol}{
\Vol_{(\cA^B_{\ell m},\alpha_{\ell m})}(SO(3)/D_n)
 ={}&\frac{8\pi^2}{2n}
 \frac12\sum_{\ell m}[\ell(\ell+1)-m^2]
       \big[|\mathcal A^B_{\ell m}|^2+|\alpha_{\ell m}|^2\big]\\
 &\qquad\times\left[\sum_{\ell m}m^2
       \big[|\mathcal A^B_{\ell m}|^2+|\alpha_{\ell m}|^2\big]\right]^{1/2}\,,
}
For $D_n$ the body axes are chosen so that the $n$-fold axis is the third axis, and $|D_n|=2n$. For $D_2$, choose the three twofold axes as the coordinate axes; the mixed metric components then vanish, and the same volume in complex harmonics is
\es{D2Vol}{
&\Vol_{(\cA^B_{\ell m},\alpha_{\ell m})}(SO(3)/D_2)
 =\pi^2\left[\sum_{\ell m}m^2\big(|\mathcal A^B_{\ell m}|^2+|\alpha_{\ell m}|^2\big)\right]^{1/2}\\
 &\quad\times\Bigg\{
 \left[\sum_{\ell m}[\ell(\ell+1)-m^2]\big(|\mathcal A^B_{\ell m}|^2+|\alpha_{\ell m}|^2\big)\right]^2\\
 &\qquad-\left[\operatorname{Re}\sum_{\ell=1}^{\ell_{\rm bg}}\sum_{m=-\ell}^{\ell-2}
 \sqrt{(\ell-m)(\ell-m-1)(\ell+m+1)(\ell+m+2)}\right.\\
 &\hspace{42mm}\left.\times\big(\mathcal A^{B*}_{\ell,m+2}\mathcal A^B_{\ell m}
 +\alpha_{\ell,m+2}^*\alpha_{\ell m}\big)\right]^2\Bigg\}^{1/2}\,.
}
We can then add \eqref{zGmodes} to \eqref{beta12} to get the total $\beta$-dependent terms:
\es{NSbetaTot}{
&F^{(1),\text{new}}_{q,\beta} =\frac{z_G\log\beta}{2\beta}- \frac{z_G}{2\beta}\log\left(\frac{N}{2\pi}\right)-\frac{1}{\beta}\log\Vol_{(\cA^B_{\ell m},\alpha_{\ell m})}(SO(3)/G)\\
&\quad-\frac{1}{2\beta}\log\det_{\rm rot}\Big[\frac12(\widetilde{\mathbf K}^{\rm new}_q+\widetilde{\mathbf K}_\text{CS})''(0)\\
&\qquad-(\widetilde{\mathbf K}^{\rm new}_q+\widetilde{\mathbf K}_\text{CS})'(0)(\widetilde{\mathbf K}^{\rm new}_q+\widetilde{\mathbf K}_\text{CS})^{+}(0)(\widetilde{\mathbf K}^{\rm new}_q+\widetilde{\mathbf K}_\text{CS})'(0)\Big]\,.
}

\clearpage
\subsection{Energy splitting}
\label{splitNonS}

Quantizing the degenerate orientations of a non-spherical saddle gives its energy splitting. We consider the intrinsic ground state, with a gap between occupied and empty Dirac levels and zero intrinsic angular momentum.

The required inertia tensor $\mathcal I_{ab}$ is determined by the same kernel as in \eqref{NSbetaTot}. To specify its normalization, let $\delta_a$ denote differentiation with respect to the rotation angle in radians about body axis $a$, at the chosen orientation. On either set of saddle coefficients,
\es{rotTangentNew}{
 \delta_a\alpha_{\ell m}=-i\sum_{m'}(L_a^{(\ell)})_{mm'}\alpha_{\ell m'},\qquad
 \delta_a\mathcal A^B_{\ell m}=-i\sum_{m'}(L_a^{(\ell)})_{mm'}\mathcal A^B_{\ell m'}\,.
}
Here $L_a^{(\ell)}$ are spin-$\ell$ matrices, with $(L_3^{(\ell)})_{mm'}=m\delta_{mm'}$ and\newline
$(L_+^{(\ell)})_{m+1,m}=\sqrt{(\ell-m)(\ell+m+1)}$, $L_-=L_+^\dagger$, $L_1=(L_++L_-)/2$, and $L_2=(L_+-L_-)/(2i)$. The matter kernel $\widetilde{\mathbf K}^{\rm new}_q$ and CS kernel $\widetilde{\mathbf K}_{\rm CS}$ below remain in the $E,B$ basis of \eqref{Knew} and \eqref{CSfourier2}. The static electric coefficient is $-i\alpha_{\ell m}$, so its rotational variation is $-i\delta_a\alpha_{\ell m}$. Integrating out fluctuations transverse to the orbit gives
\es{inertiaNew}{
\mathcal I_{ab}={}&\sum_{\ell m,\ell'm'}
 \begin{pmatrix}-i\delta_a\alpha_{\ell m}^*&\delta_a\mathcal A^{B*}_{\ell m}\end{pmatrix}
 \Big[\tfrac12(\widetilde{\mathbf K}^{\rm new}_q+\widetilde{\mathbf K}_{\rm CS})''(0)\\
 &\qquad-(\widetilde{\mathbf K}^{\rm new}_q+\widetilde{\mathbf K}_{\rm CS})'(0)
 (\widetilde{\mathbf K}^{\rm new}_q+\widetilde{\mathbf K}_{\rm CS})^+(0)
 (\widetilde{\mathbf K}^{\rm new}_q+\widetilde{\mathbf K}_{\rm CS})'(0)\Big]_{\ell m,\ell'm'}
 \begin{pmatrix}-i\delta_b\alpha_{\ell'm'}\\\delta_b\mathcal A^B_{\ell'm'}\end{pmatrix}\,.
}
 For angular velocity $\Omega_a$ in a body frame the kinetic energy and its conjugate angular momentum give
\es{HsplitNew}{
 L_a=N\sum_b\mathcal I_{ab}\Omega_b,\qquad
 H_{\rm split}^{\rm new}
 =\sum_a\Omega_aL_a-\frac N2\sum_{a,b}\mathcal I_{ab}\Omega_a\Omega_b
 =\frac1{2N}\sum_{a,b}(\mathcal I^{-1})_{ab}L_aL_b\,,
}
where here the inverse includes only broken directions. We can then compute $\Delta_f^{\rm split,new}$ by taking the eigenvalues of this Hamiltonian.

The stabilizer selects the allowed body-frame states. For $g\in G$, let $\widehat g$ act on the occupied spinors, including the monopole gauge compensation. Its phase on the filled state is
$\big\{\det[\int_{S^2}d\Omega\,\Psi_r^\dagger\widehat g\Psi_s]_{n_r=n_s=1}\big\}^{N}$.
Retain the subspace of the spin-$\ell$ representation transforming with the inverse phase. This is found by a finite group average for discrete $G$, or a normalized group integral for $O(2)$. Each eigenvalue on this subspace has $2\ell+1$ space-frame magnetic states. These filled-orbital states are $SU(N)$ flavor singlets. 

As a check, the thermal trace of this Hamiltonian is
\es{rotTraceNew}{
 \Tr e^{-\beta H_{\rm split}^{\rm new}}
 &=\sum_{\ell=0}^\infty(2\ell+1)\,
 \operatorname{Tr}
 \exp\!\left[-\frac\beta{2N}\sum_{a,b}(\mathcal I^{-1})_{ab}L_a^{(\ell)}L_b^{(\ell)}\right],\\
 -\frac1\beta\log\Tr e^{-\beta H_{\rm split}^{\rm new}}
 &=\begin{cases}
 \dfrac1\beta\log\dfrac{\beta}{N\mathcal I_{11}},&G=O(2),\\[4pt]
 \dfrac1\beta\left[\dfrac32\log\dfrac\beta N
 +\log\dfrac{|G|}{2\sqrt{2\pi}\sqrt{\det\mathcal I}}\right],&G\text{ discrete}
 \end{cases}\,.
}
For the axial case the axis is chosen along direction 3, so $\mathcal I_{11}=\mathcal I_{22}$. The metric volumes here are $2\pi\mathcal I_{11}$ and $8\pi^2\sqrt{\det\mathcal I}/|G|$. By \eqref{inertiaNew}, each equals the orbit volume defined by \eqref{metNS} times the square root of the restricted determinant in \eqref{NSbetaTot}. Thus $-\beta^{-1}\log\Tr e^{-\beta H_{\rm split}^{\rm new}}$ recovers \eqref{NSbetaTot}.

For example, $q=1$, $|\kappa|=3/2$ has $G=O(2)$, zero body-axis spin, and an even-spin ground band. We find $\mathcal I_{11}=\mathcal I_{22}=8.48194049$, which gives
\es{rotExampleNew}{
 \Delta_f^{\rm split,new}=\frac{\ell(\ell+1)}{2N(8.48194049)}\,,
}
where recall that all these monopoles are $SU(N)$ singlets.

\section{Checks of dualities}
\label{num}

We now present our results for the subleading calculation for various $\kappa$, and combine these with the leading order results and the energy splitting to check the various dualities. 

For the spherical fermionic saddle we used $j'_\text{cut}\approx 500$, $\Lambda=64$, $a_\text{cut}=10$, and $b_\text{cut}=11$. For the spherical sQED$_3$ calculation we used the analogous algorithm in \cite{Chester:2022wur}, with $j'_\text{cut}\approx 800$, $\Lambda=96$, $a_\text{cut}=10$, and $b_\text{cut}=11$.

For the non-spherical saddles, the numerical cutoffs are listed in Table~\ref{tab:nonspherical-cutoffs}. The third column gives $(j_{\max},\ell_\text{bg},\Lambda)$ for the leading energy in \eqref{rawenergy}. The last column gives $(j_{\max},\Lambda;\ell_{\max})$ for the Gaussian calculation. The internal cutoff $j_{\max}$ can have different numerical values in the two calculations. Multiple entries after the semicolon indicate the external angular cutoffs used for the extrapolation. Here $\Lambda$ is the internal spectral regulator in \eqref{rawenergy}, not the external relativistic cutoff used above. At each frequency we retained both angular indices of the kernel subject to $\omega^2+\ell(\ell+1)<(\ell_{\max}+1)(\ell_{\max}+2)$, and the same condition on $\ell'$. These are the external integration limits used for the saved non-spherical results.

\begin{table}[!htbp]
\centering
\small
\setlength{\tabcolsep}{10pt}
\begin{tabular}{cccc}
\toprule
$m$ & $q$ & $(j_{\max},\ell_\text{bg},\Lambda)$ & $(j_{\max},\Lambda;\ell_{\max})$ \\
\midrule
1 & $1/2$ & $(14,8,6)$ & $(40,10;8,12,16)$ \\
1 & $1$ & $(14.5,8,6)$ & $(64.5,24;8,12,16),\ (64.5,8;20)$ \\
1 & $3/2$ & $(14,10,6)$ & $(10,5;6),\ (14,6;8,10)$ \\
1 & $2$ & $(14.5,10,6)$ & $(10.5,5;6),\ (14.5,6;8,10)$ \\
1 & $5/2$ & $(14,10,6)$ & $(10,5;6),\ (14,6;8,10)$ \\
1 & $3$ & $(14.5,10,6)$ & $(14.5,6;8,10)$ \\
1 & $7/2$ & $(14,10,6)$ & $(14,6;8,10)$ \\
1 & $4$ & $(16.5,11,7)$ & $(14.5,6;9),\ (16.5,7;11)$ \\
1 & $9/2$ & $(17,11,7)$ & $(16,7;10),\ (17,7;11)$ \\
1 & $5$ & $(18.5,12,7)$ & $(17.5,7;11),\ (18.5,7;12)$ \\
2 & $1/2$ & $(14,8,6)$ & $(40,10;8,12,16)$ \\
2 & $3/2$ & $(14,10,6)$ & $(14,6;8,10)$ \\
2 & $2$ & $(14.5,8,6)$ & $(48.5,12;8,12,16)$ \\
2 & $5/2$ & $(14,10,6)$ & $(14,6;8,10)$ \\
2 & $3$ & $(14.5,10,6)$ & $(14.5,6;9,10)$ \\
3 & $1$ & $(14.5,8,6)$ & $(48.5,12;8,12,16)$ \\
3 & $3/2$ & $(14,10,6)$ & $(14,6;8,10)$ \\
3 & $2$ & $(14.5,8,6)$ & $(48.5,12;8,12,16)$ \\
3 & $5/2$ & $(14,10,6)$ & $(14,6;8,10)$ \\
\bottomrule
\end{tabular}
\caption{Numerical cutoffs for the non-spherical saddles at $|\kappa|=m+1/2$. The third column gives the Dirac, background-harmonic, and spectral-regulator cutoffs for the leading energy. The fourth gives the internal Dirac and spectral-regulator cutoffs followed by the external angular cutoffs used in the Gaussian extrapolation. Entries after a semicolon share the same internal parameters.}
\label{tab:nonspherical-cutoffs}
\end{table}

The largest static residuals in the saddle equations were below $10^{-12}$. We extrapolated the subtracted Gaussian integrals in inverse external radius $1/\sqrt{(\ell_{\max}+1)(\ell_{\max}+2)}$, comparing the last available cutoffs and linear fits, with a quadratic fit when four cutoffs were available. The resulting cutoff diagnostics range from $0.003$ to $0.039$ in $\Delta_f^{(1)}$; the displayed digits do not constitute four-decimal accuracy estimates.

For the rotational splitting, the six axial sectors used internal spin cutoffs up to $j_{\max}=96$ for $q=1/2$ and $96.5$ for integer $q$, with $\ell_\text{bg}=8$. For discrete saddles, the closed projected subspaces used the static cutoffs above. Additional ground-band copies were evaluated with $(j_{\max},\ell_\text{bg})=(10,6),(10.5,6),(10,6),(10.5,7),(10,8),(10.5,9),(10,10),(10.5,11)$ for $m=1$ and $q=3/2,2,5/2,3,7/2,4,9/2,5$, respectively; $(10,8),(10.5,9)$ for $m=2$, $q=5/2,3$; and $(10,6)$ for $m=3$, $q=3/2$. These lower cutoffs apply to the splitting calculation, not to the leading and Gaussian baselines.

For the spherical saddle energy splitting term, it is easy to get good accuracy, but the main challenge is diagonalizing the energy splitting Hamiltonian, which grows with size as $q$ increases. This was the main limitation on the largest values of $q$ that we looked at. Note that in the duality examples below, we will encounter cases where at large $N$ there are multiple degenerate monopoles, as we are dressing with order $N$ modes, but only one of them survives to $N=1$. Since the energy splitting is computed at large $N$, we will find a nonzero energy splitting even if for $N=1$ there is a unique state, and indeed in some cases this contribution is crucial to get a good match with the duality.

In Tables \ref{tab:duality-m1}, \ref{tab:duality-m2}, and \ref{tab:duality-m3}, we give the results for the dualities between QED$^{1,-(m+1/2)}_3$ and sQED$^{1,\frac{m+1}{m}}_3$ with $q_s=mq_f$ for $m=1,2,3$. Recall that for $m>1$, sQED$^{1,\frac{m+1}{m}}_3$ is an effective description of the original 2-node quiver gauge theory. For $m=1$, as discussed in the introduction we expect $q=1/2$ to be the extra $SO(3)$ currents with $\Delta=2$, while $q=1$ was estimated by the fuzzy sphere analysis \cite{Zhou:2025rmv} to have $\Delta= 5.1079$ for the scalar, and $\Delta=5.3435$ for the spin 2 operator. These values are close to both our QED$^{1,-(m+1/2)}_3$ and sQED$^{1,\frac{m+1}{m}}_3$ estimates. The match for $m=1,2,3$ seems to work roughly the same for all $q$ and all spins. The matches are so accurate, that we can even distinguish between the energy splitting of monopoles with the same $q$.

 In Table \ref{tab:duality-m0}, we give the values for duality between QED$^{1,-1/2}_3$ and the critical $O(2)$ model, where the CFT data of the latter is taken from conformal bootstrap \cite{Chester:2019ifh,Liu:2020tpf} for $q\leq2$ and lattice simulations \cite{Banerjee:2017fcx} for $q>2$. This can be thought of as the $m\to0$ limit of the above dualities, as $k\to\infty$ decouples the gauge field on the scalar side. In this case, we do not find a good match for any $q$. All the values of $q$ correspond to the completely filled zero energy shell in this case, so there is no energy splitting term, which might partly explain why the match is worse. We note that the subleading term is also much bigger than the leading term, unlike the other cases we considered, so it could also simply be the case that the large $N$ expansion diverges badly in this case. Curiously, we find that the subleading term alone matches the critical $O(2)$ values to high accuracy, but the leading term ruins the match.

In Tables \ref{tab:seed} and \ref{tab:son}, we compare the dualities between the free fermion theory and sQED$^{1,1}_3$ and QED$_3^{1,0}$, respectively, where in the latter case $q$ is restricted to integer values. The operators in the free fermion theory can be efficiently constructed as described in Appendix B of \cite{Chester:2022wur}. The match in these free fermion dualities is worse than the $m>0$ dualities, though it is still pretty good for Son's duality. Note that in that duality, the splitting term is larger than the other dualities.

\begin{table}[htbp]
\centering
\small
\setlength{\tabcolsep}{5pt}
\renewcommand{\arraystretch}{1.0}
\resizebox{\textwidth}{!}{%
\begin{tabular}{@{}cccrrrrrrrrr@{}}
\toprule
 & & \multicolumn{5}{c}{Fermion} & \multicolumn{4}{c}{Scalar} & \\
\cmidrule(lr){3-7}\cmidrule(lr){8-11}
$q$ & $\ell$ & Saddle & $\Delta_f^{(0)}$ & $\Delta_f^{(1)}$ & $\Delta_f^{\text{split}}$ & $\Delta_f$ & $\Delta_s^{(0)}$ & $\Delta_s^{(1)}$ & $\Delta_s^{\text{split}}$ & $\Delta_s$ & Error (\%) \\
\midrule
$1/2$ & $1$ & $O(2)$ & $1.6437$ & $0.2564$ & $0.1433$ & \textcolor{blue}{$2.0435$} & $2.1101$ & $-0.4978$ & $0.3871$ & \textcolor{blue}{$1.9995$} & $2.15$ \\
\specialrule{0.3pt}{1.5pt}{1.5pt}
$1$ & $0$ & $O(2)$ & $4.0881$ & $0.8619$ & $0.0000$ & \textcolor{blue}{$4.9501$} & $5.5782$ & $-1.1987$ & $0.4589$ & \textcolor{blue}{$4.8384$} & $2.26$ \\
$1$ & $2$ & $O(2)$ & $4.0881$ & $0.8619$ & $0.3537$ & \textcolor{blue}{$5.3037$} & $5.5782$ & $-1.1987$ & $0.8681$ & \textcolor{blue}{$5.2476$} & $1.06$ \\
\specialrule{0.3pt}{1.5pt}{1.5pt}
$3/2$ & $1$ & $D_{3}$ & $7.1372$ & $1.8411$ & $0.0678$ & \textcolor{blue}{$9.0460$} & $10.0000$ & $-2.0665$ & $1.0796$ & \textcolor{blue}{$9.0131$} & $0.36$ \\
$3/2$ & $3$ & $D_{3}$ & $7.1372$ & $1.8411$ & $0.4068$ & \textcolor{blue}{$9.3850$} & $10.0000$ & $-2.0665$ & $1.5279$ & \textcolor{blue}{$9.4614$} & $0.81$ \\
\specialrule{0.3pt}{1.5pt}{1.5pt}
$2$ & $0$ & $T$ & -- & -- & -- & -- & $15.2027$ & $-3.076$ & $1.7333$ & \textcolor{blue}{$13.8603$} & -- \\
$2$ & $2$ & $T$ & $10.6685$ & $2.8762$ & $0.1939$ & \textcolor{blue}{$13.7385$} & $15.2027$ & $-3.076$ & $1.5064$ & \textcolor{blue}{$13.633$} & $0.77$ \\
$2$ & $4$ & $T$ & $10.6685$ & $2.8762$ & $0.6463$ & \textcolor{blue}{$14.1909$} & $15.2027$ & $-3.076$ & $1.9885$ & \textcolor{blue}{$14.1155$} & $0.53$ \\
\specialrule{0.3pt}{1.5pt}{1.5pt}
$5/2$ & $1$ & $D_{3}$ & $14.6556$ & $4.1659$ & $0.0407$ & \textcolor{blue}{$18.8622$} & $21.0831$ & $-4.2093$ & $2.2060$ & \textcolor{blue}{$19.0798$} & $1.15$ \\
$5/2$ & $3$ & $D_{3}$ & $14.6556$ & $4.1659$ & $0.2454$ & \textcolor{blue}{$19.0669$} & $21.0831$ & $-4.2093$ & $2.3101$ & \textcolor{blue}{$19.1839$} & $0.61$ \\
$5/2$ & $5$ & $D_{3}$ & $14.6556$ & $4.1659$ & $0.6135$ & \textcolor{blue}{$19.4350$} & $21.0831$ & $-4.2093$ & $2.7396$ & \textcolor{blue}{$19.6134$} & $0.92$ \\
\specialrule{0.3pt}{1.5pt}{1.5pt}
$3$ & $0$ & $O$ & $19.0158$ & $5.5445$ & $0.0000$ & \textcolor{blue}{$24.5603$} & $27.5707$ & $-5.4553$ & $2.4533$ & \textcolor{blue}{$24.5687$} & $0.03$ \\
$3$ & $2$ & $O$ & -- & -- & -- & -- & $27.5707$ & $-5.4553$ & $2.9469$ & \textcolor{blue}{$25.0623$} & -- \\
$3$ & $4$ & $O$ & $19.0158$ & $5.5445$ & $0.3789$ & \textcolor{blue}{$24.9392$} & $27.5707$ & $-5.4553$ & $2.8230$ & \textcolor{blue}{$24.9384$} & $0.00$ \\
$3$ & $6$ & $O$ & $19.0158$ & $5.5445$ & $0.7957$ & \textcolor{blue}{$25.3560$} & $27.5707$ & $-5.4553$ & $3.2402$ & \textcolor{blue}{$25.3556$} & $0.00$ \\
\specialrule{0.3pt}{1.5pt}{1.5pt}
$7/2$ & $1$ & $D_{5}$ & $23.7563$ & $7.0792$ & $0.0285$ & \textcolor{blue}{$30.8640$} & $34.6132$ & $-6.8045$ & $3.3975$ & \textcolor{blue}{$31.2063$} & $1.11$ \\
$7/2$ & $3$ & $D_{5}$ & $23.7563$ & $7.0792$ & $0.1708$ & \textcolor{blue}{$31.0064$} & $34.6132$ & $-6.8045$ & $3.5562$ & \textcolor{blue}{$31.3650$} & $1.16$ \\
$7/2$ & $5$ & $D_{5}$ & $23.7563$ & $7.0792$ & $0.4270$ & \textcolor{blue}{$31.2626$} & $34.6132$ & $-6.8045$ & $3.6824$ & \textcolor{blue}{$31.4911$} & $0.73$ \\
$7/2$ & $7$ & $D_{5}$ & $23.7563$ & $7.0792$ & $0.7971$ & \textcolor{blue}{$31.6327$} & $34.6132$ & $-6.8045$ & $4.1023$ & \textcolor{blue}{$31.9111$} & $0.88$ \\
\specialrule{0.3pt}{1.5pt}{1.5pt}
$4$ & $0$ & $D_{4}$ & $28.8221$ & $8.6938$ & $0.0000$ & \textcolor{blue}{$37.5160$} & $42.1701$ & $-8.2495$ & $4.1639$ & \textcolor{blue}{$38.0845$} & $1.52$ \\
$4$ & $2$ & $D_{4}$ & $28.8221$ & $8.6938$ & $0.0763$ & \textcolor{blue}{$37.5923$} & $42.1701$ & $-8.2495$ & $4.1001$ & \textcolor{blue}{$38.0206$} & $1.14$ \\
$4$ & $4$ & $D_{4}$ & $28.8221$ & $8.6938$ & $0.2543$ & \textcolor{blue}{$37.7702$} & $42.1701$ & $-8.2495$ & $4.1244$ & \textcolor{blue}{$38.045$} & $0.73$ \\
$4$ & $6$ & $D_{4}$ & $28.8221$ & $8.6938$ & $0.5340$ & \textcolor{blue}{$38.0500$} & $42.1701$ & $-8.2495$ & $4.3940$ & \textcolor{blue}{$38.3146$} & $0.70$ \\
$4$ & $8$ & $D_{4}$ & $28.8221$ & $8.6938$ & $0.9154$ & \textcolor{blue}{$38.4314$} & $42.1701$ & $-8.2495$ & $4.8482$ & \textcolor{blue}{$38.7687$} & $0.88$ \\
\specialrule{0.3pt}{1.5pt}{1.5pt}
$9/2$ & $1$ & $D_{3}$ & $34.2031$ & $10.4398$ & $0.0214$ & \textcolor{blue}{$44.6643$} & $50.2083$ & $-9.785$ & $4.9898$ & \textcolor{blue}{$45.4136$} & $1.68$ \\
$9/2$ & $3$ & $D_{3}$ & $34.2031$ & $10.4398$ & $0.1286$ & \textcolor{blue}{$44.7715$} & $50.2083$ & $-9.785$ & $4.8460$ & \textcolor{blue}{$45.2697$} & $1.11$ \\
$9/2$ & $5$ & $D_{3}$ & $34.2031$ & $10.4398$ & $0.3215$ & \textcolor{blue}{$44.9643$} & $50.2083$ & $-9.785$ & $5.1342$ & \textcolor{blue}{$45.5580$} & $1.32$ \\
$9/2$ & $7$ & $D_{3}$ & $34.2031$ & $10.4398$ & $0.6001$ & \textcolor{blue}{$45.2429$} & $50.2083$ & $-9.785$ & $5.2343$ & \textcolor{blue}{$45.6580$} & $0.92$ \\
$9/2$ & $9$ & $D_{3}$ & $34.2031$ & $10.4398$ & $0.9644$ & \textcolor{blue}{$45.6073$} & $50.2083$ & $-9.785$ & $5.6512$ & \textcolor{blue}{$46.0749$} & $1.03$ \\
\specialrule{0.3pt}{1.5pt}{1.5pt}
$5$ & $0$ & $D_{4}$ & $39.8806$ & $12.2750$ & $0.0000$ & \textcolor{blue}{$52.1557$} & $58.7007$ & $-11.4046$ & $5.8431$ & \textcolor{blue}{$53.1392$} & $1.89$ \\
$5$ & $2$ & $D_{4}$ & $39.8806$ & $12.2750$ & $0.0522$ & \textcolor{blue}{$52.2079$} & $58.7007$ & $-11.4046$ & $5.6397$ & \textcolor{blue}{$52.9358$} & $1.39$ \\
$5$ & $4$ & $D_{4}$ & $39.8806$ & $12.2750$ & $0.1743$ & \textcolor{blue}{$52.3300$} & $58.7007$ & $-11.4046$ & $5.5769$ & \textcolor{blue}{$52.8730$} & $1.04$ \\
$5$ & $6$ & $D_{4}$ & $39.8806$ & $12.2750$ & $0.3660$ & \textcolor{blue}{$52.5217$} & $58.7007$ & $-11.4046$ & $5.8871$ & \textcolor{blue}{$53.1832$} & $1.26$ \\
$5$ & $8$ & $D_{4}$ & $39.8806$ & $12.2750$ & $0.6274$ & \textcolor{blue}{$52.7831$} & $58.7007$ & $-11.4046$ & $5.9646$ & \textcolor{blue}{$53.2607$} & $0.90$ \\
$5$ & $10$ & $D_{4}$ & $39.8806$ & $12.2750$ & $0.9586$ & \textcolor{blue}{$53.1142$} & $58.7007$ & $-11.4046$ & $6.4331$ & \textcolor{blue}{$53.7292$} & $1.16$ \\
\bottomrule
\end{tabular}
}
\caption{Comparison for $m=1$ of charge $q$ monopoles in QED$_3^{1,-3/2}$ and charge $q$ monopoles in sQED$_3^{1,2}$. The common Lorentz spin is $\ell$, and the splitting contributions are $\Delta_f^{\text{split}}$ and $\Delta_s^{\text{split}}$. At $N=1$, $\Delta_x=\Delta_x^{(0)}+\Delta_x^{(1)}+\Delta_x^{\text{split}}$ for $x=f,s$, and the error is $100|\Delta_f-\Delta_s|/|\Delta_f|$. The total dimensions $\Delta_f$ and $\Delta_s$ are shown in blue to guide the eye. The Saddle column specifies the unbroken rotation group of the fermionic saddle. Dashes mark fermionic states that cannot be obtained from the non-spherical ground state.}
\label{tab:duality-m1}
\end{table}
	
	\begin{table}[htbp]
\centering
\small
\setlength{\tabcolsep}{5pt}
\renewcommand{\arraystretch}{1.0}
\resizebox{\textwidth}{!}{%
\begin{tabular}{@{}cccrrrrrrrrr@{}}
\toprule
 & & \multicolumn{5}{c}{Fermion} & \multicolumn{4}{c}{Scalar} & \\
\cmidrule(lr){3-7}\cmidrule(lr){8-11}
$q$ & $\ell$ & Saddle & $\Delta_f^{(0)}$ & $\Delta_f^{(1)}$ & $\Delta_f^{\text{split}}$ & $\Delta_f$ & $\Delta_s^{(0)}$ & $\Delta_s^{(1)}$ & $\Delta_s^{\text{split}}$ & $\Delta_s$ & Error (\%) \\
\midrule
$1/2$ & $1$ & $O(2)$ & $3.0746$ & $0.2599$ & $0.0935$ & \textcolor{blue}{$3.4280$} & $4.0000$ & $-1.2019$ & $0.5864$ & \textcolor{blue}{$3.3846$} & $1.26$ \\
\specialrule{0.3pt}{1.5pt}{1.5pt}
$1$ & $0$ & $SO(3)$ & $7.6014$ & $0.9694$ & $0.0000$ & \textcolor{blue}{$8.5708$} & $10.8343$ & $-3.1173$ & $1.1646$ & \textcolor{blue}{$8.8816$} & $3.63$ \\
\specialrule{0.3pt}{1.5pt}{1.5pt}
$3/2$ & $3$ & $D_{6}$ & $14.3230$ & $1.6928$ & $0.0867$ & \textcolor{blue}{$16.1025$} & $19.6037$ & $-5.5489$ & $2.5451$ & \textcolor{blue}{$16.5999$} & $3.09$ \\
\specialrule{0.3pt}{1.5pt}{1.5pt}
$2$ & $0$ & $O(2)$ & $22.0468$ & $2.5906$ & $0.0000$ & \textcolor{blue}{$24.6374$} & $29.9488$ & $-8.4057$ & $3.9938$ & \textcolor{blue}{$25.5368$} & $3.65$ \\
$2$ & $2$ & $O(2)$ & $22.0468$ & $2.5906$ & $0.0473$ & \textcolor{blue}{$24.6847$} & $29.9488$ & $-8.4057$ & $3.7291$ & \textcolor{blue}{$25.2721$} & $2.38$ \\
$2$ & $4$ & $O(2)$ & $22.0468$ & $2.5906$ & $0.1576$ & \textcolor{blue}{$24.7950$} & $29.9488$ & $-8.4057$ & $4.0723$ & \textcolor{blue}{$25.6153$} & $3.31$ \\
$2$ & $6$ & $O(2)$ & $22.0468$ & $2.5906$ & $0.3310$ & \textcolor{blue}{$24.9684$} & $29.9488$ & $-8.4057$ & $4.1944$ & \textcolor{blue}{$25.7374$} & $3.08$ \\
\specialrule{0.3pt}{1.5pt}{1.5pt}
$5/2$ & $1$ & $D_{6}$ & -- & -- & -- & -- & $41.6583$ & $-11.6324$ & $5.5755$ & \textcolor{blue}{$35.6014$} & -- \\
$5/2$ & $3$ & $D_{6}$ & $30.7277$ & $3.5967$ & $0.0670$ & \textcolor{blue}{$34.3914$} & $41.6583$ & $-11.6324$ & $5.1926$ & \textcolor{blue}{$35.2185$} & $2.40$ \\
$5/2$ & $3^{*}$ & $D_{6}$ & -- & -- & -- & -- & $41.6583$ & $-11.6324$ & $5.5909$ & \textcolor{blue}{$35.617$} & -- \\
$5/2$ & $4$ & $D_{6}$ & $30.7277$ & $3.5967$ & $0.1252$ & \textcolor{blue}{$34.4497$} & $41.6583$ & $-11.6324$ & $5.3936$ & \textcolor{blue}{$35.4195$} & $2.82$ \\
$5/2$ & $5$ & $D_{6}$ & $30.7277$ & $3.5967$ & $0.1981$ & \textcolor{blue}{$34.5225$} & $41.6583$ & $-11.6324$ & $5.7146$ & \textcolor{blue}{$35.7405$} & $3.53$ \\
$5/2$ & $6$ & $D_{6}$ & $30.7277$ & $3.5967$ & $0.2854$ & \textcolor{blue}{$34.6099$} & $41.6583$ & $-11.6324$ & $5.8503$ & \textcolor{blue}{$35.8762$} & $3.66$ \\
$5/2$ & $7$ & $D_{6}$ & $30.7277$ & $3.5967$ & $0.3874$ & \textcolor{blue}{$34.7118$} & $41.6583$ & $-11.6324$ & $5.6417$ & \textcolor{blue}{$35.6676$} & $2.75$ \\
$5/2$ & $9$ & $D_{6}$ & $30.7277$ & $3.5967$ & $0.4715$ & \textcolor{blue}{$34.7960$} & $41.6583$ & $-11.6324$ & $5.9916$ & \textcolor{blue}{$36.0175$} & $3.51$ \\
\specialrule{0.3pt}{1.5pt}{1.5pt}
$3$ & $0$ & $T$ & $40.2485$ & $4.7381$ & $0.0000$ & \textcolor{blue}{$44.9866$} & $54.5888$ & $-15.1909$ & $6.5785$ & \textcolor{blue}{$45.9764$} & $2.20$ \\
$3$ & $0^{*}$ & $T$ & -- & -- & -- & -- & $54.5888$ & $-15.1909$ & $7.3795$ & \textcolor{blue}{$46.7773$} & -- \\
$3$ & $2$ & $T$ & -- & -- & -- & -- & $54.5888$ & $-15.1909$ & $7.0205$ & \textcolor{blue}{$46.418$} & -- \\
$3$ & $2^{*}$ & $T$ & -- & -- & -- & -- & $54.5888$ & $-15.1909$ & $7.2012$ & \textcolor{blue}{$46.5991$} & -- \\
$3$ & $3$ & $T$ & $40.2485$ & $4.7381$ & $0.0723$ & \textcolor{blue}{$45.0590$} & $54.5888$ & $-15.1909$ & $6.9738$ & \textcolor{blue}{$46.3716$} & $2.91$ \\
$3$ & $4$ & $T$ & $40.2485$ & $4.7381$ & $0.1206$ & \textcolor{blue}{$45.1072$} & $54.5888$ & $-15.1909$ & $7.0741$ & \textcolor{blue}{$46.4719$} & $3.03$ \\
$3$ & $4^{*}$ & $T$ & -- & -- & -- & -- & $54.5888$ & $-15.1909$ & $7.1757$ & \textcolor{blue}{$46.5736$} & -- \\
$3$ & $4^{**}$ & $T$ & -- & -- & -- & -- & $54.5888$ & $-15.1909$ & $7.3325$ & \textcolor{blue}{$46.7303$} & -- \\
$3$ & $5$ & $T$ & -- & -- & -- & -- & $54.5888$ & $-15.1909$ & $7.1841$ & \textcolor{blue}{$46.5820$} & -- \\
$3$ & $6$ & $T$ & $40.2485$ & $4.7381$ & $0.2532$ & \textcolor{blue}{$45.2398$} & $54.5888$ & $-15.1909$ & $6.9351$ & \textcolor{blue}{$46.3330$} & $2.42$ \\
$3$ & $6^{*}$ & $T$ & $40.2485$ & $4.7381$ & $0.2532$ & \textcolor{blue}{$45.2398$} & $54.5888$ & $-15.1909$ & $7.2362$ & \textcolor{blue}{$46.6340$} & $3.08$ \\
$3$ & $6^{**}$ & $T$ & -- & -- & -- & -- & $54.5888$ & $-15.1909$ & $7.4360$ & \textcolor{blue}{$46.8339$} & -- \\
$3$ & $7$ & $T$ & $40.2485$ & $4.7381$ & $0.3376$ & \textcolor{blue}{$45.3242$} & $54.5888$ & $-15.1909$ & $7.2885$ & \textcolor{blue}{$46.686$} & $3.00$ \\
$3$ & $8$ & $T$ & $40.2485$ & $4.7381$ & $0.4341$ & \textcolor{blue}{$45.4207$} & $54.5888$ & $-15.1909$ & $7.3677$ & \textcolor{blue}{$46.7655$} & $2.96$ \\
$3$ & $8^{*}$ & $T$ & -- & -- & -- & -- & $54.5888$ & $-15.1909$ & $7.6355$ & \textcolor{blue}{$47.033$} & -- \\
$3$ & $9$ & $T$ & $40.2485$ & $4.7381$ & $0.5426$ & \textcolor{blue}{$45.5292$} & $54.5888$ & $-15.1909$ & $7.6015$ & \textcolor{blue}{$46.9994$} & $3.23$ \\
$3$ & $10$ & $T$ & $40.2485$ & $4.7381$ & $0.6632$ & \textcolor{blue}{$45.6498$} & $54.5888$ & $-15.1909$ & $7.6345$ & \textcolor{blue}{$47.0323$} & $3.03$ \\
$3$ & $12$ & $T$ & $40.2485$ & $4.7381$ & $0.9405$ & \textcolor{blue}{$45.9271$} & $54.5888$ & $-15.1909$ & $7.9005$ & \textcolor{blue}{$47.2984$} & $2.99$ \\
\bottomrule
\end{tabular}
}
\caption{Comparison for $m=2$ of charge $q$ monopoles in QED$_3^{1,-5/2}$ and charge $2q$ monopoles in effective sQED$_3^{1,3/2}$. The common Lorentz spin is $\ell$, and the splitting contributions are $\Delta_f^{\text{split}}$ and $\Delta_s^{\text{split}}$. At $N=1$, $\Delta_x=\Delta_x^{(0)}+\Delta_x^{(1)}+\Delta_x^{\text{split}}$ for $x=f,s$, and the error is $100|\Delta_f-\Delta_s|/|\Delta_f|$. One or two asterisks on $\ell$ denote the second or third operator with the same charge and spin; the levels on each side are ordered by energy. The total dimensions $\Delta_f$ and $\Delta_s$ are shown in blue to guide the eye. The Saddle column specifies the unbroken rotation group of the fermionic saddle. Dashes mark fermionic states that cannot be obtained from the non-spherical ground state.}
\label{tab:duality-m2}
\end{table}
	
	\begin{table}[htbp]
\centering
\small
\setlength{\tabcolsep}{5pt}
\renewcommand{\arraystretch}{1.15}
\resizebox{\textwidth}{!}{%
\begin{tabular}{@{}cccrrrrrrrrr@{}}
\toprule
 & & \multicolumn{5}{c}{Fermion} & \multicolumn{4}{c}{Scalar} & \\
\cmidrule(lr){3-7}\cmidrule(lr){8-11}
$q$ & $\ell$ & Saddle & $\Delta_f^{(0)}$ & $\Delta_f^{(1)}$ & $\Delta_f^{\text{split}}$ & $\Delta_f$ & $\Delta_s^{(0)}$ & $\Delta_s^{(1)}$ & $\Delta_s^{\text{split}}$ & $\Delta_s$ & Error (\%) \\
\midrule
$1/2$ & $0$ & $SO(3)$ & $4.5077$ & $0.2993$ & $0.0000$ & \textcolor{blue}{$4.8071$} & $6.2552$ & $-2.0937$ & $0.8165$ & \textcolor{blue}{$4.9780$} & $3.56$ \\
\specialrule{0.3pt}{1.5pt}{1.5pt}
$1$ & $0$ & $O(2)$ & $13.2012$ & $0.7494$ & $0.0000$ & \textcolor{blue}{$13.9506$} & $17.1381$ & $-5.5844$ & $2.1229$ & \textcolor{blue}{$13.6766$} & $1.96$ \\
$1$ & $2$ & $O(2)$ & $13.2012$ & $0.7494$ & $0.0765$ & \textcolor{blue}{$14.0270$} & $17.1381$ & $-5.5844$ & $2.8213$ & \textcolor{blue}{$14.3751$} & $2.48$ \\
$1$ & $4$ & $O(2)$ & $13.2012$ & $0.7494$ & $0.2549$ & \textcolor{blue}{$14.2054$} & $17.1381$ & $-5.5844$ & $2.7625$ & \textcolor{blue}{$14.3162$} & $0.78$ \\
\specialrule{0.3pt}{1.5pt}{1.5pt}
$3/2$ & $0$ & $D_{2}$ & $23.7722$ & $1.4312$ & $0.0000$ & \textcolor{blue}{$25.2033$} & $31.1397$ & $-10.0496$ & $4.6702$ & \textcolor{blue}{$25.760$} & $2.21$ \\
$3/2$ & $2$ & $D_{2}$ & $23.7722$ & $1.4312$ & $0.0296$ & \textcolor{blue}{$25.2329$} & $31.1397$ & $-10.0496$ & $4.3787$ & \textcolor{blue}{$25.4688$} & $0.93$ \\
$3/2$ & $3$ & $D_{2}$ & $23.7722$ & $1.4312$ & $0.0865$ & \textcolor{blue}{$25.2898$} & $31.1397$ & $-10.0496$ & $4.7662$ & \textcolor{blue}{$25.8563$} & $2.24$ \\
$3/2$ & $4$ & $D_{2}$ & $23.7722$ & $1.4312$ & $0.0978$ & \textcolor{blue}{$25.3012$} & $31.1397$ & $-10.0496$ & $4.5595$ & \textcolor{blue}{$25.6496$} & $1.38$ \\
$3/2$ & $6$ & $D_{2}$ & $23.7722$ & $1.4312$ & $0.2033$ & \textcolor{blue}{$25.4066$} & $31.1397$ & $-10.0496$ & $4.8355$ & \textcolor{blue}{$25.9256$} & $2.04$ \\
\specialrule{0.3pt}{1.5pt}{1.5pt}
$2$ & $0$ & $O(2)$ & $35.9381$ & $2.2584$ & $0.0000$ & \textcolor{blue}{$38.1966$} & $47.6755$ & $-15.312$ & $6.6578$ & \textcolor{blue}{$39.0217$} & $2.16$ \\
$2$ & $2$ & $O(2)$ & $35.9381$ & $2.2584$ & $0.0254$ & \textcolor{blue}{$38.2220$} & $47.6755$ & $-15.312$ & $6.6364$ & \textcolor{blue}{$39.000$} & $2.04$ \\
$2$ & $4$ & $O(2)$ & $35.9381$ & $2.2584$ & $0.0847$ & \textcolor{blue}{$38.2812$} & $47.6755$ & $-15.312$ & $6.7428$ & \textcolor{blue}{$39.1067$} & $2.16$ \\
$2$ & $6$ & $O(2)$ & $35.9381$ & $2.2584$ & $0.1778$ & \textcolor{blue}{$38.3744$} & $47.6755$ & $-15.312$ & $6.7479$ & \textcolor{blue}{$39.1118$} & $1.92$ \\
\specialrule{0.3pt}{1.5pt}{1.5pt}
$5/2$ & $4$ & $D_{8}$ & $49.5328$ & $3.2210$ & $0.0625$ & \textcolor{blue}{$52.8163$} & $66.4037$ & $-21.2644$ & $9.0922$ & \textcolor{blue}{$54.2315$} & $2.68$ \\
\bottomrule
\end{tabular}
}
\caption{Comparison for $m=3$ of charge $q$ monopoles in QED$_3^{1,-7/2}$ and charge $3q$ monopoles in effective sQED$_3^{1,4/3}$. The common Lorentz spin is $\ell$, and the splitting contributions are $\Delta_f^{\text{split}}$ and $\Delta_s^{\text{split}}$. At $N=1$, $\Delta_x=\Delta_x^{(0)}+\Delta_x^{(1)}+\Delta_x^{\text{split}}$ for $x=f,s$, and the error is $100|\Delta_f-\Delta_s|/|\Delta_f|$. The total dimensions $\Delta_f$ and $\Delta_s$ are shown in blue to guide the eye. The Saddle column specifies the unbroken rotation group of the fermionic saddle. }
\label{tab:duality-m3}
\end{table}

\begin{table}[htbp]
\centering
\small
\setlength{\tabcolsep}{6pt}
\renewcommand{\arraystretch}{1.15}
\begin{tabular}{@{}ccrrrrr@{}}
\toprule
$q$ & Saddle & $\Delta_f^{(0)}$ & $\Delta_f^{(1)}$ & $\Delta_f$ &
$\Delta_{O(2)}$ & Error (\%) \\
\midrule
$1/2$ & $SO(3)$ & $0.265096$ & $0.49283$ & $0.75792$ & $0.519130434$ & $46.00$ \\
\specialrule{0.3pt}{1.5pt}{1.5pt}
$1$ & $SO(3)$ & $0.673153$ & $1.20516$ & $1.87832$ & $1.23648971$ & $51.91$ \\
\specialrule{0.3pt}{1.5pt}{1.5pt}
$3/2$ & $SO(3)$ & $1.186434$ & $2.08310$ & $3.26953$ & $2.1086(3)$ & $55.06$ \\
\specialrule{0.3pt}{1.5pt}{1.5pt}
$2$ & $SO(3)$ & $1.786901$ & $3.09988$ & $4.88679$ & $3.11535(73)$ & $56.86$ \\
\specialrule{0.3pt}{1.5pt}{1.5pt}
$5/2$ & $SO(3)$ & $2.463451$ & $4.23867$ & $6.70212$ & $4.265(6)$ & $57.14$ \\
\specialrule{0.3pt}{1.5pt}{1.5pt}
$3$ & $SO(3)$ & $3.208372$ & $5.48756$ & $8.69594$ & $5.509(7)$ & $57.85$ \\
\specialrule{0.3pt}{1.5pt}{1.5pt}
$7/2$ & $SO(3)$ & $4.015906$ & $6.83761$ & $10.85352$ & $6.841(8)$ & $58.65$ \\
\specialrule{0.3pt}{1.5pt}{1.5pt}
$4$ & $SO(3)$ & $4.881539$ & $8.28174$ & $13.16328$ & $8.278(9)$ & $59.02$ \\
\specialrule{0.3pt}{1.5pt}{1.5pt}
$9/2$ & $SO(3)$ & $5.801615$ & $9.81417$ & $15.61579$ & $9.796(9)$ & $59.41$ \\
\specialrule{0.3pt}{1.5pt}{1.5pt}
$5$ & $SO(3)$ & $6.773088$ & $11.43009$ & $18.20318$ & $11.399(10)$ & $59.69$ \\
\bottomrule
\end{tabular}
\caption{Comparison for $m=0$ for charge $q$ monopoles in QED$_3^{1,-1/2}$. Here $\Delta_f=\Delta_f^{(0)}+\Delta_f^{(1)}$ is the extrapolation to $N=1$,
and the error is $100|\Delta_{O(2)}-\Delta_f|/\Delta_{O(2)}$.
The $O(2)$ values for $q\leq2$ were computed from conformal bootstrap \cite{Chester:2019ifh}, and for $q>2$ from a lattice simulation \cite{Banerjee:2017fcx}.
The Saddle column specifies the unbroken rotation group of the fermionic saddle; all entries use the spherical $SO(3)$ saddle. Digits reflect numerical precision. The zero-energy shell is completely filled for every $q$, so these monopoles have $\ell=0$ and $\Delta_f^{\text{split}}=0$.}
\label{tab:duality-m0}
\end{table}

\begin{table}[htbp]
\centering
\small
\setlength{\tabcolsep}{5pt}
\renewcommand{\arraystretch}{1.15}
\begin{tabular}{@{}ccrrrrrr@{}}
\toprule
 & & \multicolumn{4}{c}{Scalar} & Free fermion & \\
\cmidrule(lr){3-6}\cmidrule(lr){7-7}
$q$ & $\ell$ & $\Delta_s^{(0)}$ & $\Delta_s^{(1)}$ & $\Delta_s^{\text{split}}$ & $\Delta_s$ & $\Delta_{\mathrm{free}}$ & Error (\%) \\
\midrule
$1/2$ & $1/2$ & $1.0000$ & $-0.4688$ & $0.2770$ & $0.8082$ & $1$ & $19.18$ \\
\specialrule{0.3pt}{1.5pt}{1.5pt}
$1$ & $0$ & $2.5833$ & $-1.1820$ & $0.3261$ & $1.7274$ & $2$ & $13.63$ \\
$1$ & $2$ & $2.5833$ & $-1.1820$ & $1.1629$ & $2.5641$ & $3$ & $14.53$ \\
\specialrule{0.3pt}{1.5pt}{1.5pt}
$3/2$ & $3/2$ & $4.5873$ & $-2.0729$ & $1.0590$ & $3.5734$ & $4$ & $10.66$ \\
$3/2$ & $5/2$ & $4.5873$ & $-2.0729$ & $1.5288$ & $4.043$ & $5$ & $19.1$ \\
$3/2$ & $9/2$ & $4.5873$ & $-2.0729$ & $3.0322$ & $5.5466$ & $6$ & $7.56$ \\
\specialrule{0.3pt}{1.5pt}{1.5pt}
$2$ & $0$ & $6.9380$ & $-3.1123$ & $1.7519$ & $5.5776$ & $6$ & $7.04$ \\
$2$ & $2$ & $6.9380$ & $-3.1123$ & $1.4266$ & $5.2524$ & $6$ & $12.46$ \\
$2$ & $4$ & $6.9380$ & $-3.1123$ & $2.4507$ & $6.2764$ & $7$ & $10.34$ \\
$2$ & $5$ & $6.9380$ & $-3.1123$ & $3.1821$ & $7.0079$ & $8$ & $12.40$ \\
$2$ & $6$ & $6.9380$ & $-3.1123$ & $4.0599$ & $7.8856$ & $9$ & $12.38$ \\
$2$ & $8$ & $6.9380$ & $-3.1123$ & $6.2543$ & $10.0800$ & $10$ & $0.80$ \\
\specialrule{0.3pt}{1.5pt}{1.5pt}
$5/2$ & $1/2$ & $9.5904$ & $-4.282$ & $2.1434$ & $7.4520$ & $9$ & $17.20$ \\
$5/2$ & $3/2$ & $9.5904$ & $-4.282$ & $1.8231$ & $7.1316$ & $8$ & $10.85$ \\
$5/2$ & $5/2$ & $9.5904$ & $-4.282$ & $2.6323$ & $7.9409$ & $9$ & $11.77$ \\
$5/2$ & $7/2$ & $9.5904$ & $-4.282$ & $2.3716$ & $7.6802$ & $9$ & $14.66$ \\
$5/2$ & $9/2$ & $9.5904$ & $-4.282$ & $2.7328$ & $8.0414$ & $9$ & $10.65$ \\
$5/2$ & $11/2$ & $9.5904$ & $-4.282$ & $3.8121$ & $9.1207$ & $10$ & $8.79$ \\
$5/2$ & $13/2$ & $9.5904$ & $-4.282$ & $4.0357$ & $9.3442$ & $11$ & $15.05$ \\
$5/2$ & $15/2$ & $9.5904$ & $-4.282$ & $4.7620$ & $10.0706$ & $11$ & $8.45$ \\
$5/2$ & $17/2$ & $9.5904$ & $-4.282$ & $5.8008$ & $11.1094$ & $12$ & $7.42$ \\
$5/2$ & $19/2$ & $9.5904$ & $-4.282$ & $6.9619$ & $12.2705$ & $13$ & $5.61$ \\
$5/2$ & $21/2$ & $9.5904$ & $-4.282$ & $8.2452$ & $13.554$ & $14$ & $3.19$ \\
$5/2$ & $25/2$ & $9.5904$ & $-4.282$ & $11.1784$ & $16.4869$ & $15$ & $9.91$ \\
\bottomrule
\end{tabular}
\caption{Comparison of charge $q$ monopoles in sQED$_3^{1,1}$ with free-fermion primaries of charge $Q=2q$ and the same Lorentz spin $\ell$, for $q\leq5/2$. At each spin we show the lowest scalar level and the lowest free primary. The scalar estimate includes the splitting term, $\Delta_s=\Delta_s^{(0)}+\Delta_s^{(1)}+\Delta_s^{\text{split}}$, and the error is $100|\Delta_s-\Delta_{\mathrm{free}}|/\Delta_{\mathrm{free}}$. Digits reflect numerical precision.}
\label{tab:seed}
\end{table}

\begin{table}[htbp]
\centering
\small
\setlength{\tabcolsep}{5pt}
\renewcommand{\arraystretch}{1.15}
\begin{tabular}{@{}cccrrrrrr@{}}
\toprule
 & & \multicolumn{5}{c}{Fermion} & Free fermion & \\
\cmidrule(lr){3-7}\cmidrule(lr){8-8}
$q$ & $\ell$ & Saddle & $\Delta_f^{(0)}$ & $\Delta_f^{(1)}$ & $\Delta_f^{\text{split}}$ & $\Delta_f$ & $\Delta_{\mathrm{free}}$ & Error (\%) \\
\midrule
$1$ & $1/2$ & $SO(3)$ & $0.6732$ & $-0.1934$ & $0.4335$ & $0.9132$ & $1$ & $8.68$ \\
\specialrule{0.3pt}{1.5pt}{1.5pt}
$2$ & $0$ & $SO(3)$ & $1.7869$ & $-0.705$ & $0.9368$ & $2.019$ & $2$ & $0.9$ \\
$2$ & $2$ & $SO(3)$ & $1.7869$ & $-0.705$ & $1.7066$ & $2.789$ & $3$ & $7.0$ \\
\specialrule{0.3pt}{1.5pt}{1.5pt}
$3$ & $3/2$ & $SO(3)$ & $3.2084$ & $-1.408$ & $2.2234$ & $4.02$ & $4$ & $0.6$ \\
$3$ & $5/2$ & $SO(3)$ & $3.2084$ & $-1.408$ & $2.7780$ & $4.578$ & $5$ & $8.4$ \\
$3$ & $9/2$ & $SO(3)$ & $3.2084$ & $-1.408$ & $4.0927$ & $5.893$ & $6$ & $1.8$ \\
\specialrule{0.3pt}{1.5pt}{1.5pt}
$4$ & $0$ & $SO(3)$ & $4.8815$ & $-2.26$ & $3.6716$ & $6.291$ & $6$ & $4.85$ \\
$4$ & $2$ & $SO(3)$ & $4.8815$ & $-2.26$ & $3.3193$ & $5.94$ & $6$ & $1.0$ \\
$4$ & $2^{*}$ & $SO(3)$ & $4.8815$ & $-2.26$ & $3.9006$ & $6.520$ & $7$ & $6.86$ \\
$4$ & $4$ & $SO(3)$ & $4.8815$ & $-2.26$ & $4.5417$ & $7.161$ & $7$ & $2.3$ \\
$4$ & $4^{*}$ & $SO(3)$ & $4.8815$ & $-2.26$ & $4.8023$ & $7.42$ & $8$ & $7.23$ \\
$4$ & $5$ & $SO(3)$ & $4.8815$ & $-2.26$ & $5.0884$ & $7.71$ & $8$ & $3.7$ \\
$4$ & $6$ & $SO(3)$ & $4.8815$ & $-2.26$ & $6.1162$ & $8.74$ & $9$ & $2.94$ \\
$4$ & $8$ & $SO(3)$ & $4.8815$ & $-2.26$ & $7.8537$ & $10.473$ & $10$ & $4.73$ \\
\bottomrule
\end{tabular}
\caption{Comparison for Son's duality between charge $q$ monopoles in QED$_3^{1,0}$ and free-fermion primaries of charge $Q=q$ and the same Lorentz spin $\ell$. We include every retained-shell multiplet for integer $q=1,\ldots,4$, with $\Delta_f=\Delta_f^{(0)}+\Delta_f^{(1)}+\Delta_f^{\text{split}}$. The error is $100|\Delta_f-\Delta_{\mathrm{free}}|/\Delta_{\mathrm{free}}$. An asterisk on $\ell$ denotes the second operator with the same charge and spin; the spectra are paired in increasing energy order. The Saddle column specifies the unbroken rotation group of the fermionic saddle; all entries use the spherical $SO(3)$ saddle. Digits reflect numerical precision.}
\label{tab:son}
\end{table}

\clearpage
\section{Conclusion}\label{sec:conclusion}

We computed the scaling dimensions of charge $q\in\mathbb{Z}/2$ monopole operators in QED$_3$ in the large $N,k$ limit at fixed $\kappa=k/N$ to subleading order. We also computed the energy splitting term of general large $N$ degenerate monopoles for both QED$_3$ and sQED$_3$. For fully filled Landau levels or $|k|\leq1/2$, we used the same spherical saddle considered in earlier work. Otherwise, we found this saddle was unstable and so instead considered non-spherical saddles. We then used these results extrapolated to $N=1$ to check various dualities, as summarized in the introduction. For the $m=1$ duality between sQED$_3^{1,2}$ and QED$_3^{1,-3/2}$, the $q=1/2$ monopoles match the expected extra currents for the $SO(3)$ enhancement of the naive $U(1)$ symmetry, for $q=1$ the two monopoles at spins $0$ and $2$ match non-perturbative calculations from the fuzzy sphere \cite{Zhou:2025rmv}, and for general $q$ they match each other to high precision. For $m=2,3$, we find accurate matches between the dual theories for all $q$ and spins. For Son's duality between QED$_3^{1,0}$ and a free fermion, we also found a good match for all $q$ and spins. For the seed duality between a free fermion and sQED$_3^{1,1}$ we found a somewhat less accurate match, while for the $m=0$ duality we did not find a match against scaling dimensions in the dual critical $O(2)$ model. 

One limitation of the non-spherical saddles is that they do not describe the full set of states that spherical saddles do. For instance, for $k=-3/2$, which is relevant to the $m=1$ duality, we could not compute the states with $q=2,\ell=0$ or $q=3,\ell=2$, which would appear in the spherical saddle, and which appear in the dual $k=2$ sQED$_3$ spherical saddle. These states can in principle be found by considering excitations above the non-spherical saddle vacuum, which it would be interesting to consider in future work. Excited states could also be used to check the $SO(3)$ symmetry by pairing these states with ground states in the corresponding $SO(3)$ multiplet.

It would be nice if the fuzzy sphere \cite{Zhou:2025rmv} could be used to extract the scaling dimension of operators with $q>1$, so that we can compare against the predictions made in this paper. This is somewhat challenging at present, as these scaling dimensions are large, while the fuzzy sphere works best for smaller scaling dimensions. It would be interesting if fuzzy sphere could also be applied to some of the other gauge theories considered in this work.

It would be interesting to better understand why the $m=0$ duality does not match the critical $O(2)$ model for any $q$. Curiously, we observed that if we drop the leading term and only consider the subleading correction, then this remarkably matches the dual critical $O(2)$ values to high accuracy. It would be nice to find an explanation for this.

\begin{figure}[!htb]
\centering
\includegraphics[width=0.92\textwidth]{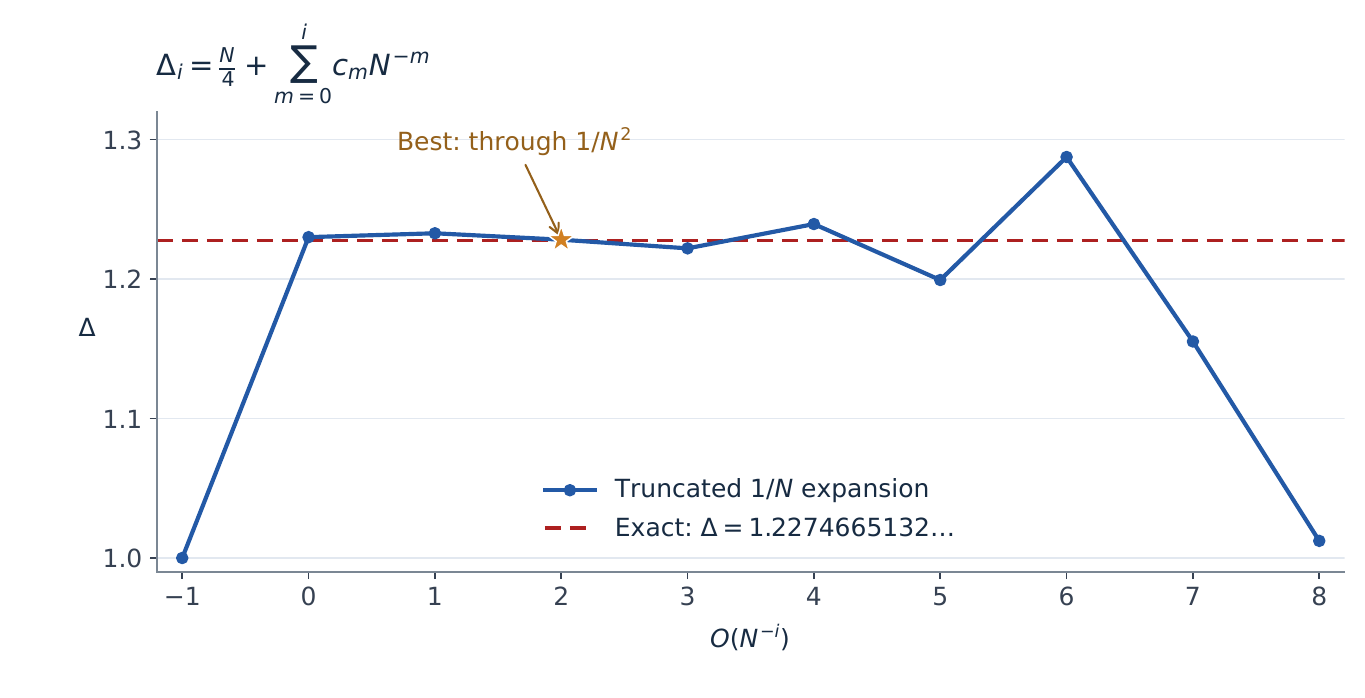}
\caption{Scaling dimension of the BPS $q=1/2$ monopole in $\mathcal N=2$ SQED$_3$ with $N=4$, $k=1$, three chiral multiplets of charge $+1$, and one of charge $-1$. The large $N$ expansion is taken at fixed matter fractions and $k/N=1/4$. The blue points include terms through $O(N^{-i})$, with $i=-1$ denoting the leading result $N/4$. The dashed line is the exact localization result, evaluated numerically; the star marks the closest truncation shown, through $1/N^2$. Both the partial sums and the exact result are computed from the localized partition function described in the text.}
\label{fig:susy-monopole-truncation}
\end{figure}

The energy splitting term appears at order $1/N$, alongside the standard state-independent correction. Computing the latter would let us assess the complete $1/N$ correction to the duality comparisons. This is motivated by the BPS monopole example in Figure~\ref{fig:susy-monopole-truncation}, whose finite $N$ dimension follows from localization and $F$-maximization \cite{Jafferis:2010un,Klebanov:2011td}.\footnote{Specifically, we numerically maximize $F=-\log|Z|$ using Eq.~(4.28) of \cite{Klebanov:2011td}, with their $(N,\widetilde N,k)=(3,1,-1)$ and the choice of vanishing topological mixing parameter used in that equation. Their $N$ and $\widetilde N$ count positively and negatively charged chiral multiplets separately, and their localization level has the opposite sign to our $k$. In terms of their extremized matter $R$-charges $\Delta$ and $\widetilde\Delta$, the gauge-invariant BPS bare monopole has dimension $\Delta_{1/2}=\frac12[3(1-\Delta)+(1-\widetilde\Delta)]\simeq1.2274665132$ \cite{Borokhov:2002cg}.} The first few corrections substantially improve the leading approximation, and the result through $1/N^2$ is closer to the exact value than the one through $N^0$. Eventually higher orders worsen the result as expected for an asymptotic series.

One can also consider the large $q$ expansion of the monopole scaling dimensions. The large charge expansion predicts a universal EFT given by a superfluid that is supposed to describe all parity preserving $U(1)$ CFTs in 3d, and in particular predicts a universal $q^0$ \cite{Hellerman:2015nra}. For the $k=0$ QED$_3$  \cite{Dupuis:2021flq}  and sQED$_3$  \cite{DeLaFuente:2018uee} , the large $N$ expansion recovers this $q^0$ term from $N^0$ calculation. The large $q$ expansion of energy splitting was also recently discussed in \cite{Dumitrescu:2025vfp}. The $k\neq0$ calculations in this work seem to give the same $q^0$ term, even though parity is broken. We leave a deeper explanation of this to future work.

Finally, the subleading calculation can also be generalized to QCD$_3$ with $U(N)$ gauge group and $k\neq0$, which was already studied for $k=0$ in \cite{Dyer:2013fja}, as well as $\mathcal{N}=1$ QED$_3$. In particular, we can check if extrapolating to small $N,k$ might also provide evidence for dualities in those cases \cite{Aharony:2016jvv,Benini:2018umh}.

\clearpage
\section*{Acknowledgments} 
We thank Adar Sharon, Mark Mezei, Silviu Pufu, Zohar Komargodski, Rishi Mouland, William Witczak-Krempa, Rufus Boyack, Chong Wang, Yin-Chen He, Zheng Zhou, and Ofer Aharony for useful discussions. SMC is supported by the Royal Society under the grant URF\textbackslash R1\textbackslash 221310 and the UK Engineering and Physical Sciences Research Council grant number EP/Z000106/1. The computations presented here were conducted on the facilities provided by the Imperial College Research Computing Service (\url{https://doi.org/10.14469/hpc/2232}). We acknowledge the use of Claude (Anthropic) and ChatGPT (OpenAI) as assistive tools for coding.

\appendix \addtocontents{toc}{\protect\setcounter{tocdepth}{1}}

\section{Asymptotic expansion }\label{asymp}

In this appendix we will discuss the asymptotic expansion at large $\ell,\omega$ for the integrand $L_\ell^{q,\kappa}(\omega)$ in \eqref{freeFinalReg2}. We will closely follow the strategy introduced for sQED$_3$ in Appendix C of \cite{Dyer:2015zha}. In particular, we will first expand the CP preserving Green's function $G(x,x')$ at small distance. We will then plug this into the expression for the CP preserving kernels and take Fourier transforms to get their large $\ell,\omega$ expansion. Finally, we will combine these with the explicit expansion of the CP violating kernels to get the asymptotic of $L_\ell^{q,\kappa}(\omega)$ to several orders. Throughout this appendix, we will work at $\beta\to\infty$.

\subsection{Small distance expansion of CP preserving Green's function}
\label{smallGreen}

The single fermion CP preserving Green's function in \eqref{TempGreen} satisfies the differential equation \eqref{GreenDefF}, where we set the holonomy to zero, i.e. $\mathcal{A}=q(1-\cos\theta)d\phi$. It is convenient to change variables to
\es{st}{
t=\sinh^2\frac\tau2\,,\qquad s=\sin^2\frac\theta2\,,\qquad X=\sqrt{s+t}\,,
}
and then solve the differential equation at small $s,t$. The singular terms to the lowest few orders are
\es{eq:uv-gcp-master}{
G(x,0)=\sum_{a=0}^{3}G^{(a)}(x,0)
  +\mathcal C_q  \frac{i}{8\pi}
 \begin{pmatrix}
  2\sqrt t\,\sgn\tau&-\sqrt s\,e^{-i\phi}\\
  -\sqrt s\,e^{i\phi}&-2\sqrt t\,\sgn\tau
 \end{pmatrix} +O(X^2)\,,
}
where the singular particular solutions are
\begingroup
\small
\begin{align}
 G^{(0)}={}&\frac{i}{16\pi X^3}
 \begin{pmatrix}
  \sqrt t\,\sgn\tau&\sqrt s\,e^{-i\phi}\\
  \sqrt s\,e^{i\phi}&-\sqrt t\,\sgn\tau
 \end{pmatrix},
 \label{eq:uv-gcp0}\\[3pt]
 G^{(1)}={}&\frac{i\sqrt s}{16\pi X^3}
 \begin{pmatrix}
  -\sqrt s&\sqrt t\,e^{-i\phi}\sgn\tau\\
  \sqrt t\,e^{i\phi}\sgn\tau&\sqrt s
 \end{pmatrix},
 \label{eq:uv-gcp1}\\[3pt]
 G^{(2)}={}&\frac{i}{32\pi X^3}
 \begin{pmatrix}
  2\sqrt t\,\sgn\tau\big[(2q-1)s+2qt\big]
  &-\sqrt s\,e^{-i\phi}(s-t)\\
  -\sqrt s\,e^{i\phi}(s-t)
  &2\sqrt t\,\sgn\tau\big[(2q+1)s+2qt\big]
 \end{pmatrix},
 \label{eq:uv-gcp2}\\[3pt]
 G^{(3)}={}&\frac{i}{32\pi X^3}
 \begin{pmatrix}
  -st&-\sqrt{st}\,e^{-i\phi}\sgn\tau
       \big[(4q+1)s+4qt\big]\\
  \sqrt{st}\,e^{i\phi}\sgn\tau
       \big[(4q-1)s+4qt\big]&st
 \end{pmatrix}\,,
 \label{eq:uv-gcp3}
\end{align}
\endgroup
while the coefficient $\mathcal{C}_q$ of the homogeneous solution is not fixed by the differential equation. At higher order, more such unfixed coefficients will appear. To fix them, we consider the spectral decomposition of the Green's function in \eqref{TempGreen}, which for $s=0$ can be written as
\es{specG}{
G(\tau,0)=\frac{i\sgn\tau}{4\pi}
 \Bigg[
  \frac q2(\mathbf1+\sigma_3)
  +\sigma_3\sum_{j=q+1/2}^{\infty}
   \left(j+\frac12\right)e^{-\lambda_j|\tau|}
 \Bigg]\,.
}
We can expand this expression at small $\tau$ and compare to \eqref{eq:uv-gcp-master} to fix the coefficients of the homogeneous terms. However, if the exponential in \eqref{specG} is expanded first, the coefficient of
$|\tau|^n$ contains
$\sum_j(j+\tfrac12)\lambda_j^n$, which diverges.  Consequently neither the
finite coefficient nor its sign can be obtained by naively interchanging the
small-$\tau$ expansion and the spectral sum.  A convenient subtraction,
sufficient through $t^{3/2}$, is
\begingroup
\small
\begin{align}
 \mathcal S_q(|\tau|)={}&\sum_{j=q+1/2}^{\infty}
  \left(j+\frac12\right)e^{-\lambda_j |\tau|}\,,\qquad 
 \mathcal S_q^{\rm sub}(|\tau|)={}&
 \sum_{j=q+1/2}^{\infty}e^{-(j+1/2)|\tau|}
 \left[
  j+\frac12+\frac{q^2|\tau|}{2}
  +\frac{q^4|\tau|\left(1+(j+\frac12)|\tau|\right)}
  {8(j+\frac12)^2}
 \right].
 \label{eq:uv-Sq-sub}
\end{align}
\endgroup
Equation~\eqref{eq:uv-Sq-sub} is just the large-$j$ expansion of the exact
summand before the small-$|\tau|$ expansion. The difference
$\mathcal S_q-\mathcal S_q^{\rm sub}$, together with the derivatives needed
through order $|\tau|^3$, is absolutely summable at $\tau=0$, so that difference may be
expanded term by term.  Adding the two pieces and rewriting
$|\tau|=2\operatorname{arsinh}\sqrt t$ gives
\begin{equation}
 \mathcal S_q(|\tau|)
 =\frac{1}{4t}-\frac q2-\frac{q^2}{3}t
  +\mathcal F_q^{\infty}\sqrt t
  +\mathcal F_q^{(2)}t^{3/2}+O(t^2),
 \label{eq:uv-Sq-small-t}
\end{equation}
where the first two globally determined coefficients are
\begin{align}
 \mathcal F_q^{\infty}
 ={}&-\sum_{j=q+1/2}^{\infty}(2j+1)\lambda_j\Big|_{\zeta},
 \label{eq:Fsea}\\
 \mathcal F_q^{(2)}
 ={}&\frac13\sum_{j=q+1/2}^{\infty}
 \left(j+\frac12\right)\lambda_j
 \left(1-4\lambda_j^2\right)\Big|_{\zeta}.
 \label{eq:Fsea2}
\end{align}
The $\zeta$ symbol here is shorthand for the analytic value defined by the
large-$j$ subtraction just described, which in practice is just zeta regularization.  Keeping more terms in the large-$j$ subtraction
fixes every later odd-order spectral moment in the same way. We can plug \eqref{eq:uv-Sq-small-t} into the exact decomposition
\eqref{specG} to cancel the $-q\sigma_3/2$ from the excited
levels against the $q\sigma_3/2$ in the lowest level and get
\begin{equation}
 G_q^{\CP}(\tau,0)=\frac{i\sgn\tau}{4\pi}
 \left[\frac{\sigma_3}{4t}+\frac q2\mathbf1
 -\frac{q^2t}{3}\sigma_3
 +\sqrt t\,\mathcal F_q^{\infty}\sigma_3+O(t^{3/2})\right].
 \label{eq:uv-g-north}
\end{equation}
Comparison with \eqref{eq:uv-gcp-master} then fixes
\begin{equation}
 \mathcal C_q=\mathcal F_q^{\infty}\,.
 \label{eq:uv-gcp-hom}
\end{equation}
It also fixes the allowed order-two homogeneous constant to zero; at order
five the new regular coefficient is $\mathcal F_q^{(2)}$.  That later moment
first contributes beyond the determinant order retained here.  This also supplies an independent check of the signs of the
off-diagonal spinor entries in \eqref{eq:uv-gcp0}--\eqref{eq:uv-gcp3}.

\subsection{UV asymptotics of kernels}
\label{UVkernels}

We can now plug the small distance expansion of the CP preserving Green's function into \eqref{Ks} and \eqref{fourierKernscal} (in the large $\beta$ limit) to compute the Fourier space kernels at large $\omega$ and $\ell$. The Fourier transforms of each order in $s,t$ are described in C.2 of \cite{Dyer:2015zha}, to which we direct the reader for further detail. The resulting nonzero kernels are:
\es{Kasymp}{
 \widetilde{K}_{q,\ell,CP}^{UU}(\omega)
=&\frac{\ell[(\ell+1)^2+\omega^2]}{2(2\ell+1)}D_\ell
+\frac{\mathcal F_q^\infty}{4\pi ((\ell+1/2)^2+\omega^2)}
+\frac{\mathcal F_q^\infty((\ell+1/2)^2-\omega^2)}{8\pi(\ell+1/2) ((\ell+1/2)^2+\omega^2)^2}\\
&+\frac{q^2((\ell+1/2)^2-4\omega^2)}{32((\ell+1/2)^2+\omega^2)^{5/2}}+O\big([(\ell+1/2)^2
 +\omega^2]^{-2}\big)\,,\\
 \widetilde{K}_{q,\ell,CP}^{VV}(\omega)
=&\frac{\ell^2+\omega^2}{2}D_{\ell-1}
+\frac{\mathcal F_q^\infty(\omega^2-2(\ell+1/2)^2)}{2\pi ((\ell+1/2)^2+\omega^2)^2}
+\frac{q^2(3(\ell+1/2)^2-4\omega^2)}{16((\ell+1/2)^2+\omega^2)^{5/2}}
 +O\big([(\ell+1/2)^2
 +\omega^2]^{-2}\big)\,,
}
where we expand $D_\ell(\omega)$ in \eqref{D} as
\begin{equation}
 \begin{aligned}
 D_\ell(\omega)={}&
 \frac1{8\sqrt{\left(\ell+\frac12\right)^2+\omega^2}}
 +\frac{\omega^2-\left(\ell+\frac12\right)^2}
 {64\left[\left(\ell+\frac12\right)^2+\omega^2\right]^{5/2}}\\
 &+\frac{11\left(\ell+\frac12\right)^4
 -62\left(\ell+\frac12\right)^2\omega^2+11\omega^4}
 {1024\left[\left(\ell+\frac12\right)^2+\omega^2\right]^{9/2}}
 +O\big([(\ell+\frac12)^2
 +\omega^2]^{-7/2}\big)\,.
 \end{aligned}
\end{equation}

For the CP violating kernels, since they are already finite sums, we can just expand them directly in $\omega,\ell$ without need to first expand the Green's function. For the lowest level terms in \eqref{NCPfin1}, we find
\es{zeroExp}{
\widetilde{K}^{UV,0}_{q,\ell,NCP}(\omega)
=&
 \frac{q}{2\sqrt{2}\pi}
 \frac{\ell+\frac12+i\omega}
 {\left(\ell+\frac12\right)^2+\omega^2}
 +\frac{q}{8\sqrt{2}\pi}
 \frac{\ell+\frac12-i\omega}
 {\left(\ell+\frac12\right)
  \left[\left(\ell+\frac12\right)^2+\omega^2\right]}
\\
&+\frac{q}
 {64\sqrt{2}\pi
  \left[\left(\ell+\frac12\right)^2+\omega^2\right]^3}
 \Bigg[
 3\left(\ell+\frac12\right)^3
 +(30-64q)\omega^2\left(\ell+\frac12\right)
 -5\frac{\omega^4}{\ell+\frac12}
\\
&\qquad\qquad
 +7i\omega\left(\ell+\frac12\right)^2
 -i\omega^3(64q-38)
 -i\frac{\omega^5}{\left(\ell+\frac12\right)^2}
 \Bigg]
 +O\big([(\ell+\frac12)^2
 +\omega^2]^{-2}\big)\,.
}
For the cross term between higher level CP violating Green's function and the lowest level of the CP preserving Green's function \eqref{NCPfin2}, the sum over $j'$ must satisfy $j'-q+1/2\leq \ell\leq q+j'-1/2$, so the large $\ell,\omega$ expansion vanishes for fixed $j',q$. Finally, we can consider the sum in \eqref{NCPff} over the cross term between two higher level CP violating Green's functions given in \eqref{NCPfin3}, for which we find
\es{NCPexp}{
&\sum_{j=q+1/2}^{j'+\ell+1}\widetilde{K}^{UU,(j,j')}_{q,\ell,NCP}(\omega)=\frac{\left(j'+\frac12\right)\lambda_{j'}}
 {2\pi\left[\left(\ell+\frac12\right)^2+\omega^2\right]}
 +\frac{\left(j'+\frac12\right)\lambda_{j'}
 \left[\left(\ell+\frac12\right)^2-\omega^2\right]}
 {4\pi\left(\ell+\frac12\right)
  \left[\left(\ell+\frac12\right)^2+\omega^2\right]^2}
 +O\big([(\ell+\frac12)^2
 +\omega^2]^{-2}\big)\,,\\
 &\sum_{j=q+1/2}^{j'+\ell+1}\widetilde{K}^{VV,(j,j')}_{q,\ell,NCP}(\omega)= \frac{\left(j'+\frac12\right)\lambda_{j'}}{\pi}
 \frac{\omega^2-2\left(\ell+\frac12\right)^2}
 {\left[\left(\ell+\frac12\right)^2+\omega^2\right]^2}
 +O\big([(\ell+\frac12)^2
 +\omega^2]^{-2}\big)\,,\\
  &\sum_{j=q+1/2}^{j'+\ell+1}\widetilde{K}^{UV,(j,j')}_{q,\ell,NCP}(\omega)= -\frac{\sqrt2\,q\left(j'+\frac12\right)}{\pi}
 \frac{\omega^2\left(\ell+\frac12+i\omega\right)}
 {\left[\left(\ell+\frac12\right)^2+\omega^2\right]^3}
 +O\big([(\ell+\frac12)^2
 +\omega^2]^{-2}\big)\,.
}

We finally take all these expanded kernels and plug them into the determinant in \eqref{freeFinal} as written in \eqref{freeFinalReg2} to get 
\es{LasApp}{
 L_\ell^{q,\kappa}(\omega)=&
 \frac{128\kappa q(2\xi_{q-1/2}-1)}
 {(\pi^2+64\kappa^2)
  \left[\left(\ell+\frac12\right)^2+\omega^2\right]}
 +\frac{8\pi\mathcal F_q^{\infty}}{\pi^2+64\kappa^2}
 \frac{2\omega^2-(\ell+1/2)^2}
 {
  \left[\left(\ell+\frac12\right)^2+\omega^2\right]^{5/2}}\\
 &+\frac{8\pi \sum_{j'=q+1/2}^{\widetilde j}
 \xi_{j'}(2j'+1)\lambda_{j'}}{\pi^2+64\kappa^2}
 \frac{2\omega^2-\left(\ell+\frac12\right)^2}
 {\left[\left(\ell+\frac12\right)^2+\omega^2\right]^{5/2}}+
 \frac{256\kappa q}{\pi^2+64\kappa^2}
 \frac{\omega^2\sum_{j'=q+1/2}^{\widetilde j}
 \xi_{j'}(2j'+1)}
 {\left[\left(\ell+\frac12\right)^2+\omega^2\right]^3}\\
 &+L^{(4)}_{\ell,0}(\omega) +O\big([(\ell+\frac12)^2
 +\omega^2]^{-5/2}\big)\,,
}
where
\es{ug}{
&L_{\ell,0}^{(4)}(\omega)
 ={}\frac{q}{(\pi^2+64\kappa^2)^2
 \left[\left(\ell+\frac12\right)^2+\omega^2\right]^3}\\[-2pt]
 &\quad\times\Bigg[\left(\ell+\frac12\right)^2\Big[
 q\big(4\pi^2(\pi^2+64\kappa^2)
 +64(2\xi_{q-1/2}-1)^2(\pi^2-64\kappa^2)\big)
 +32\kappa(2\xi_{q-1/2}-1)(\pi^2+64\kappa^2)\Big]\\
 &\quad+4\omega^2\Big[-q\big(2\pi^2(\pi^2+64\kappa^2)
 +16(2\xi_{q-1/2}-1)^2(64\kappa^2-\pi^2)\big)
 +8\kappa(2\xi_{q-1/2}-1)(\pi^2+64\kappa^2)(5-8q)\Big]\Bigg].
}
We can find explicit higher orders for this asymptotic, as well as the asymptotic used in the sQED$_3$ case in \cite{Chester:2022wur}, in the attached \texttt{Mathematica} notebook.

\bibliographystyle{JHEP}
\bibliography{QED3CS}

\end{document}